\RequirePackage[T1]{fontenc}
\documentclass[12pt]{article}
\usepackage[height=8.85in,width=6.45in]{geometry}

\usepackage[utf8]{inputenc}

\usepackage{amsmath}
\usepackage{amssymb}
\usepackage{mathtools}
\usepackage{amsthm}

\usepackage{times}
\usepackage[scaled]{couriers}
\usepackage{mathrsfs}

\usepackage{graphicx}
\usepackage{subcaption}

\usepackage{tikz}
\usepackage{tikz-cd}
\usetikzlibrary{calc}
\usepackage[all]{xy}

\usepackage{colortbl}
\definecolor{lightyellow}{rgb}{1.0, 0.95, 0.7}

\usepackage[svgnames]{xcolor}
\usepackage[colorlinks,linktocpage=true,citecolor=DarkGreen,linkcolor=FireBrick]{hyperref}
\usepackage{cite}

\usepackage{bm}
\usepackage{slashed}
\usepackage{braket}

\numberwithin{equation}{section}

\theoremstyle{plain}

\theoremstyle{definition}

\numberwithin{thm}{section}

\def\i{{\mathsf i}}

\def\Im{\mathop{\mathrm{Im}}}
\def\Ker{\mathop{\mathrm{Ker}}}

\def\Hom{\mathop{\mathrm{Hom}}}
\def\Ext{\mathop{\mathrm{Ext}}}
\def\Sqtwo{\mathop{\mathrm{Sq}^{2}}}

\def\cN{{\cal N}}

\def\bA{{\mathbb A}}

\def\bC{{\mathbb C}}

\def\bP{{\mathbb P}}
\def\bQ{{\mathbb Q}}
\def\bR{{\mathbb R}}

\def\bZ{{\mathbb Z}}

\def\so{{\mathsf o}}

\def\su{{\mathsf u}}

\def\U{\mathrm{U}}
\def\SU{\mathrm{SU}}

\def\Spin{\mathrm{Spin}}

\def\u{\mathfrak{u}}
\def\su{\mathfrak{su}}

\def\so{\mathfrak{so}}

\def\e{\mathfrak{e}}
\def\g{\mathfrak{g}}
\def\h{\mathfrak{h}}
\def\t{\mathfrak{t}}

\newcommand{\UE}[3]{{}^{G}E_{#1}^{#2,#3}}
\newcommand{\TE}[3]{{}^{T}E_{#1}^{#2,#3}}
\newcommand{\KE}[3]{{}^{K}E_{#1}^{#2,#3}}
\newcommand{\HE}[3]{{}^{H}E_{#1}^{#2,#3}}
\newcommand{\Ud}[3]{{}^{G}d_{#1}^{#2,#3}}
\newcommand{\Td}[3]{{}^{T}d_{#1}^{#2,#3}}
\newcommand{\Kd}[3]{{}^{K}d_{#1}^{#2,#3}}
\newcommand{\Hd}[3]{{}^{H}d_{#1}^{#2,#3}}
\newcommand{\ssf}[3]{f_{#1}^{#2,#3}}
\newcommand{\ssg}[3]{g_{#1}^{#2,#3}}
\newcommand{\ssh}[3]{h_{#1}^{#2,#3}}
\newcommand{\CE}[3]{{}^{C}E^{#1}_{#2,#3}}
\newcommand{\PE}[3]{{}^{P}E^{#1}_{#2,#3}}
\newcommand{\DE}[3]{{}^{D}E^{#1}_{#2,#3}}
\newcommand{\QE}[3]{{}^{Q}E^{#1}_{#2,#3}}
\newcommand{\Cd}[3]{{}^{C}d^{#1}_{#2,#3}}
\newcommand{\Pd}[3]{{}^{P}d^{#1}_{#2,#3}}
\newcommand{\Dd}[3]{{}^{D}d^{#1}_{#2,#3}}
\newcommand{\Qd}[3]{{}^{Q}d^{#1}_{#2,#3}}

\newcommand{\ce}[1]{c_{#1}^{(7)}}
\newcommand{\cfi}[1]{c_{#1}^{(5)}}
\newcommand{\cth}[1]{c_{#1}^{(3)}}
\newcommand{\yp}[1]{y'_{#1}}
\newcommand{\zp}[1]{z'_{#1}}
\newcommand{\afi}[1]{a_{#1}}
\newcommand{\bth}[1]{b_{#1}}
\newcommand{\acfi}[1]{\check{a}_{#1}}
\newcommand{\cpfi}[1]{c_{#1}^{\prime(5)}}
\newcommand{\cpth}[1]{c_{#1}^{\prime(3)}}
\newcommand{\coroot}[1]{\alpha^{\vee}_{#1}}
\newcommand{\fcoroot}[1]{\alpha^{\su_{5}\vee}_{#1}}
\newcommand{\tcoroot}[1]{\alpha^{\su_{3}\vee}_{#1}}
\newcommand{\ocoroot}{\alpha^{\u_{1}\vee}}
\newcommand{\fweight}[1]{\omega^{\su_{5}}_{#1}}
\newcommand{\tweight}[1]{\omega^{\su_{3}}_{#1}}

\newcommand{\ocoroota}{\alpha^{\u_{1}^{(1)}\vee}}
\newcommand{\ocorootb}{\alpha^{\u_{1}^{(2)}\vee}}
\newcommand{\ocorootc}{\alpha^{\u_{1}^{(3)}\vee}}
\newcommand{\ch}[1]{\mathrm{ch}_{#1}}

\begin{document}

\begin{titlepage}

\begin{flushright}
TU-1302
\end{flushright}

\vskip 3cm

\begin{center}

{\Large \bfseries Anomalies in family unification models\\ from bordism classification}

\vskip 1cm
Tsubasa Sugeno and Hiroki Wada
\vskip 1cm

\begin{tabular}{ll}
Department of Physics, Tohoku University, Sendai 980-8578, Japan
\end{tabular}

\vskip 1cm

\end{center}

\noindent
We study anomalies in family unification models within the framework of the bordism classification of invertible field theories.
These models are based on four-dimensional $\cN=1$ supersymmetric nonlinear sigma models, in which the three generations of quarks and leptons arise as superpartners of the sigma model fields.
We focus on models whose target spaces are constructed from the exceptional group $E_{7}$ and its subgroups.
For the consistency of the theory, sigma model anomalies must be cancelled.
We show the absence of global sigma model anomalies, which are encoded in the torsion part of the relevant bordism groups, by explicitly computing these groups using the Atiyah-Hirzebruch spectral sequence.
In constructing family unification models, symmetries acting on the coset spaces are gauged, which may introduce additional anomalies.
We identify the relevant bordism groups in this setting and demonstrate that no global anomalies arise when the isotropy subgroup of the coset space is gauged.

\end{titlepage}

\setcounter{tocdepth}{3}


\tableofcontents

\section{Introduction and summary}
Anomaly cancellation is one of the most fundamental constraints in the formulation of a consistent quantum field theory.
Even when a theory appears to be well-defined, for instance, through a local Lagrangian, its partition function or correlation functions may fail to be well-defined due to anomalies.
To avoid such inconsistencies, one must ensure the absence of anomalies in the theory.
A prototypical example of such an inconsistency is a gauge anomaly arising from fermions coupled to a gauge field~\cite{Georgi:1972bb,Gross:1972pv}.
The path integral of fermions coupled to a gauge field may not be defined in a gauge invariant manner.
In such cases, gauge symmetry cannot be regarded as a redundancy of the theory, implying that the gauge theory itself is ill-defined.
It is well known that perturbative gauge anomalies, which are associated with infinitesimal gauge transformations, can be characterized by the anomaly polynomial~\cite{Alvarez-Gaume:1983ihn,Alvarez-Gaume:1984zlq}.

Over the past decade, our understanding of anomalies has advanced significantly.
In particular, global anomalies have been systematically incorporated into the modern framework of anomalies, alongside the conventional perturbative anomalies.
It is now understood that fermions possibly coupled to background fields can be realized as boundary modes of massive fermions defined on a higher-dimensional manifold~\cite{Witten:2015aba,Witten:2019bou}.
Within this framework, anomalies are interpreted as the dependence of the partition function on the choice of higher-dimensional manifolds.
More precisely, an anomaly in a $d$-dimensional theory is characterized by an invertible field theory in $(d+1)$-dimensions~\cite{Freed:2004yc}.
It has been established that such invertible field theories are classified by appropriate bordism groups~\cite{Kapustin:2014tfa,Kapustin:2014dxa,Freed:2016rqq,Yonekura:2018ufj,Yamashita:2021cao}.
In other words, possible anomalies in the theory are classified by the corresponding bordism groups.

The study of anomalies plays a crucial role in constructing models beyond the Standard Model (BSM).
For a candidate BSM theory to be consistent, both perturbative and global anomalies must be cancelled.
If an anomaly is present, the model must either be ruled out or extended by introducing additional degrees of freedom to achieve anomaly cancellation.
In this way, anomaly cancellation conditions impose strong constraints on viable BSM theories.
Along these lines, global anomalies have been investigated in various models with applications to particle physics in mind.
For related works, see~\cite{Hsieh:2018ifc,Garcia-Etxebarria:2018ajm,Wang:2018jkc,Wan:2018bns,Davighi:2019rcd,Wan:2019gqr,Wan:2019soo,Wan:2019oax,Wang:2020xyo,Wang:2020gqr,Wang:2020mra,Wang:2021hob,Wang:2021ayd,Wang:2022eag,Davighi:2022icj,Kawasaki:2023mjm,Cheng:2024awi,Wang:2025oow,Wan:2025lad,Debray:2025kfg,Wan:2025ymd,Yonekura:2026uiq}.

In this paper, we study anomalies in family unification models~\cite{Buchmuller:1982tf,Buchmuller:1983iu,Ong:1983uj,Kugo:1983ai} within the framework of the bordism classification of invertible field theories.
A key ingredient in these models is a four-dimensional $\cN=1$ supersymmetric nonlinear sigma model whose target space is a coset constructed from an exceptional Lie group and its subgroup.
A distinctive feature of these models is that the three generations of quarks and leptons arise naturally as superpartners of the sigma model fields.
Although various versions of family unification models have been proposed, we focus on two models whose coset spaces are constructed from the group $E_{7}$.
The target spaces of these models are given by $E_{7}/G$ and $E_{7}/H$, where
\begin{align}
  G&=\frac{\SU(5)\times\SU(3)\times\U(1)}{\bZ_{5}\times\bZ_{3}},&
  H&=\frac{\SU(5)\times\U(1)^{3}}{\bZ_{5}\times\bZ_{3}}.
\end{align}
Details of these subgroups of $E_{7}$ will be explained in Section~\ref{sec:family_unification}.
The model with $E_{7}/G$ was originally proposed in~\cite{Kugo:1983ai}, while the model with $E_{7}/H$ was discussed in~\cite{Yanagida:1985jc,Sato:1997hv}.~\footnote{
See e.g.~\cite{Irie:1983cd,Ibanez:1984ec,Ong:1984ej,Buchmuller:1985rc,Barr:1987pu}, for other family unification models studied in early works.
Phenomenological aspects of family unification models have been explored in~\cite{Yanagida:1998jk,Watari:2002tf,Hellerman:2014npa,Harigaya:2015iva,Yanagida:2016kag,Ho:2019ayl,Yanagida:2019evh,Sato:2021fpo}.
}

In~\cite{Mizoguchi:2014gva,Mizoguchi:2015kza}, it was suggested that the family unification model with $E_{7}/H$ can be realized as an F-theory model~\cite{Vafa:1996xn}.
If this scenario is realized, the generation structure may be naturally explained within the framework of string theory.
This provides one of the motivations for focusing on the $E_{7}/H$ model in addition to the original family unification model with $E_{7}/G$.~\footnote{
See~\cite{Mizoguchi:2014dra,Kan:2020lbe,Kuramochi:2020jzz,Mizoguchi:2024kpg}, for the related discussions of realizing family unification models in F-theory.
}

Family unification models contain fermions coupled to sigma model fields.
It is known that such fermions give rise to anomalies, referred to as sigma model anomalies~\cite{Yamashita:2021cao}.%
\footnote{
Although we focus on sigma model anomalies in four dimensions, this subject has also been extensively studied in two-dimensional theories. See e.g.~\cite{Witten:1999eg,Freed:1999vc,Kapustin:1999di,Gaiotto:2019asa,Yonekura:2022reu,Choi:2025ukq}.
}
Based on the pioneering works~\cite{Moore:1984dc,Moore:1984ws}, sigma model anomalies in the $E_{7}/G$ and $E_{7}/H$ models were studied in~\cite{Yanagida:1985jc}.%
\footnote{
Sigma model anomalies have also been discussed in other models. See e.g.~\cite{Yasui:1986ch,Buchmuller:1986zp,Kotcheff:1988ji} for early works.
}
In the present paper, we revisit sigma model anomalies from the viewpoint of bordism classification.
From this perspective, the anomalies discussed in~\cite{Moore:1984dc,Moore:1984ws} should be regarded as perturbative anomalies, in the sense that they are encoded in the free part of the sixth bordism group and are captured by characteristic classes of appropriate vector bundles.
This raises the possibility that family unification models may additionally suffer from global anomalies.
We show that, for both the $E_{7}/G$ and $E_{7}/H$ models, there is no invertible field theory that gives rise to a global anomaly.
This result is obtained by explicitly computing the relevant bordism groups using the Atiyah-Hirzebruch spectral sequence.
We also comment on perturbative sigma model anomalies in these models.
Specifically, we derive the conditions under which perturbative sigma model anomalies vanish on arbitrary four-dimensional spin manifolds equipped with sigma model fields.
As noted in~\cite{Yanagida:1985jc}, the fermions that are superpartners of the sigma model fields exhibit perturbative anomalies in these models.
One way to cancel these anomalies is to introduce additional fermions coupled to the sigma model fields.
Taking into account the global structure of the groups $G$ and $H$, we analyze the contributions of these additional fermions to the perturbative sigma model anomalies.

In order to construct a family unification model from the $\cN=1$ sigma model, we gauge a symmetry acting on the target space.
Under this gauging procedure, the model may suffer from additional anomalies.
We discuss the relevant bordism groups that classify the possible anomalies in this setting.
Furthermore, we show that both the $E_{7}/G$ and $E_{7}/H$ models do not suffer from global anomalies even after gauging $G$ and $H$, respectively.

The rest of the paper is organized as follows.
In Section~\ref{sec:general_anomaly}, we identify the bordism groups whose Anderson duals classify sigma model anomalies and anomalies in gauged sigma models.
We begin with a review of fermion anomalies, with particular emphasis on sigma model anomalies, and then discuss the bordism groups relevant to gauged sigma models.
In Section~\ref{sec:family_unification}, we introduce the $E_{7}/G$ and $E_{7}/H$ models as concrete realizations of family unification models, after specifying the global structures of the groups $G$ and $H$.
We also derive the constraints on linear representations of $G$ and $H$, which are imposed due to their nontrivial global structures.
In Section~\ref{sec:anomaly_family_unification}, we analyze anomalies in these models.
We first examine sigma model anomalies and then study additional anomalies that arise from gauging the symmetries of the coset spaces.
In Section~\ref{sec:discussion}, we outline possible future directions.
The appendices provide the necessary topological data and detailed computations.
In particular, we compute the cohomology and bordism groups of $BG$ and $BH$, as well as those of the coset spaces $E_{7}/G$ and $E_{7}/H$.
For the corresponding Borel constructions, we compute the relevant bordism groups.

\section{Anomalies in sigma model}
\label{sec:general_anomaly}
In this section, we identify the bordism groups that arise in the classification of anomalies in family unification models.
We begin with a brief review of fermion anomalies, focusing on sigma model anomalies, in Section~\ref{subsec:general_sigma}.
Fermion anomalies in gauged sigma models are discussed in Section~\ref{subsec:general_gauged_sigma}.
We will see that the Borel construction of the target space plays a crucial role in the classification of anomalies.

\subsection{Review of sigma model anomalies}
\label{subsec:general_sigma}
Let us consider a sigma model on a $d$-dimensional spacetime $M_{d}$ with a target space $X$.
We denote the sigma model field in the theory by $\phi:M_{d}\to X$.
Throughout this paper, we assume that $M_{d}$ is a closed spin manifold, so that fermions can be defined on $M_{d}$.
For a given vector bundle $V$ on $X$, one can formulate the fermion which takes value in the pullback bundle $\phi^{\ast}V$.
More precisely, this fermion is a section of the bundle $S\otimes\phi^{*}V$, where $S$ is a spinor bundle on $M_{d}$.
An important example is the fermion constructed from the tangent bundle $TX$ of the target space.

Let us consider the path integral of the fermion associated with $V$.
Here, the sigma model field~$\phi$ is treated as a background field.
In general, this fermion path integral suffers from an anomaly; namely, the partition function on $M_{d}$ is not defined as a complex number.
To formulate the anomaly, it is convenient to introduce a $(d+1)$-dimensional spin manifold $N_{d+1}$ with boundary $\partial N_{d+1}=M_{d}$ and to extend the sigma model field to $N_{d+1}$.
Then, the fermion on $M_{d}$ can be realized as the boundary mode of a massive fermion on $N_{d+1}$ by imposing an appropriate boundary condition.
After taking the large mass limit using a suitable regularization, the partition function of the fermion on $N_{d+1}$ can be identified with that of the original fermion on $M_{d}$~\cite{Witten:2015aba,Witten:2019bou}.
In this formulation, the anomaly is encoded in the dependence of the partition function on the choice of $N_{d+1}$ and on the extension of the background fields.
The Dai-Freed theorem~\cite{Dai:1994kq,Yonekura:2016wuc} implies that the ratio of partition functions for two different extensions is given by a phase expressed in terms of the eta invariant on a closed $(d+1)$-dimensional manifold.

This phase on a closed $(d+1)$-dimensional manifold is itself the partition function of an invertible field theory~\cite{Freed:2004yc} in $(d+1)$-dimensions, whose Hilbert space is one-dimensional.
This fact implies that the fermion anomaly of the $d$-dimensional theory is characterized by the $(d+1)$-dimensional invertible field theory.\footnote{More precisely, one needs to consider the deformation class of invertible field theories. See~\cite{Freed:2016rqq} for details.}
If the invertible field theory is nontrivial, the partition function of the original $d$-dimensional theory should be regarded as an element of the one-dimensional Hilbert space of the invertible field theory, rather than as a complex number.
Possible invertible field theories are known to be classified in terms of appropriate bordism groups~\cite{Kapustin:2014tfa,Kapustin:2014dxa,Freed:2016rqq,Yonekura:2018ufj,Yamashita:2021cao}.
In the present context, the relevant bordism group is $\Omega^{\Spin}_{\ast}(X)$, and the anomaly is classified by the Anderson dual $(I_{\bZ}\Omega^{\Spin})^{d+2}(X)$~\cite{Yamashita:2021cao}.
We refer to the anomaly encoded in $(I_{\bZ}\Omega^{\Spin})^{d+2}(X)$ as the sigma model anomaly.
In particular, the sigma model field cannot be promoted to a dynamical field if the theory has a sigma model anomaly.

In order to obtain some intuitions, it is useful to recall that the group $(I_{\bZ}\Omega^{\Spin})^{d+2}(X)$ fits into the short exact sequence:
\begin{align}
  0 \longrightarrow \Ext(\Omega^{\Spin}_{d+1}(X),\bZ)
    \longrightarrow (I_{\bZ}\Omega^{\Spin})^{d+2}(X)
    \longrightarrow \Hom(\Omega^{\Spin}_{d+2}(X),\bZ)
    \longrightarrow 0.
\end{align}
Here we assume that the bordism groups are finitely generated.
Under this assumption, the group $\Ext(\Omega^{\Spin}_{d+1}(X),\bZ)$ is the torsion part of $\Omega^{\Spin}_{d+1}(X)$ and encodes global anomalies, whereas $\Hom(\Omega^{\Spin}_{d+2}(X),\bZ)$ corresponds to the free part of $\Omega^{\Spin}_{d+2}(X)$ and captures perturbative anomalies \cite{Lee:2020ojw,Yamashita:2021cao,Yonekura:2022reu}.\footnote{The sigma model anomalies discussed in~\cite{Moore:1984dc,Moore:1984ws,Manohar:1984zj} are encoded in $\Hom(\Omega^{\Spin}_{d+2}(X),\bZ)$. Although such anomalies are sometimes referred to as global obstructions, we call them perturbative anomalies in this paper.}

\subsection{Classification of anomalies in gauged sigma models}
\label{subsec:general_gauged_sigma}
Suppose that a group $K$ acts from the left on the target space $X$, and that this action is a symmetry of the sigma model.
Once the absence of the sigma model anomaly is ensured, one can attempt to gauge this symmetry $K$.
To this end, we introduce a principal $K$-bundle $P$ over $M_{d}$.
The sigma model field $\phi : M_{d} \to X$ is then promoted to a section of the associated bundle $P \times_{K} X$, where the symbol $\times_{K}$ denotes the quotient by the equivalence relation $(pk,x)\sim(p,kx)$ for $(p,x)\in P\times X$ and $k\in K$.

Let $X_{K}=EK\times_{K}X$ denote the Borel construction of $X$, where $EK$ is the universal $K$-bundle over the classifying space $BK$.
Since the principal bundle $P$ is realized as the pullback of $EK$ by a map $f_{P}:M_{d}\to BK$, the associated bundle $P\times_{K}X$ can be identified with the pullback bundle $f_{P}^{\ast}X_{K}$.
The situation is summarized by the following commutative diagram:
\begin{align}\label{diag:Borel}
\xymatrix{
P\times_{K}X=f_{P}^{\ast}X_{K} \ar[r]^-{\overline{f_{P}}}\ar[d] & X_{K}=EK\times_{K}X \ar[d]\\
M_{d} \ar[r]_-{f_{P}} & BK
}
\end{align}
Recall that the vertical projection $X_{K}\to BK$ is given by $(e,x)\mapsto\pi(e)$ for $(e,x)\in EK\times_{K}X$, where $\pi:EK\to BK$ is the projection.
The pullback bundle is defined as
\begin{align}
    f_{P}^{\ast}X_{K}=\{(m,(e,x))\in M_{d}\times X_{K}\,|\,f_{P}(m)=\pi(e)\},
\end{align}
with the projection to $M_{d}$ given by the first factor.
The horizontal map $\overline{f_{P}}:f_{P}^{\ast}X_{K}\to X_{K}$ is then defined by $\overline{f_{P}}(m,(e,x))=(e,x)$.
From these definitions, it is straightforward to verify that the diagram~\eqref{diag:Borel} commutes.

For a given section $\Phi\in\Gamma(P\times_{K}X)$, one obtains a lift of $f_{P}$ to $X_{K}$ by the composition $\overline{f_{P}}\circ\Phi$.
Conversely, a lift $\widetilde{\phi}:M_{d}\to X_{K}$ determines a section of $P\times_{K}X$ via $m\mapsto(m,\widetilde{\phi}(m))$ for $m\in M_{d}$.
These observations imply that, after gauging the symmetry $K$, the sigma model field is regarded as a map $\widetilde{\phi}:M_{d}\to X_{K}$.

In order to incorporate the fermion associated with the $K$-equivariant vector bundle $V$ over $X$, we need to introduce the bundle $V_{K}=EK\times_{K}V$.
After the gauging of symmetry $K$, the fermion field becomes a section of $S\otimes\widetilde{\phi}^{\ast}V_{K}$.

As in ordinary gauge theory, the gauging procedure may generate additional anomalies, even when the sigma model anomaly discussed in Section~\ref{subsec:general_sigma} is absent.
We now consider the path integral of the fermion field, which is a section of $S\otimes\widetilde{\phi}^{\ast}V_{K}$, in the presence of a background $K$ gauge field and a background sigma model field~$\widetilde{\phi}$.
In the spirit of the framework of~\cite{Witten:2015aba,Witten:2019bou}, the anomaly is encoded in an invertible field theory defined on closed $(d+1)$-dimensional spin manifolds equipped with a map to $X_{K}$.
The group relevant for such invertible field theories is $\Omega^{\Spin}_{d}(X_{K})$, and the anomaly in the gauged sigma model is classified by the Anderson dual $(I_{\bZ}\Omega^{\Spin})^{d+2}(X_{K})$.
See also~\cite{Shiozaki:2018yyj,Freed:2019jzd} for related discussions in the context of condensed matter physics.
As before, the group $(I_{\bZ}\Omega^{\Spin})^{d+2}(X_{K})$ fits into the short exact sequence:
\begin{align}
  0 \to \Ext(\Omega^{\Spin}_{d+1}(X_{K}),\bZ)
    \to (I_{\bZ}\Omega^{\Spin})^{d+2}(X_{K})
    \to \Hom(\Omega^{\Spin}_{d+2}(X_{K}),\bZ)
    \to 0.
\end{align}

\section{Family unification model}
\label{sec:family_unification} 
In this section, we introduce the $E_{7}/G$ and $E_{7}/H$ models as specific realizations of family unification models.
As a preliminary, in Section~\ref{subsec:general_gauged_sigma}, we summarize the conventions for Lie algebras and Lie groups that are necessary to describe these models.
In Section~\ref{subsec:family_unification}, we review the construction of the models and the resulting quantum numbers of the fermions that arise as superpartners of the sigma model fields.
We also discuss how to introduce additional fermions into the theory.
To determine the possible quantum numbers of such fermions, we study the constraints on linear representations of $G$ and $H$ in Section~\ref{subsub:charge_constraint}.
While the necessity of imposing these constraints is well known, to the best of our knowledge, their explicit analysis in the context of family unification models has not been presented in the literature.

\subsection{Conventions of Lie algebra and Lie group}\label{subsec:conventions}
Here, we confirm the conventions for Lie algebras and Lie groups used in this paper.
For Lie algebras, we largely follow the conventions in Chapter~13 of~\cite{Francesco:1997}.
In Section~\ref{subsec:algebra_e7}, we introduce the notation for the Lie algebra $\e_{7}$ and describe two of its subalgebras.
We then discuss the corresponding subgroups of $E_{7}$ in Section~\ref{subsec:subgroup_GH}.
In particular, we pay attention to their global structure in order to correctly describe the coset spaces that arise as target spaces in family unification models.

\subsubsection{Lie algebra \texorpdfstring{$\e_{7}$}{} and its subalgebras}
\label{subsec:algebra_e7}
\begin{figure}
	\begin{center}
	\begin{tikzpicture}[transform shape, scale=0.8, >=stealth]
	\coordinate (O) at (0,0);
	\draw[thick] ($(O)$) circle [radius=0.3];
    \draw[thick] ($(O)+(2,0)$) circle [radius=0.3];
    \draw[thick] ($(O)+(4,0)$) circle [radius=0.3];
    \draw[thick] ($(O)+(6,0)$) circle [radius=0.3];
    \draw[thick] ($(O)+(8,0)$) circle [radius=0.3];
    \draw[thick] ($(O)+(10,0)$) circle [radius=0.3];
    \draw[thick] ($(O)+(4,-2)$) circle [radius=0.3];
    \draw[thick] ($(O)+(0.3,0)$)--($(O)+(1.7,0)$);
    \draw[thick] ($(O)+(2.3,0)$)--($(O)+(3.7,0)$);
    \draw[thick] ($(O)+(4.3,0)$)--($(O)+(5.7,0)$);
    \draw[thick] ($(O)+(6.3,0)$)--($(O)+(7.7,0)$);
    \draw[thick] ($(O)+(8.3,0)$)--($(O)+(9.7,0)$);
    \draw[thick] ($(O)+(4,-0.3)$)--($(O)+(4,-1.7)$);
    \node[above=10pt,font=\large] at ($(O)$) {$\alpha_{1}$};
    \node[above=10pt,font=\large] at ($(O)+(2,0)$) {$\alpha_{2}$};
    \node[above=10pt,font=\large] at ($(O)+(4,0)$) {$\alpha_{3}$};
    \node[above=10pt,font=\large] at ($(O)+(6,0)$) {$\alpha_{4}$};
    \node[above=10pt,font=\large] at ($(O)+(8,0)$) {$\alpha_{5}$};
    \node[above=10pt,font=\large] at ($(O)+(10,0)$) {$\alpha_{6}$};
    \node[right=10pt,font=\large] at ($(O)+(4,-2)$) {$\alpha_{7}$};
\end{tikzpicture}
\end{center}
\caption{The Dynkin diagram of $\e_{7}$. The numbering of the simple roots is the same as~\cite{Francesco:1997}.}\label{fig:Dynkin_e7}
\end{figure}
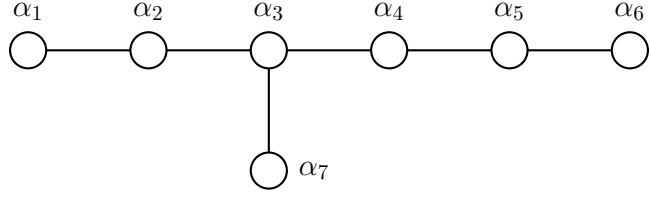

The Dynkin diagram of $\e_{7}$ is shown in Fig.~\ref{fig:Dynkin_e7}.
Let $\t$ denote the Cartan subalgebra of $\e_{7}$, and let $\alpha_{1},\dots,\alpha_{7}\in\t^{\ast}=\Hom(\t,\bR)$ be the simple roots.
The coroots $\coroot{i}\in\t$ for $(i=1,\dots,7)$ are chosen so that $\alpha_{i}(\coroot{j})=A_{ij}$, where $A_{ij}$ is the Cartan matrix of $\e_{7}$.
The fundamental weights $\omega_{1},\dots,\omega_{7}\in\t^{\ast}$ are defined by the conditions $\omega_{i}(\coroot{j})=\delta_{ij}$.

We consider the following chain of Lie subalgebras:
\begin{align}\label{eq:alg_embed}
  \h=\su_{5}\oplus\u_{1}\oplus\u_{1}\oplus\u_{1}
  \xhookrightarrow{\iota_{\h}}\g=\su_{5}\oplus\su_{3}\oplus\u_{1}
  \xhookrightarrow{\iota_{\g}}\e_{7}.
\end{align}
In order to specify the embedding $\iota_{\g}$ explicitly, we decompose the Cartan subalgebra $\t_{\g}$ of $\g$ as $\t_{\g}=\t_{\su_{5}}\oplus\t_{\su_{3}}\oplus\u_{1}$,
where $\t_{\su_{5}}$ and $\t_{\su_{3}}$ are the Cartan subalgebras of $\su_{5}$ and $\su_{3}$, respectively.
Let $\alpha_{j}^{\su_{5}}$ $(j=1,2,3,4)$ and $\alpha_{k}^{\su_{3}}$ $(k=1,2)$ denote the simple roots of $\su_{5}$ and $\su_{3}$, respectively.
The corresponding coroots and fundamental weights of $\su_{5}$ are denoted by $\fcoroot{j}\in\t_{\su_{5}}$ and $\fweight{j}\in\t^{\ast}_{\su_{5}}$, while those of $\su_{3}$ are denoted by $\tcoroot{k}\in\t_{\su_{3}}$ and $\tweight{k}\in\t^{\ast}_{\su_{3}}$.
We choose a generator of the $\u_{1}$ factor in $\g$, denoted by $\ocoroot$.
Then, the set of vectors $\fcoroot{j}\,(j=1,2,3,4)$, $\,\tcoroot{k}\,(k=1,2)$, and $\ocoroot$ forms a basis of $\t_{\g}$.
For the purpose of defining the embedding $\iota_{\g}$, it is sufficient to specify the images of this basis of $\t_{\g}$ under $\iota_{\g}$.
These images are given by
\begin{align}\begin{aligned}\label{eq:image_iotag}
  \iota_{\g}(\fcoroot{j}) &= \coroot{j}\quad(j=1,2,3),\\
  \iota_{\g}(\fcoroot{4}) &= \coroot{7},\\
  \iota_{\g}(\tcoroot{k}) &= \coroot{k+4}\quad(k=1,2),\\
  \iota_{\g}(\ocoroot) &= 6\coroot{1}+12\coroot{2}+18\coroot{3}+15\coroot{4}+10\coroot{5}+5\coroot{6}+9\coroot{7}.
\end{aligned}\end{align}
In other words, the subalgebra $\g$ is obtained by deleting the node corresponding to $\alpha_{4}$ from the diagram of $\e_{7}$ in Fig.~\ref{fig:Dynkin_e7}.
The generator $\ocoroot$ of $\u_{1}$ factor is chosen so that its image under $\iota_{\g}$ lies in the coroot lattice of $\e_{7}$.
In order to specify the embedding $\iota_{\h}$, we identify the first two $\u_{1}$ factors in $\h$ with $\t_{\su_{3}}$.
Then, $\iota_{\h}$ is defined via the inclusion $\t_{\su_{3}}\hookrightarrow\su_{3}$ in the obvious way.
More explicitly, when we denote the three $\u_1$ generators in $\h$ by $\ocoroota$, $\ocorootb$, and $\ocorootc$, the images of embedding $\iota_{\h}$ for $\u_1$ generators are given as 
\begin{align}\label{eq:image_iotah}
  \iota_{\h}(\ocoroota) = \tcoroot{1},\quad  
  \iota_{\h}(\ocorootb) = \tcoroot{2},\quad
  \iota_{\h}(\ocorootc) = \ocoroot.
\end{align}

\subsubsection{Subgroups \texorpdfstring{$G$}{} and \texorpdfstring{$H$}{} in \texorpdfstring{$E_{7}$}{}}
\label{subsec:subgroup_GH}
Let $G$ and $H$ be the subgroups of $E_{7}$ corresponding to the subalgebras $\g$ and $\h$ in Eq.~\eqref{eq:alg_embed}, respectively.
Our goal is to determine these groups.

We begin with $G$.
The lattice in $\t_{\g}$ generated by $\fcoroot{j}\,(j=1,2,3,4)$, $\tcoroot{k}\,(k=1,2)$, and $\ocoroot$ is mapped into the coroot lattice of $\e_{7}$ by $\iota_{\g}$.
From Eq.~\eqref{eq:image_iotag}, one finds that the index of this sublattice is $15$.
This fact implies that
\begin{align}
  G=\frac{\SU(5)\times\SU(3)\times\U(1)}{\bZ_{5}\times\bZ_{3}},
  \label{eq:subgroup_G}
\end{align}
as mentioned in the introduction.
More explicitly, we can choose generators of the $\bZ_{5}$ and $\bZ_{3}$ subgroups of $\SU(5)\times\SU(3)\times\U(1)$ as
\begin{align}
  z^\g_{5}&=\exp\!\left[2\pi\i\left(\omega^{\su_{5}\vee}_{1}
  +\tfrac{1}{5}\ocoroot\right)\right],&
  z^\g_{3}&=\exp\!\left[2\pi\i\left(\omega^{\su_{3}\vee}_{1}
  +\tfrac{1}{3}\ocoroot\right)\right],
  \label{eq:generators_Z5Z3_g}
\end{align}
where $\exp:\t_{\g}\to \SU(5)\times\SU(3)\times\U(1)$ denotes the exponential map.
Here, $\omega^{\su_{5}\vee}_{1}\in\t_{\su_{5}}$ and $\omega^{\su_{3}\vee}_{1}\in\t_{\su_{3}}$ are the fundamental coweights corresponding to $\fweight{1}$ and $\tweight{1}$, respectively.
One readily verifies that $\iota_{\g}(\omega^{\su_{5}\vee}_{1}+\ocoroot/5)$ and $\iota_{\g}(\omega^{\su_{3}\vee}_{1}+\ocoroot/3)$ lie in the coroot lattice of $\e_{7}$.

Turning to $H$, it follows from the above discussion and the definition of $\h$ that
\begin{align}
  H=\frac{\SU(5)\times\U(1)^{3}}{\bZ_{5}\times\bZ_{3}}.
  \label{eq:subgroup_H}
\end{align}
The generators of $\bZ_{5}$ and $\bZ_{3}$ are written as
\begin{align}
  z^\h_{5}&=\exp\!\left[2\pi\i\left(\omega^{\su_{5}\vee}_{1}
  +\tfrac{1}{5}\ocorootc\right)\right],&
  z^\h_{3}&=\exp\!\left[2\pi\i\left(\tfrac{2}{3}\ocoroota+\tfrac{1}{3}\ocorootb
  +\tfrac{1}{3}\ocorootc\right)\right],
  \label{eq:generators_Z5Z3_h}
\end{align}
since the two $\u_1$ generators are embedded into $\su_3$ as in Eq.~\eqref{eq:image_iotah}.

\subsection{Construction of the family unification model}\label{subsec:family_unification}
We review the construction of the family unification models.
In Section~\ref{subsubsec:model_G}, we begin by introducing the $E_{7}/G$ model~\cite{Kugo:1983ai}.
Following~\cite{Bando:1983ab,Bando:1984cc,Bando:1984fn,Itoh:1985ha,Itoh:1985jz}, we briefly review the construction of the $\cN=1$ supersymmetric sigma model, specializing to the $E_{7}/G$ case.
This construction allows us to extract the properties of the superfields containing the sigma model fields.
In Section~\ref{subsubsec:model_H}, we present the corresponding discussion for the $E_{7}/H$ model.
In Section~\ref{subsub:charge_constraint}, we analyze the constraints on linear representations of $G$ and $H$ that arise from the global structure of these groups.
These constraints are essential for consistently introducing additional chiral superfields.

\subsubsection{\texorpdfstring{$E_7/G$}{} model}
\label{subsubsec:model_G}
As mentioned in the introduction, family unification models are constructed from $\cN=1$ supersymmetric nonlinear sigma models~\cite{Zumino:1979et,Alvarez-Gaume:1981exv,Lee:1983pj,Lerche:1983qa,Kugo:1983ma,Bando:1983ab,Bando:1984cc,Bando:1984fn,Itoh:1985ha,Itoh:1985jz}.
In an $\cN=1$ supersymmetric theory, the target space of the sigma model must be a K\"{a}hler manifold.
Since $G$ is the centralizer of a torus subgroup $\U(1)$ of $E_7$, the coset space $E_7/G$ is a K\"{a}hler manifold.
Thus, we consider the sigma model with target space $X=E_7/G$.
In this model, the sigma model field $\phi$ is a map from the four-dimensional spacetime $M_4$ to $E_{7}/G$.
Its superpartner is a section of the bundle $S\otimes\phi^*T(E_7/G)$, where $S$ is the spinor bundle on $M_4$ and $T(E_7/G)$ is the holomorphic tangent bundle of the target space.

As a practical method for treating these fields, we briefly summarize the construction of $\cN=1$ supersymmetric nonlinear sigma models following~\cite{Bando:1983ab,Bando:1984cc,Bando:1984fn,Itoh:1985ha,Itoh:1985jz}.
Since chiral superfields are complex, they are regarded as complex coordinates on $E_7^\bC/\widehat G$, where $E_7^\bC$ is the complexification of $E_7$ and $\widehat G$ is a suitable subgroup of $E_7^\bC$.
The construction of $\widehat G$ proceeds as follows.
First, we decompose the Lie algebra $\e_7$ into $\g=\su_5\oplus\su_3\oplus\u_1$ and its complement $\g^\perp$.
The space $\g^\perp$ is further decomposed into subspaces with positive and negative eigenvalues under the adjoint action of the $\u_1$ generator.
We denote these by $\g^\perp_+$ and $\g^\perp_-$, respectively.
Then, the generators in $\g^\perp_+$ together with those in $\g$ form a closed Lie algebra, whose corresponding complexified Lie group is defined as $\widehat G$.
The sigma model field is identified with the complex coordinates of $E_7^\bC/\widehat G$ via the exponential map $(\g^\perp_-)^\bC \to E_7^\bC/\widehat G$.
When $X=E_7/G$ is K\"{a}hler, there exists an explicit map between $E_7/G$ and $E_7^\bC/\widehat G$, establishing a homeomorphism and ensuring that no extra degrees of freedom are required to describe the model.
For details, see~\cite{Bando:1983ab,Bando:1984cc,Bando:1984fn,Itoh:1985ha,Itoh:1985jz}.

The representations of the superpartners of the field $\phi$ under $\SU(5)\times\SU(3)\times\U(1)$ are shown in Table~\ref{tab:repE7G}.
By identifying the $\SU(5)$ factor in $G$ with the grand unified gauge group, the fermions in the $(\mathbf{10},\mathbf{\bar{3}})(2)$ and $(\mathbf{\bar{5}},\mathbf{3})(4)$ representations correspond to the three generations of quarks and leptons.
Here, the entries in the first parentheses denote representations of $\SU(5)$ and $\SU(3)$, while those in the second parentheses denote the $\U(1)$ charge.
In~\cite{Kugo:1983ai}, the fermions in the $(\mathbf{5},\mathbf{1})(6)$ representation are interpreted as Higgsinos.
Although $\g$ is combined with $\g^{\perp}_{+}$ to define $\widehat{G}$, one can alternatively construct another subgroup $\widehat{G}'$ from $\g$ and $\g^{\perp}_{-}$.
In this case, one obtains fermions in the complex conjugate representations of those shown in Table~\ref{tab:repE7G}.

In addition to the superfield containing the field $\phi$, we may introduce additional chiral superfields.
Suppose we introduce a matter field transforming in a representation $(\rho,W)$ of $G$, where $W$ is a vector space and $\rho$ is a homomorphism from $G$ to the automorphism group of $W$.
To this end, note that there is a fibration
\begin{align}
    G\to E_{7}\to E_{7}/G.
\end{align}
This implies that $E_{7}$ can be regarded as a principal $G$-bundle over $E_{7}/G$.
One can then construct an associated vector bundle $E_{7}\times_{\rho}W$ over $E_{7}/G$.
As explained in Section~\ref{subsec:general_sigma}, this allows us to introduce a fermion field on $M_{4}$ as a section of $S\otimes\phi^{\ast}(E_{7}\times_{\rho}W)$.

The $\cN=1$ supersymmetric sigma model constructed above typically possesses a global symmetry originating from the isotropy group of the target space.
In the $E_{7}/G$ model, a subgroup of the isotropy group is gauged to obtain a gauge theory.
For phenomenological applications, a subgroup containing the $\SU(5)$ factor is usually gauged.
In this paper, we focus on the case in which the entire isotropy group $G$ is gauged.
As discussed in Section~\ref{subsec:general_gauged_sigma}, after gauging $G$, the sigma model field is regarded as a map $\widetilde{\phi}:M_4\to EG\times_{G}E_{7}/G$, while the fermion is a section of $S\otimes\widetilde{\phi}^*(EG\times_G T(E_7/G))$.

\begin{table}[t]
\begin{center}
\begin{tabular}{ccc}
$\SU(5)$ & $\SU(3)$ & $\U(1)$\\
\hline
{$\mathbf{10}$}&$\mathbf{\bar{3}}$&$2$\\
{$\mathbf{\bar{5}}$}&$\mathbf{3}$&$4$\\
{$\mathbf{5}$}&$\mathbf{1}$&$6$\\
\end{tabular}
\caption{Representations of the superfields for $G$ in the model with $X=E_7/G$.}
\label{tab:repE7G}
\end{center}
\end{table}

\subsubsection{\texorpdfstring{$E_7/H$}{} model}
\label{subsubsec:model_H}
The overall structure is similar in the $X=E_7/H$ model.
The coset space $E_7/H$ is again a K\"{a}hler manifold, since $H$ is the centralizer of a torus subgroup $\U(1)^3$ of $E_7$.
In this case, the superpartner of the sigma model field $\phi$ is a section of $S\otimes\phi^*T(E_7/H)$, where $T(E_7/H)$ is a holomorphic tangent bundle of the target space.

The sigma model field is identified with the complex coordinates of $E_7^\bC/\widehat H$.
The construction of the subgroup $\widehat H \subset E_7^\bC$ is analogous to that of $\widehat G$ in the $E_7/G$ model.
First, we decompose the Lie algebra $\e_{7}$ into $\h=\su_5\oplus\u_1^{\oplus 3}$ and its complement $\h^\perp$.
To proceed, we choose a specific $\u_1$ subalgebra in the factor $\u_1^{\oplus 3}\subset\h$.
We then decompose $\h^\perp$ into subspaces with positive and negative eigenvalues under the adjoint action of the chosen $\u_1$ generator.
We denote these by $\h^\perp_+$ and $\h^\perp_-$, respectively.
Then, the generators in $\h^\perp_+$ together with those in $\h$ form a closed Lie algebra, whose corresponding complexified Lie group is defined as $\widehat H$.
The sigma model field is identified with the complex coordinates of $E_7^\bC/\widehat H$ via the exponential map $(\h^\perp_-)^\bC \to E_7^\bC/\widehat H$.

One must be careful in choosing the $\u_1$ generator used for the decomposition of $\h^\perp$.
The choice of the $\u_1$ subalgebra in $\u_1^{\oplus 3}$ is not unique.
Different choices of the $\u_1$ generator lead to different subgroups $\widehat H$, corresponding to different invariant complex structures on the target space.
For details, see~\cite{Bando:1983ab,Bando:1984cc,Bando:1984fn,Itoh:1985ha,Itoh:1985jz}.

For a particular choice of the $\u_1$ generator, the representations of the superpartners of $\phi$ under $\SU(5)$ and the three $\U(1)$ charges are shown in Table~\ref{tab:repE7H}.~\footnote{
In Table~2 of~\cite{Sato:1997hv}, different $\U(1)$ charge assignments are presented due to a different choice of the three $\u_1$ generators in $\h$.
Let us denote the three $\u_1$ generators in $\h$, corresponding to $\U(1)_I~(I=1,2,3)$ in~\cite{Sato:1997hv}, by $\alpha'^{\u_1^{(I)}\vee}$.
Then, the embedding $\iota_{\h}:\h\xhookrightarrow{}\g$ for the $\u_1$ generators is given by
\begin{align}\begin{aligned}
  \iota_{\g}\circ\iota_{\h}(\alpha'^{\u_1^{(1)}\vee}) &= 2\coroot{1}+4\coroot{2}+6\coroot{3}+5\coroot{4}+4\coroot{5}+3\coroot{6}+3\coroot{7}, \\
  \iota_{\g}\circ\iota_{\h}(\alpha'^{\u_1^{(2)}\vee}) &= 2\coroot{1}+4\coroot{2}+6\coroot{3}+5\coroot{4}+4\coroot{5}+3\coroot{7}, \\
  \iota_{\g}\circ\iota_{\h}(\alpha'^{\u_1^{(3)}\vee}) &= 2\coroot{1}+4\coroot{2}+6\coroot{3}+5\coroot{4}+3\coroot{7},
\end{aligned}\end{align}
where the coroots $\coroot{i}$~$(i=1,\ldots,7)$ of $\e_7$ on the right-hand sides, as well as the embedding $\iota_{\g}:\g\xhookrightarrow{}\e_7$, follow the conventions adopted in the present paper.
For the $\su_5$ generators, the images under the embedding coincide with those given in Eq.~\eqref{eq:image_iotag}.
If we denote the $\U(1)$ charges in the present paper and in~\cite{Sato:1997hv} by $(n_1,n_2,n_3)$ and $(n'_1,n'_2,n'_3)$, respectively, their relation is given by
\begin{align}\begin{aligned}
  n_1 = \frac{1}{4}(n'_2 - n'_3), \quad
  n_2 = \frac{1}{3}(n'_1 - n'_2), \quad
  n_3 = \frac{1}{6}(10n'_1 + 5n'_2 + 3n'_3).
\end{aligned}\end{align}
}
The three $\U(1)$ factors in $H$ are generated by $\ocoroota$, $\ocorootb$, and $\ocorootc$ in the Cartan subalgebra of $\h$, following the notation introduced in Subsection~\ref{subsec:algebra_e7}.
In the $E_7/H$ model, there are three $\mathbf{10}$, $\mathbf{\bar{5}}$, and $\mathbf{1}$ representations, and one $\mathbf{5}$ representation of $\SU(5)$.
In particular, these fields accommodate three generations of quarks and leptons.

As before, the space $E_{7}$ can be regarded as a principal $H$-bundle over $E_{7}/H$.
From a linear representation of $H$, one can construct an associated vector bundle over $E_{7}/H$, and introduce fermions on $M_{4}$ valued in this bundle.

In constructing a gauge theory, one may in principle gauge any subgroup of $H$ that contains the $\SU(5)$ factor.
In this paper, however, we focus on the case in which the entire group $H$ is gauged.
One motivation for this choice comes from the suggestion that the $E_{7}/H$ model may admit a realization in F-theory~\cite{Mizoguchi:2014gva,Mizoguchi:2015kza}.
If such a realization exists, it is natural to expect that the full group $H$ appears as the gauge symmetry.

\begin{table}[t]
\begin{center}
\begin{tabular}{cccc}
$\SU(5)$ & $\U(1)^{(1)}$ & $\U(1)^{(2)}$ & $\U(1)^{(3)}$\\
\hline
{$\mathbf{10}$}& $-1$ & $0$ & $2$\\
{$\mathbf{10}$}& $1$ & $-1$ & $2$ \\
{$\mathbf{10}$}& $0$ & $1$ & $2$\\
{$\mathbf{\bar{5}}$} & $0$ & $-1$ & $4$\\
{$\mathbf{\bar{5}}$} & $-1$ & $1$ & $4$\\
{$\mathbf{\bar{5}}$} & $1$ & $0$ & $4$\\
{$\mathbf{1}$} & $2$ & $-1$ & $0$\\
{$\mathbf{1}$} & $1$ & $1$ & $0$\\
{$\mathbf{1}$} & $-1$ & $2$ & $0$\\
{$\mathbf{5}$} & $0$ & $0$ & $6$\\
\end{tabular}
\caption{Representations of the superfields for $H$ in the model with $X=E_7/H$.}
\label{tab:repE7H}
\end{center}
\end{table}

\subsubsection{Constraints on representations and charges}
\label{subsub:charge_constraint}
The groups $G$ and $H$ are obtained as quotient groups of $\SU(5)\times\SU(3)\times\U(1)$ and $\SU(5)\times\U(1)^3$ by the subgroup $\bZ_{5}\times\bZ_{3}$.
From this viewpoint, fields in the models must be invariant under the action of $\bZ_5\times\bZ_3$.
This requirement imposes constraints on the allowed representations and charges of the fields.~\footnote{
See~\cite{Tong:2017oea} for a related discussion in the context of the Standard Model.
}

In the $X=E_7/G$ model, we consider the action of the element $\left((z_5^\g)^k,(z_3^\g)^l\right)$ with $k,l\in\bZ$ on the fields, where $z^\g_5$ and $z^\g_3$ are the generators of $\bZ_5$ and $\bZ_3$ defined in Eq.~\eqref{eq:generators_Z5Z3_g}.
Suppose that a field transforms in representations with highest weights $\lambda^{\su_5} = \sum_{i=1}^4\lambda^{\su_5}_i\omega^{\su_5}_i$ of $\SU(5)$ and $\lambda^{\su_3} = \sum_{j=1}^2\lambda^{\su_3}_j\omega^{\su_3}_j$ of $\SU(3)$, and carries $\U(1)$ charge $n$.
Then, the action of $\left((z_5^\g)^k,(z_3^\g)^l\right)$ on the field is given by multiplication by the phase
\begin{align}
  \exp
  \left[2\pi\i k \left(\lambda^{\su_5}(\omega_1^{\su_5\vee}) + \frac{n}{5}\right)
  + 2\pi \i l \left(\lambda^{\su_3}(\omega_1^{\su_3\vee}) +\frac{n}{3}\right)\right].
\end{align}
By explicit computation, one finds that
\begin{align}
    \lambda^{\su_5}(\omega_1^{\su_5\vee}) = -\frac{1}{5}q^{\su_5}(\lambda^{\su_5}), \quad
    \lambda^{\su_3}(\omega_1^{\su_3\vee}) = -\frac{1}{3}q^{\su_3}(\lambda^{\su_3}),
\end{align}
where $q^{\su_5}(\lambda^{\su_5})$ and $q^{\su_3}(\lambda^{\su_3})$ denote the congruence classes of the representations.~\footnote{
For the algebra $\su_N$, the congruence class of an irreducible representation with a highest weight $\lambda$ and Dynkin labels $\{\lambda_i\}~(i=1,\ldots,N-1)$ is given by $q^{\su_N}(\lambda) = \sum_{i=1}^{N-1} i \lambda_i \mod N$.
}
Requiring invariance under $\bZ_5\times\bZ_3$ implies
\begin{align}
  \frac{k}{5} \left(-q^{\su_5}(\lambda^{\su_5}) + n\right) 
  +\frac{l}{3} \left( -q^{\su_3}(\lambda^{\su_3}) + n\right) \in \bZ,
  \label{eq:constraint_Z5Z3}
\end{align}
for all $k,l\in\bZ$.
It is necessary and sufficient that this condition holds for $k=l=1$, which is equivalent to
\begin{align}
    3q^{\su_5}(\lambda^{\su_5})+5q^{\su_3}(\lambda^{\su_3}) - 8 n \equiv 0 \mod 15.
    \label{eq:constraint_Z5Z3_final}
\end{align}
The sigma model fields and fermions listed in Table~\ref{tab:repE7G} satisfy this constraint.~\footnote{
In certain exotic situations, one may introduce fermions that do not satisfy constraints of this type.
As discussed in~\cite{Kaidi:2023tqo}, such a situation arises in the $\so(32)$ heterotic string compactified on $\bC\bP^{1}$.
We thank K.~Yonekura for pointing this out.
}

In the $X=E_7/H$ model, the condition for invariance under $\bZ_5\times\bZ_3$ is obtained analogously by considering the action of $((z_5^\h)^k,(z_3^\h)^l)$ with $k,l\in\bZ$, where $z^\h_5$ and $z^\h_3$ are defined in Eq.~\eqref{eq:generators_Z5Z3_h}.
The constraint on the representations and charges becomes
\begin{align}
   3q^{\su_5}(\lambda^{\su_5}) -10 n_1 -5 n_2 - 8 n_3 \equiv 0 \mod 15,
\end{align}
where $n_1$, $n_2$, and $n_3$ are the $\U(1)^{(1)}$, $\U(1)^{(2)}$, and $\U(1)^{(3)}$ charges, respectively.
The sigma model fields and fermions listed in Table~\ref{tab:repE7H} satisfy this constraint, and any additional matter fields must also obey it.

\section{Anomalies in family unification models}\label{sec:anomaly_family_unification}
In this section, we discuss anomalies in the family unification models.
Although the $\cN=1$ supersymmetric sigma models plays a crucial role in their construction, the key point is that fermions coupled to the sigma model fields may give rise to sigma model anomalies, which are classified by the Anderson dual of the bordism groups of the target space, as reviewed in Section~\ref{subsec:general_sigma}.
To establish the consistency of the sigma models, it is necessary to verify that both perturbative and global sigma model anomalies are absent.
Furthermore, additional anomalies may arise from the gauging procedure in the construction of the family unification models, even when the sigma model anomalies are absent.
We also analyze these anomalies based on the classification discussed in Section~\ref{subsec:general_gauged_sigma}.
In the remainder of this section, we study the $E_{7}/G$ and $E_{7}/H$ models in turn.

\subsection{Anomalies in \texorpdfstring{$E_7/G$}{} model}
\label{subsec:anomaly_E7G}

\subsubsection{Sigma model anomalies in \texorpdfstring{$E_7/G$}{} model}
\label{subsubsec:sigma_anomaly_E7G}
In the four-dimensional sigma model with the target space $E_7/G$, sigma model anomalies arising from fermions coupled to the sigma model field are classified by the Anderson dual of the spin bordism group $\Omega^{\Spin}_*(E_7/G)$, as reviewed in Section~\ref{subsec:general_sigma}. 
Global anomalies are encoded in the torsion part of $\Omega^{\Spin}_{5}(E_7/G)$, while the free part of $\Omega^{\Spin}_{6}(E_7/G)$ corresponds to perturbative anomalies.

Let us begin with the discussion of global anomalies.
In Appendix~\ref{sec:E7G}, we explicitly compute the bordism group $\Omega^{\Spin}_5(E_7/G)$ using the Atiyah-Hirzebruch spectral sequence and find that
\begin{align}
    \Omega^{\Spin}_5(E_7/G) = 0.
\end{align}
This result implies that there is no invertible field theory that gives rise to a global sigma model anomaly in the $E_7/G$ model.
In other words, even if we introduce fermions transforming in arbitrary linear representations of $G$ into the sigma model, no global sigma model anomaly arises as long as perturbative anomalies are absent.
Hence, it remains to analyze the perturbative sigma model anomalies.

Perturbative anomalies, encoded in the free part of $\Omega^{\Spin}_{6}(E_{7}/G)$, are captured by characteristic classes of vector bundles over the target space~\cite{Moore:1984dc,Moore:1984ws}.
In analogy with gauge anomalies, we refer to such characteristic classes as anomaly polynomials.
Let us consider the perturbative sigma model anomaly associated with a fermion valued in $\phi^{\ast}(E_{7}\times_{\rho}W)$, where $\phi$ is the sigma model map and $E_{7}\times_{\rho}W$ is the vector bundle associated with a linear representation $(\rho,W)$ of $G$.
In this case, the anomaly polynomial is given by the third Chern character of the bundle $E_{7}\times_{\rho}W$.~\footnote{
There is also a mixed anomaly involving the first Pontryagin class and the second Chern character of the vector bundle.
}
It is well known that the splitting principle is useful for computing anomaly polynomials in this context~\cite{MR102800}.
By restricting the representation $(\rho,W)$ to the maximal torus $T$ of $G$, the vector bundle $E_{7}\times_{\rho}W$ can be decomposed into a direct sum of line bundles corresponding to the weights of the representation.
The total Chern class of $E_{7}\times_{\rho}W$ is then given by
\begin{align}
    c(E_{7}\times_{\rho}W)=\prod_{w:\text{weight}}(1+c_{1}(w)),
    \label{eq:splitting}
\end{align}
where $c_{1}(w)$ denotes the first Chern class of the line bundle associated with the weight $w$.
Here, the product is taken with respect to the ring structure of the cohomology ring $H^{\ast}(E_{7}/T;\bR)$.
More precisely, the total Chern class of the bundle $E_{7}\times_{\rho}W$ takes a value in the cohomology ring $H^{\ast}(E_{7}/G;\bR)$.
In the application of the splitting principle, one considers the image of $H^{\ast}(E_{7}/G;\bR)$ under the natural pullback to $H^{\ast}(E_{7}/T;\bR)$.
Relevant properties of this cohomology ring are reviewed in Appendix~\ref{subsubsec:cohom_E7G_Q}.~\footnote{
In Appendix~\ref{subsubsec:cohom_E7G_Q}, the cohomology rings with rational coefficients are shown. The cohomology rings with real coefficients are obtained by tensoring with $\bR$.
}
The $n$-th Chern class $c_n(E_{7}\times_{\rho}W)$ of the bundle $E_{7}\times_{\rho}W$ is defined as an $n$-th order polynomial of $c_1(w)$ in the right-hand side of Eq.~\eqref{eq:splitting}.
Then, the third Chern character $\ch{3}(E_{7}\times_{\rho}W)$ is given as
\begin{align}
    \ch{3}(E_{7}\times_{\rho}W) = \frac{1}{6}({c_1^3 + 3c_3 -3c_1c_2}),
    \label{eq:ch3}
\end{align}
where we have abbreviated the argument $E_{7}\times_{\rho}W$ of the Chern classes on the right-hand side.

The fermion that is the superpartner of the sigma model field takes values in $\phi^*T(E_7/G)$, where $T(E_7/G)$ denotes the holomorphic tangent bundle of $E_7/G$.
The contribution of this fermion to the anomaly polynomial is characterized by the third Chern character of $T(E_7/G)$.
By applying the splitting principle, the total Chern class of $T(E_7/G)$ can be computed as
\begin{align}
    c(T(E_7/G)) = \prod_{\alpha\in\Delta^{\g_+^\perp}} \left(1 + c_{1}(\alpha)\right),
\end{align}
where $\Delta^{\g_+^\perp}$ is the set of roots in $\g_+^\perp$ defined in Section~\ref{subsubsec:model_G}, and $c_{1}(\alpha)$ denotes the first Chern class of the line bundle associated with the root $\alpha$.
In this calculation, the first Chern classes $c_{1}(\alpha)$ are regarded as coordinates of the second cohomology group $H^2(BT;\bR)$ via the isomorphism $H^{2}(BT;\bR)\cong H^{2}(E_{7}/T;\bR)$.
Using the coordinates $y'_j~(j=1,\dots,5)$, $z'_k~(k=1,2,3)$, and $b_1$ introduced in Eq.~\eqref{eq:coordinate_byz}, the first Chern classes can be expressed as
\begin{align}
    c_{1}(\alpha^{(\mathbf{10},\mathbf{\bar{3}})(2)}_{j_1,j_2,k}) &= y'_{j_1} + y'_{j_2} - z'_{k} + \frac{2}{15}b_1  & & (1\leq j_1 <j_2 \leq 5, 1\leq k \leq 3), \\
    c_{1}(\alpha^{(\mathbf{\bar{5}},\mathbf{3})(4)}_{j,k}) &= -y'_{j} + z'_k + \frac{4}{15}b_1 & & (1 \leq j \leq 5, 1\leq k \leq 3), \\
    c_{1}(\alpha^{(\mathbf{5},\mathbf{1})(6)}_{j}) &= y'_{j} + \frac{6}{15}b_1 & & (1\leq j \leq 5),
\end{align}
where we decompose the roots in $\Delta^{\g_+^\perp}$ into irreducible representations of $G$, corresponding to those listed in Table~\ref{tab:repE7G}.
Here, the entries in the first parentheses denote the representations of $\SU(5)$ and $\SU(3)$, while the number in the second parentheses denotes the $\U(1)$ charge.
After a straightforward computation, we obtain the third Chern character as
\begin{align}\label{eq:thirdch_isoG}
    \ch{3}(T(E_7/G)) = -\frac{4}{15}b_1\cpth{2} - \frac{32}{675}b_1^3 + \frac{1}{2}\cpfi{3}-\frac{5}{2}\cpth{3},
\end{align}
where $\cpfi{j}~(j=1,\dots,5)$ and $\cpth{k}~(k=1,2,3)$ are the symmetric polynomials defined in Eq.~\eqref{eq:symmetric_polynomial_byz}.
In deriving this result, we used the relations $\sum_{j=1}^5 y'_j = 0$ and $\sum_{k=1}^3 z'_k = 0$, as well as
\begin{align}
    I_{2}=\frac{32}{5}\bth{1}^{2}+48\left((\cpfi{1})^{2}-2\cpfi{2}+(\cpth{1})^{2}-2\cpth{2}\right)=0
\end{align}
to eliminate $\cpfi{1}$, $\cpth{1}$, and $\cpfi{2}$.
Here, $I_2$ is the Weyl-invariant polynomial of $E_7$ introduced in Eq.~\eqref{eq:I2}.
This contribution to the anomaly polynomial is firstly calculated in~\cite{Yanagida:1985jc}.

The nontrivial third Chern character of $T(E_7/G)$ implies the presence of a perturbative sigma model anomaly when the chiral multiplet containing the sigma model field is the only matter content in the model.
Since such anomalies lead to an inconsistency of the theory, they must be cancelled by introducing additional degrees of freedom. 
One possibility is to introduce massless fermions in additional chiral multiplets.
After introducing an additional fermion transforming in a linear representation $(\rho',W')$ of $G$, the total anomaly polynomial is given by the third Chern character of the direct sum of the vector bundles $T(E_7/G)$ and $E_7\times_{\rho'} W'$:
\begin{align}
    \ch{3}(T(E_7/G)\oplus V) = \ch{3}(T(E_7/G)) + \ch{3}(E_{7}\times_{\rho'}W').
\end{align}
One simple candidate for achieving anomaly cancellation is to introduce chiral multiplets in representations that are complex conjugate to those listed in Table~\ref{tab:repE7G}.
However, this choice renders the theory non-chiral, which is not desirable from a phenomenological viewpoint.

As another candidate for cancelling the perturbative anomaly, let us reconsider the set of additional chiral multiplets discussed in~\cite{Yanagida:1985jc}.
We introduce one multiplet in the $(\mathbf{\bar{5}},\mathbf{1})(m)$ representation of $G$ and five multiplets in the $(\mathbf{1},\mathbf{3})(n)$ representation, where $m$ and $n$ denote their $\U(1)$ charges.
Due to the constraint~\eqref{eq:constraint_Z5Z3_final} on linear representations of $G$, these charges must satisfy
\begin{align}
    m \equiv 9, \quad n \equiv 10 \mod 15. 
    \label{eq:u1charges}
\end{align}
We denote the vector bundles associated with the $(\mathbf{\bar{5}},\mathbf{1})(m)$ and $(\mathbf{1},\mathbf{3})(n)$ representations by $V_{(\mathbf{\bar{5}},\mathbf{1})(m)}$ and $V_{(\mathbf{1},\mathbf{3})(n)}$, respectively.
Their Chern classes can be computed using the splitting principle as in Eq.~\eqref{eq:splitting}.
The first Chern classes corresponding to the weights are given by
\begin{align}\begin{aligned}
    c_{1}(w^{(\mathbf{\bar{5}},\mathbf{1})(m)}_{j}) &= -y'_{j} + \frac{m}{15}b_1 & & (1 \leq j \leq 5), \\ 
    c_{1}(w^{(\mathbf{1},\mathbf{3})(n)}_{k}) &= z'_{k} + \frac{n}{15}b_1 & & (1\leq k \leq 3),
\end{aligned}\end{align}
in terms of the coordinates introduced in Eq.~\eqref{eq:coordinate_byz}.
After a straightforward computation, we obtain
\begin{align}\begin{aligned}
    \ch{3}(V_{(\mathbf{\bar{5}},\mathbf{1})(m)}) &= \frac{m}{15}b_1\cpth{2} + \left(\frac{m^3}{4050}-\frac{m}{225}\right)b_1^3 -\frac{1}{2}\cpfi{3}, \\
    \ch{3}(V_{(\mathbf{1},\mathbf{3})(n)}) &= -\frac{n}{15}b_1\cpth{2} + \frac{n^3}{6750}b_1^3 + \frac{1}{2}\cpth{3}.
\end{aligned}\end{align}
The total third Chern character is then given by
\begin{align}\begin{aligned}\label{eq:ch3total}
    &\ch{3}(T(E_7/G))+\ch{3}(V_{(\mathbf{\bar{5}},\mathbf{1})(m)})+ 5\ch{3}(V_{(\mathbf{1},\mathbf{3})(n)}) \\
    &=\frac{1}{15}(m-5n-4)b_1\cpth{2} + \frac{1}{4050}(m^3-18m + 3n^3 -192)b_1^3.
\end{aligned}\end{align}
To cancel the anomaly, the $\U(1)$ charges $(m,n)$ must satisfy
\begin{align}
    m -5n -4 &= 0, \label{eq:cancel1}\\
    m^3 -18m + 3n^3 -192 &=0. \label{eq:cancel2}
\end{align}
However, one finds that there is no pair of integers $(m,n)$ satisfying both conditions~\eqref{eq:cancel1} and~\eqref{eq:cancel2}.
This implies that additional degrees of freedom beyond these multiplets are required to cancel the perturbative sigma model anomalies.\footnote{
In the pioneering work~\cite{Yanagida:1985jc}, the authors imposed the condition that the theory be consistently defined on the four-sphere.
In the present paper, we instead require that the theory be well-defined on an arbitrary four-dimensional spin manifold equipped with a sigma model field.
}

It is important to note that the contribution of the $b_1^3$ term in Eq.~\eqref{eq:ch3total} should be qualitatively distinguished from that of the $b_1\cpth{2}$ term.
The anomaly of the form $b_1^3$ may be cancelled via the Green-Schwarz mechanism~\cite{Green:1984sg} in four dimensions.~\footnote{
The Green-Schwarz mechanism in four dimensions are discussed from the view point of the bordism classification in~\cite{Saito:2025idl}.
}
Motivated by this, we relax the anomaly cancellation condition and impose only Eq.~\eqref{eq:cancel1}.
The integers $(m,n)$ satisfying both the constraint~\eqref{eq:u1charges} and the condition~\eqref{eq:cancel1} are given by
\begin{align}
    (m,n) = (75l+54,\,15l+10), \quad (l\in\bZ).
\end{align}
By introducing one multiplet in the $(\mathbf{\bar{5}},\mathbf{1})(75l+54)$ representation of $G$ and five multiplets in the $(\mathbf{1},\mathbf{3})(15l+10)$ representation, the perturbative anomalies, except for those proportional to $b_1^3$, can be cancelled.

\subsubsection{Anomalies in gauged \texorpdfstring{$E_7/G$}{} model}
\label{subsubsec:gauged_anomaly_E7G}
For phenomenological applications, it is necessary to gauge a subgroup of $G$ that contains an $\SU(5)$ factor, and the $\SU(5)$ factor is identified as the grand unified gauge group.
Even if the sigma model anomalies are successfully cancelled, additional anomalies may arise after gauging.
Since such anomalies also lead to inconsistencies in the theory, their absence must be verified.
In this paper, we focus on the case in which the entire group $G$ is gauged.
In the gauged sigma model, anomalies are classified by the Anderson dual of the spin bordism group $\Omega^{\Spin}_*((E_7/G)_G)$, where $(E_7/G)_G = EG\times_G E_7/G$ is the Borel construction of $E_7/G$, as discussed in Section~\ref{subsec:general_gauged_sigma}.
The global anomalies are encoded in the torsion part of $\Omega^{\Spin}_5((E_7/G)_G)$, while the perturbative anomalies are encoded in the free part of $\Omega^{\Spin}_6((E_7/G)_G)$.

For global anomalies, we explicitly compute the bordism group $\Omega^{\Spin}_5((E_7/G)_G)$ using the Atiyah-Hirzebruch spectral sequence in Appendix~\ref{sec:Borel_E7G}, and find that
\begin{align}
\Omega^{\Spin}_5((E_7/G)_G) = 0.
\end{align}
This result implies that no global anomalies arise after gauging $G$, provided that the perturbative anomalies are cancelled.

Perturbative anomalies, encoded in the free part of $\Omega^{\Spin}_6((E_7/G)_G)$, are captured by characteristic classes of vector bundles over $(E_7/G)_G$.
Intuitively, the anomaly polynomial can be decomposed into three contributions: the pure sigma model anomalies discussed in Section~\ref{subsubsec:sigma_anomaly_E7G}, the pure gauge anomalies encoded in the free part of $\Omega^{\Spin}_6(BG)$, and mixed anomalies between them.

For a fermion transforming in a representation $(\rho,W)$ of $G$, the anomaly polynomial is given by the third Chern character of the associated bundle $(E_7\times_\rho W)_{G} = EG\times_G (E_7\times_{\rho}W)$.
The Chern classes of $(E_7\times_\rho W)_G$ can also be computed using the splitting principle, by decomposing the bundle into a direct sum of line bundles corresponding to the weights of the representation.
The computation is carried out by mapping the characteristic classes to $H^{\ast}((E_7/T)_T;\bR)$ and applying a formula analogous to Eq.~\eqref{eq:splitting}.
The ring structure of $H^{\ast}((E_7/T)_T;\bR)$, which is required for this computation, can be found for example in~\cite{MR4305997}.
The explicit form of the anomaly polynomial, as well as the anomaly cancellation conditions after gauging, depend sensitively on the additional degrees of freedom introduced to cancel the perturbative sigma model anomalies.

\subsection{Anomalies in \texorpdfstring{$E_7/H$}{} model}
\label{subsec:anomaly_E7H}

\subsubsection{Sigma model anomalies in \texorpdfstring{$E_7/H$}{} model}
The possible sigma model anomalies in a four-dimensional sigma model with the target space $E_{7}/H$ are classified by the group $(I_{\bZ}\Omega^{\Spin})^{6}(E_{7}/H)$, which is the Anderson dual of the bordism groups $\Omega^{\Spin}_{\ast}(E_{7}/H)$.
As before, global anomalies are encoded in the torsion part of $\Omega^{\Spin}_{5}(E_{7}/H)$, whereas perturbative anomalies are captured by the free part of $\Omega^{\Spin}_{6}(E_{7}/H)$.

As shown in Appendix~\ref{sec:E7H}, the bordism group $\Omega^{\Spin}_{5}(E_{7}/H)$ is trivial.
This implies that there is no global sigma model anomaly in the $E_{7}/H$ model.
Therefore, we only need to consider perturbative sigma model anomalies.

Let us derive the anomaly polynomial for the fermions listed in Table~\ref{tab:repE7H} using the splitting principle.
The computation proceeds in close analogy with that in Section~\ref{subsubsec:sigma_anomaly_E7G}.
To describe characteristic classes in $H^{\ast}(E_{7}/T)$, it is convenient to use $\bth{1}$, $y'_{j}\,(j=1,\dots,5)$, and
\begin{align}
    e'_{1}&=-\omega_{5}+2\omega_{6},&
    e'_{2}&=-\omega_{4}+2\omega_{5}-\omega_{6},
\end{align}
as coordinates of $H^{2}(BT;\bR)\cong H^{2}(E_{7}/T;\bR)$.
The definitions of $y'_{j}$ and $\bth{1}$ are given in Eq.~\eqref{eq:coordinate_byz}.
In terms of these coordinates, the first Chern classes corresponding to the weights in Table~\ref{tab:repE7H} are given by
\begin{align}\begin{aligned}
    c_{1}(\alpha^{(\mathbf{10})(-1,0,2)}_{j_1,j_2}) &= y'_{j_1}+y'_{j_2} -\frac{1}{3}e'_{1}+\frac{2}{15}\bth{1},&
    c_{1}(\alpha^{(\mathbf{10})(1,-1,2)}_{j_1,j_2}) &= y'_{j_1}+y'_{j_2} +\frac{1}{3}e'_{1}-\frac{1}{3}e'_{2} + \frac{2}{15}\bth{1}, \\
    c_{1}(\alpha^{(\mathbf{10})(0,1,2)}_{j_1,j_2}) &= y'_{j_1}+y'_{j_2} +\frac{1}{3}e'_{2}+\frac{2}{15}\bth{1},&
    c_{1}(\alpha^{(\mathbf{\bar{5}})(0,-1,4)}_{j}) &= -y'_{j}-\frac{1}{3}e'_{2} + \frac{4}{15}\bth{1},\\
    c_{1}(\alpha^{(\mathbf{\bar{5}})(-1,1,4)}_{j}) &= -y'_{j} -\frac{1}{3}e'_{1}+\frac{1}{3}e'_{2} + \frac{4}{15}\bth{1},&
    c_{1}(\alpha^{(\mathbf{\bar{5}})(1,0,4)}_{j}) &= -y'_{j} +\frac{1}{3}e'_{1} + \frac{4}{15}\bth{1},\\
    c_{1}(\alpha^{(\mathbf{1})(2,-1,0)}) &= \frac{2}{3}e'_{1} - \frac{1}{3}e'_{2},&
    c_{1}(\alpha^{(\mathbf{1})(1,1,0)}) &= \frac{1}{3}e'_{1} +\frac{1}{3}e'_{2},\\
    c_{1}(\alpha^{(\mathbf{1})(-1,2,0)}) &= -\frac{1}{3}e'_{1} +\frac{2}{3}e'_{2},&
    c_{1}(\alpha^{(\mathbf{5})(0,0,6)}_{j}) &= y'_{j} + \frac{6}{15}\bth{1},
\end{aligned}\end{align}
where $j=1,\dots,5$ and $1\le j_{1}<j_{2}\le 5$.
In the superscripts, the first parentheses denote $\SU(5)$ representations, while the second parentheses specify the charges under the three $\U(1)$ factors.

Using the formulae~\eqref{eq:splitting} and~\eqref{eq:ch3}, and taking into account the ring structure of $H^{\ast}(E_{7}/T;\bR)$ reviewed in Appendix~\ref{subsubsec:cohom_E7G_Q}, we obtain
\begin{align}\begin{aligned}
    \ch{3}(T(E_{7}/H))=
    -\frac{32}{675}\bth{1}^{3}
    &+\frac{4}{45}\cpfi{1}\left((e'_{1})^{2}+e'_{1}e'_{2}+(e'_{2})^{2}\right)\\
    &+\frac{1}{27}\left(14(e'_{1})^{3}+21(e'_{1})^{2}e'_{2}+6e'_{1}(e'_{2})^{2}+4(e'_{2})^{3}\right)
    +\frac{1}{2}\cpfi{3}.
\end{aligned}\end{align}
This similar result has been obtained in~\cite{Yanagida:1985jc}.
Since the third Chern character does not vanish, additional degrees of freedom must be introduced to cancel the perturbative sigma model anomaly.

As in the $E_{7}/G$ case, the contribution proportional to $\cpfi{3}$ should be qualitatively distinguished from the remaining terms, which are cubic in degree-two classes.
The latter contributions may be cancelled by a Green-Schwarz mechanism in four dimensions.
The remaining contribution proportional to $\cpfi{3}$ can be cancelled by introducing massless fermions in the $\mathbf{\bar{5}}$ representation.
This prescription was proposed in~\cite{Yanagida:1985jc} to consistently define the $E_{7}/H$ model on $S^{4}$.

\subsubsection{Anomalies in gauged \texorpdfstring{$E_7/H$}{} model}
As mentioned in Section~\ref{subsubsec:model_H}, we focus on the case in which the entire isotropy group $H$ is gauged, motivated by the possibility that the $E_{7}/H$ model can be realized in F-theory~\cite{Mizoguchi:2014gva,Mizoguchi:2015kza}.
In this situation, global anomalies are encoded in the torsion part of $\Omega^{\Spin}_5((E_7/H)_H)$, while perturbative anomalies are encoded in the free part of $\Omega^{\Spin}_6((E_7/H)_H)$.

The computation in Appendix~\ref{sec:Borel_E7H} shows that the bordism group $\Omega^{\Spin}_5((E_7/H)_H)$ is trivial.
This result implies that no global anomalies arise even after gauging the symmetry $H$.

Perturbative anomalies, encoded in the free part of $\Omega^{\Spin}_6((E_7/H)_{H})$, are captured by the third Chern character of vector bundles on $(E_{7}/H)_{H}$.
The computation of the anomaly polynomial using the splitting principle proceeds in a manner similar to that in Section~\ref{subsubsec:gauged_anomaly_E7G}.

There are several possible ways to introduce additional degrees of freedom in order to construct phenomenologically viable models.
For instance, in~\cite{Sato:1997hv}, it is proposed that the mass spectrum and mixing angles of quarks and leptons can be explained by introducing one Higgs multiplet in the $\mathbf{5}$ representation, one in the $\mathbf{\bar{5}}$ representation, and two singlets of $H$ in the $E_7/H$ model.
A detailed analysis of anomalies in each specific scenario is left for future work.

\section{Discussion}
\label{sec:discussion}

In this paper, we have focused on anomalies arising from fermions coupled to the sigma model fields.
As mentioned in Section~\ref{sec:anomaly_family_unification}, two-form gauge fields, which are dual to periodic scalar fields in four dimensions, can also contribute to anomaly cancellation in family unification models via the Green-Schwarz mechanism.
In addition, the Wess-Zumino-Witten term~\cite{Wess:1971yu,Witten:1983tw} may play an important role in the analysis of anomalies in such models~\cite{Figueroa-OFarrill:1994uwr,Figueroa-OFarrill:1994vwl,Kobayashi:2019lep,Yonekura:2019vyz,Lee:2020ojw,Yonekura:2020upo,Saito:2024zxx}.
It would be interesting to investigate whether the anomalies in the $E_{7}/G$ and $E_{7}/H$ models can be cancelled by a suitable combination of these ingredients together with contributions from fermions, in a phenomenologically viable manner.

Although in this paper we have focused on the $E_{7}/G$ and $E_{7}/H$ models, other types of family unification models have also been proposed~\cite{Yanagida:1985jc,Irie:1983cd,Ibanez:1984ec,Ong:1984ej,Buchmuller:1985rc,Barr:1987pu}.
It would be interesting to study anomalies in these models from the viewpoint of bordism classification.
In particular, anomalies in models whose coset spaces are constructed from $E_{6}$ or $E_{8}$ may exhibit qualitatively different features from those considered in the present work.

As mentioned in the introduction, it has been suggested that the $E_{7}/H$ model can be realized by F-theory compactification~\cite{Mizoguchi:2014gva,Mizoguchi:2015kza}.
If string theory is consistent, any quantum field theory realized in the string landscape must be anomaly free.
From this perspective, the perturbative anomalies found in Section~\ref{subsec:anomaly_E7H} should be cancelled by some mechanism if the model is indeed realized within the F-theory framework.
Clarifying this mechanism may provide useful insights into identifying the Calabi-Yau fourfold associated with the $E_{7}/H$ model, and into understanding anomaly cancellation mechanisms, such as anomaly inflow, in the context of Type~IIB string theory or F-theory.

In the present work, we have focused on four-dimensional sigma models.
As discussed in Section~\ref{subsec:general_gauged_sigma}, bordism groups of the Borel construction of the target space play a crucial role when nontrivial symmetry actions on the target space are taken into account.
This perspective can also be applied to two-dimensional sigma models.
In particular, it may be useful for extracting some features of string backgrounds obtained via orbifold or orientifold constructions, by analyzing the corresponding world-sheet theories along the lines developed in~\cite{Kaidi:2019pzj,Kaidi:2019tyf,Witten:2023snr,BoyleSmith:2026oay}.

\section*{Acknowledgements}
The authors would like to thank Naoto~Kan, Masashi~Kawahira, and Shotaro~Kawanago for helpful discussions.
The authors are especially grateful to Kazuya~Yonekura for valuable comments.
The authors thank RFCMP2025-S at Yukawa Institute for Theoretical Physics; the current paper is influenced by Masashi~Kawahira's talk in this workshop.
The work of HW is supported in part by JSPS KAKENHI (Grant-in-Aid for JSPS fellows Grant Number 24KJ1603) and  JST FOREST Program (Grant Number JPMJFR2030, Japan).
The work of TS is supported by Graduate Program on Physics for the Universe (GP-PU) and AGS LEAP Program from Tohoku University. 

\appendix

\section{Topological data of \texorpdfstring{$BG$}{}}
\label{sec:BG}
In this appendix, we compute the cohomology groups of the classifying space $BG$ with integer coefficients up to degree six.
In Section~\ref{subsec:isom_G}, we first establish a useful group isomorphism that provides a convenient description of $G$ for the subsequent computations.
Using this isomorphism, the cohomology groups $H^{\ast}(BG;\bZ)$ are computed in Section~\ref{subsec:cohom_BG}.
We then use this result to compute the bordism group $\Omega^{\Spin}_{5}(BG)$ in Section~\ref{subsec:bordism_BG}.
This bordism group encodes the global anomaly of the four-dimensional theory on spin manifolds equipped with a $G$-bundle.

\subsection{Group isomorphism involving \texorpdfstring{$G$}{}}
\label{subsec:isom_G}
Here, we verify the following group isomorphism:
\begin{align}\label{eq:iso_a4a2t1}
  G=\frac{\SU(5)\times\SU(3)\times\U(1)}{\bZ_{5}\times\bZ_{3}}
  \cong \frac{\U(5)\times\U(3)}{\U(1)}.
\end{align}
This isomorphism is useful for computing the cohomology ring of the classifying space $BG$.

To this end, we consider the homomorphism
\begin{align}
\begin{array}{cccc}
    f_{G}: & \SU(5)\times\SU(3)\times\U(1) & \longrightarrow & \U(5)\times\U(3)\\
    &
    \rotatebox{90}{$\in$}&& \rotatebox{90}{$\in$} \\
    & (U_{5},V_{3},z) & \longmapsto & (z^{-6}U_{5},z^{5}V_{3}).
\end{array}
\end{align}
An element $(U_{5},V_{3},z)$ lies in $\Ker f_{G}$ if and only if
\begin{align}
  z^{-6}U_{5}=I_{5}, \qquad z^{5}V_{3}=I_{3},
\end{align}
where $I_{5}$ and $I_{3}$ denote the identity matrices of rank $5$ and $3$, respectively.
Since $\det U_{5}=\det V_{3}=1$, these conditions imply
\begin{align}
  z^{15}=1.
\end{align}
Thus,
\begin{align}
  \Ker f_{G}
  =\{(\zeta_{5}^{k}I_{5},\,\zeta_{3}^{l}I_{3},\,\zeta_{5}^{k}\zeta_{3}^{l})
  \mid k,l\in\bZ\}
  \cong \bZ_{5}\times\bZ_{3},
\end{align}
where $\zeta_{5}$ and $\zeta_{3}$ denote primitive fifth and third roots of unity, respectively.
Hence,
\begin{align}
  \Im f_{G}
  \cong \frac{\SU(5)\times\SU(3)\times\U(1)}{\bZ_{5}\times\bZ_{3}}.
\end{align}

Next, we define a homomorphism,
\begin{align}
\begin{array}{cccc}
    g_{G}: & \Im{f_{G}}\times\U(1) & \longrightarrow & \U(5)\times\U(3)\\
    &
    \rotatebox{90}{$\in$}&& \rotatebox{90}{$\in$} \\
    & ((\widetilde{U}_{5},\widetilde{V}_{3}),w) & \longmapsto & (w^{5}\widetilde{U}_{5},w^{-4}\widetilde{V}_{3}),
\end{array}
\end{align}
where $(\widetilde{U}_{5},\widetilde{V}_{3})\in \Im f_{G}$.
We claim that $g_{G}$ is an isomorphism.

To see this, we explicitly construct the inverse of $g_{G}$ as
\begin{align}
\begin{array}{cccc}
    h_{G}: & \U(5)\times\U(3) & \longrightarrow & \Im{f_{G}}\times\U(1)\\
    &
    \rotatebox{90}{$\in$}&& \rotatebox{90}{$\in$} \\
    & (U'_{5},V'_{3}) & \longmapsto & ((U'_{5}(\det{U'_{5}})^{-5}(\det{V'_{3}})^{-10},V'_{3}(\det{U'_{5}})^{4}(\det{V'_{3}})^{8}),\det{U'_{5}}(\det{V'_{3}})^{2}).
\end{array}
\end{align}
The first component lies in $\Im f_{G}$, since an element
$(\widetilde{U}_{5},\widetilde{V}_{3})\in\U(5)\times\U(3)$ belongs to $\Im f_{G}$
if and only if
\begin{align}
  \det\widetilde{U}_{5}\,(\det\widetilde{V}_{3})^{2}=1,
\end{align}
and one can check that
\begin{align}
  \det\!\left[U'_{5}(\det U'_{5})^{-5}(\det V'_{3})^{-10}\right]
  (\det\!\left[V'_{3}(\det U'_{5})^{4}(\det V'_{3})^{8}\right])^{2}=1.
\end{align}
Thus, $h_{G}$ is well-defined.
A direct computation shows that $h_{G}\circ g_{G}=\mathrm{id}_{\Im{f_{G}}\times\U(1)}$ and $g_{G}\circ h_{G}=\mathrm{id}_{\U(5)\times\U(3)}$.
Therefore, $g_{G}$ is an isomorphism, and we conclude that
\begin{align}
    \frac{\SU(5)\times\SU(3)\times\U(1)}{\bZ_{5}\times\bZ_{3}}\cong\Im{f_{G}}\cong\frac{\Im{f_{G}}\times\U(1)}{\U(1)}\cong\frac{\U(5)\times\U(3)}{\U(1)}.
\end{align}

\subsection{Cohomology groups \texorpdfstring{$H^{\ast}(BG;\bZ)$}{} up to degree six}
\label{subsec:cohom_BG}
We describe the strategy for computing the cohomology groups of $BG$, imitating the method developed in Ref.~\cite{MR4284619}.~\footnote{For the background knowledge of the Serre spectral sequence required throughout the appendix, see e.g. Chapter~3.2 of~\cite{Mimura:1991}}
For our purpose, it is enough to determine the cohomology group up to degree six.
In order to achieve the goal, three fibrations are utilized.
The isomorphism~\eqref{eq:iso_a4a2t1} implies that we have the short exact sequence of Lie groups:
\begin{align}
    1 \to \U(1) \to \U(5)\times\U(3) \to G \to 1.
\end{align}
Since $B^2{\U(1)}\cong K(\bZ,3)$, the fibration
\begin{align}\label{eq:fibration_first}
    B\U(5)\times B\U(3) \to BG \to K(\bZ,3)
\end{align}
is obtained from the short exact sequence of the Lie groups.
This fibration is one of the ingredients for the computation of the cohomology groups.
We denote by $\UE{n}{p}{q}$ the element in the $E_{n}$-page of the Serre spectral sequence for the cohomology ring with integer coefficients associated with this fiber sequence.
We can also write the short exact sequence of the Lie groups,
\begin{align}
    1 \to \U(1) \to \U(1)^{5}\times\U(1)^{3} \to \frac{\U(1)^{5}\times\U(1)^{3}}{\U(1)} \to 1.
\end{align}
The second homomorphism in this short exact sequence is defined by
\begin{align}\label{eq:induce_gG}
\begin{array}{cccc}
    & \U(1) & \longrightarrow & \U(1)^{5}\times\U(1)^{3}\\
    &
    \rotatebox{90}{$\in$}&& \rotatebox{90}{$\in$} \\
    & w & \longmapsto & (w^{5},w^{5},w^{5},w^{5},w^{5},w^{-4},w^{-4},w^{-4}).
\end{array}
\end{align}
The groups $\U(1)^{5}\times\U(1)^{3}$ and $\frac{\U(1)^{5}\times\U(1)^{3}}{\U(1)}$ are regarded as the maximal tori of $\U(5)\times\U(3)$ and $G$, respectively.
From this short exact sequence, one obtains the fiber sequence:
\begin{align}
    B\U(1)^{5}\times B\U(1)^{3} \to B\frac{\U(1)^{5}\times\U(1)^{3}}{\U(1)} \to K(\bZ,3).
\end{align}
Let $\TE{n}{p}{q}$ be the element in the $E_{n}$-page of the Serre spectral sequence associated with this fiber sequence.
The third ingredient for computing the cohomology groups of $BG$ is the fibration given by
\begin{align}
    B\U(1) \to PK(\bZ,3) \to K(\bZ,3),
\end{align}
where $PK(\bZ,3)$ is the path space of $K(\bZ,3)$.
We denote by $\KE{n}{p}{q}$ the element of the Serre spectral sequence for the cohomology ring associated with this path fibration.

In order to outline the strategy of the computation, we prepare the following commutative diagram involving three fiber sequences introduced above:
\begin{align}\label{eq:com_diagram_cohom}
\xymatrix{
B\U(1) \ar[d]^{}  \ar[r]^{} & PK(\bZ,3) \ar[d]^{} \ar[r]^{} & K(\bZ,3) \ar[d]^{\rotatebox{-90}{$\cong$}}\\
B\U(1)^{5}\times B\U(1)^{3} \ar[d]^{}  \ar[r]^{} & B\frac{\U(1)^{5}\times\U(1)^{3}}{\U(1)} \ar[d]^{} \ar[r]^{} & K(\bZ,3) \ar[d]^{\rotatebox{-90}{$\cong$}}\\
B\U(5)\times B\U(3) \ar[r]^{} & BG \ar[r] & K(\bZ,3)
}
\end{align}
For the calculation of the cohomology ring $H^{\ast}(BG;\bZ)$, we need to identify the differential maps $\Ud{n}{p}{q}:\UE{n}{p}{q}\to\UE{n}{p+n}{q-n+1}$.
To this end, we first identify the differential map $\Td{n}{p}{q}:\TE{n}{p}{q}\to\TE{n}{p+n}{q-n+1}$ by comparing $\KE{n}{p}{q}$ to $\TE{n}{p}{q}$.
Note that the vertical maps from the first row to the second row in the diagram~\eqref{eq:com_diagram_cohom} induce the homomorphisms $\ssf{n}{p}{q}:\TE{n}{p}{q}\to\KE{n}{p}{q}$ such that the following diagram commutes:
\begin{align}\label{eq:com_diagram_f}
\xymatrix{
\KE{n}{p}{q} \ar[r]^-{\Kd{n}{p}{q}} & \KE{n}{p+n}{q-n+1}\\
\TE{n}{p}{q} \ar[u]^-{\ssf{n}{p}{q}} \ar[r]^-{\Td{n}{p}{q}} & \TE{n}{p+n}{q-n+1} \ar[u]_-{\ssf{n}{p+n}{q-n+1}}
}
\end{align}
Due to this commutative feature of $\ssf{n}{p}{q}$, we can investigate the differential map $\Td{n}{p}{q}$ from information about $\Kd{n}{p}{q}$.
Similarly, the vertical maps from the second row to the third row in the diagram~\ref{eq:com_diagram_cohom} induce the homomorphisms $\ssg{n}{p}{q}:\UE{n}{p}{q}\to\TE{n}{p}{q}$ such that the following diagram commutes:
\begin{align}\label{eq:com_diagram_g}
\xymatrix{
\TE{n}{p}{q} \ar[r]^-{\Td{n}{p}{q}} & \TE{n}{p+n}{q-n+1}\\
\UE{n}{p}{q} \ar[u]^-{\ssg{n}{p}{q}} \ar[r]^-{\Ud{n}{p}{q}} & \UE{n}{p+n}{q-n+1} \ar[u]_-{\ssg{n}{p+n}{q-n+1}}
}
\end{align}
Thus, once the homomorphisms $\Td{n}{p}{q}$ are identified, we can study $\Ud{n}{p}{q}$ through this commutative diagram.

\subsubsection{Cohomology ring of \texorpdfstring{$K(\bZ,3)$}{} and differential maps \texorpdfstring{$\Kd{n}{p}{q}$}{}}\label{subsub:Kdn}
We summarize the cohomology ring of $K(\bZ,3)$ and the differential maps $\Kd{n}{p}{q}$.
For the detailed derivation of the facts expressed here, see Section.~2 in Ref.~\cite{MR4284619}.

The nontrivial cohomology groups of $K(\bZ,3)$ up to degree six are given by
\begin{align}\begin{aligned}
    H^{0}(K(\bZ,3);\bZ)&\cong\bZ,&
    H^{3}(K(\bZ,3);\bZ)&\cong \bZ,&
    H^{6}(K(\bZ,3);\bZ)&\cong \bZ_{2}.
\end{aligned}\end{align}
Let $x_{1}$ and $y_{2,0}$ be the generators of $H^{3}(K(\bZ,3);\bZ)$ and $H^{6}(K(\bZ,3);\bZ)$, respectively.
The ring structure is given by $x_{1}\smallsmile x_{1}=y_{2,0}$.

The $E_{2}$-page of the Serre spectral sequence associated with the first row in the diagram~\eqref{eq:com_diagram_cohom} is given by
\begin{align}
    \KE{2}{p}{q}=H^{p}(K(\bZ,3);\bZ)\otimes H^{q}(B\U(1);\bZ).
\end{align}
The cohomology ring $H^{\ast}(B\U(1);\bZ)$ is generated by the first Chern class, which is denoted by $t$, i.e.~$H^{\ast}(B\U(1);\bZ)\cong\bZ[t]$.
Obviously, all the differential maps $\Kd{2}{p}{q}$ in the $E_{2}$-page are trivial and $\KE{3}{p}{q}\cong\KE{2}{p}{q}$.
The differential map $\Kd{3}{0}{2}:\KE{3}{0}{2}\to\KE{3}{3}{0}$ is given by $\Kd{3}{0}{2}(t)=x_{1}$.
The other differential maps in the $E_{3}$-pages are determined from $\Kd{3}{0}{2}$, since they are compatible with the product.
For instance, we have
\begin{align}\begin{aligned}
    \Kd{3}{0}{4}(t^{2})&=2\Kd{3}{0}{2}(t)t=2x_{1}t,\\
    \Kd{3}{0}{6}(t^{3})&=3\Kd{3}{0}{2}(t)t^{2}=2x_{1}t^{2},\\
    \Kd{3}{3}{2}(x_{1}t)&=\Kd{3}{3}{0}(x_{1})t+x_{1}\Kd{3}{0}{2}(t)=0+x_{1}^{2}=y_{2,0}.
\end{aligned}\end{align}

\subsubsection{Differential map \texorpdfstring{$\Td{3}{0}{2}$}{}}
\label{subsub:induce_f2}
In order to determine the differential map $\Td{3}{0}{2}$ required in the calculation below, we identify the induced homomorphisms $\ssf{2}{p}{q}$.
The $E_{2}$-page of the Serre spectral sequence associated with the second row in the diagram~\eqref{eq:com_diagram_cohom} is given by
\begin{align}
    \TE{2}{p}{q}=H^{p}(K(\bZ,3);\bZ)\otimes H^{p}(B\U(1)^{5}\times B\U(1)^{3};\bZ).
\end{align}
The cohomology ring of $B\U(1)^{5}\times B\U(1)^{3}$ is written as
\begin{align}\label{eq:cohom_T8}
    H^{\ast}(B\U(1)^{5}\times B\U(1)^{3};\bZ)\cong\bZ[u_{1},\dots,u_{5},v_{1},\dots,v_{3}],
\end{align}
where $u_{j}\,(j=1,2,3,4,5)$ stand for the first Chern classes of $\U(1)^{5}$, whereas $v_{k}\,(k=1,2,3)$ stand for those of $\U(1)^{3}$.

We first focus on the homomorphism $\ssf{2}{0}{2}:\TE{2}{0}{2}\to\KE{2}{0}{2}$.
The group $\TE{2}{0}{2}\cong H^{2}(B\U(1)^{5}\times B\U(1)^{3};\bZ)$ is generated by these first Chern classes $u_{i}$ and $v_{j}$.
We denote by $t$ the generator of $\KE{2}{0}{2}\cong H^{2}(B\U(1);\bZ)$ as before.
The map $B\U(1)\to B\U(1)^{5}\times B\U(1)^{3}$ in the diagram~\eqref{eq:com_diagram_cohom} is induced from the homomorphism~\eqref{eq:induce_gG}.
As a result, the homomorphism $\ssf{2}{0}{2}$ is determined as
\begin{align}\begin{aligned}\label{eq:f202}
    \ssf{2}{0}{2}(u_{i})&=5t &\quad (i&=1,2,3,4,5),\\
    \ssf{2}{0}{2}(v_{j})&=-4t &\quad (j&=1,2,3).
\end{aligned}\end{align}

Since the vertical maps between the base spaces in the diagram~\eqref{eq:com_diagram_cohom} are the identity maps, the induced map $\ssf{2}{3}{0}:\TE{2}{3}{0}\to\KE{2}{3}{0}$ is also the identity homomorphism. 

The other induced homomorphisms $\ssf{2}{p}{q}$ between the $E_{2}$-pages are determined from $\ssf{2}{0}{2}$ and $\ssf{2}{3}{0}$.
For instance, $\ssf{2}{3}{2}:\TE{2}{3}{2}\to\KE{2}{3}{2}$ is given by
\begin{align}\begin{aligned}
    \ssf{2}{3}{2}(x_{1}u_{i})&=\ssf{2}{3}{0}(x_{1})\ssf{2}{0}{2}(u_{i})=5x_{1}t &\quad (i&=1,2,3,4,5),\\
    \ssf{2}{3}{2}(x_{1}v_{j})&=\ssf{2}{3}{0}(x_{1})\ssf{2}{0}{2}(v_{j})=-4x_{1}t &\quad (j&=1,2,3).
\end{aligned}\end{align}

The elements in the $E_{3}$-page are given by $\TE{3}{p}{q}\cong\TE{2}{p}{q}$, since all the differential maps $\Td{2}{p}{q}$ are trivial.
As mentioned in Section~\ref{subsub:Kdn}, it also follows that $\KE{3}{p}{q}\cong\TE{2}{p}{q}$.
From these facts, it turns out that $\ssf{3}{p}{q}=\ssf{2}{p}{q}$.

By applying the commutativity of the diagram~\eqref{eq:com_diagram_f}, we can identify the differential map $\Td{3}{0}{2}$.
From Eq.~\eqref{eq:f202}, we have the relations,
\begin{align}\begin{aligned}
    \ssf{3}{3}{0}\circ\Td{3}{0}{2}(u_{i})&=\Kd{3}{0}{2}\circ\ssf{3}{0}{2}(u_{i})=\Kd{3}{0}{2}(5t)=5x_{1} &\quad (i&=1,2,3,4,5),\\
    \ssf{3}{3}{0}\circ\Td{3}{0}{2}(v_{j})&=\Kd{3}{0}{2}\circ\ssf{3}{0}{2}(v_{j})=\Kd{3}{0}{2}(-4t)=-4x_{1} &\quad (j&=1,2,3).
\end{aligned}\end{align}
Since $\ssf{3}{3}{0}$ is the identity map, these relations imply that
\begin{align}\begin{aligned}\label{eq:Td302}
    \Td{3}{0}{2}(u_{i})&=5x_{1} & (i&=1,2,3,4,5), &\quad
    \Td{3}{0}{2}(v_{j})&=-4x_{1} & (j&=1,2,3).
\end{aligned}\end{align}
The other differential maps in the $E_{3}$-page are determined by the Leibniz rule.
For example, $\Td{3}{3}{2}:\TE{3}{3}{2}\to\TE{3}{6}{0}$ is determined as
\begin{align}\begin{aligned}
    \Td{3}{3}{2}(x_{1}u_{i})&=\Td{3}{3}{0}(x_{1})u_{i}-x_{1}\Td{3}{0}{2}(u_{i})=0-5x_{1}^{2}\equiv y_{2,0} &\quad (i&=1,2,3,4,5),\\
    \Td{3}{3}{2}(x_{1}v_{j})&=\Td{3}{3}{0}(x_{1})v_{j}-x_{1}\Td{3}{0}{2}(v_{j})=0+4x_{1}^{2}\equiv 0 &\quad (j&=1,2,3).
\end{aligned}\end{align}

\subsubsection{Differential maps \texorpdfstring{$\Ud{3}{p}{q}$}{}}\label{subsubsec:Gd3pq}
We next identify the differential maps $\Ud{3}{p}{q}$ by comparing $\TE{3}{p}{q}$ to $\UE{3}{p}{q}$ through the commutative diagram~\eqref{eq:com_diagram_g}.
The $E_{2}$-page of the Serre spectral sequence associated with the third row in the diagram~\eqref{eq:com_diagram_cohom} is given by
\begin{align}
    \UE{2}{p}{q}=H^{p}(K(\bZ,3);\bZ)\otimes H^{q}(B\U(5)\times B\U(3);\bZ).
\end{align}
The cohomology ring of $B\U(5)\times B\U(3)$ is written as
\begin{align}
    H^{\ast}(B\U(5)\times B\U(3);\bZ)\cong\bZ[\cfi{1},\dots,\cfi{5},\cth{1},\dots,\cth{3}],
\end{align}
where $\cfi{i}\,(i=1,2,3,4,5)$ is the $i$-th Chern class of $\U(5)$, whereas $\cth{j}\,(j=1,2,3)$ is the $j$-th Chern class of $\U(3)$.
In particular, the second cohomology group $H^{2}(B\U(5)\times B\U(3);\bZ)$ is generated by $\cfi{1}$ and $\cth{1}$.

We introduce the notation $\sigma_{i}(\cdots)$ that stands for the degree $i$ symmetric polynomial of the arguments.
For instance, $\sigma_{2}(v_{1},v_{2},v_{3})=v_{1}v_{2}+v_{1}v_{3}+v_{2}v_{3}$.
The induced maps $\ssg{2}{p}{q}$ arising in the computation later are summarized as follows:
\begin{itemize}
    \item
    $\ssg{2}{0}{2}:\UE{2}{0}{2}\to\TE{2}{0}{2}$:
    \begin{align}\label{eq:g202}
      \ssg{2}{0}{2}(\cfi{1})&=\sigma_{1}(u_{1},\dots,u_{5}),&
      \ssg{2}{0}{2}(\cth{1})&=\sigma_{1}(v_{1},v_{2},v_{3}).
    \end{align}
    \item
    $\ssg{2}{0}{4}:\UE{2}{0}{4}\to\TE{2}{0}{4}$:
    \begin{align}\label{eq:g204}
      \ssg{2}{0}{4}(\cfi{2})&=\sigma_{2}(u_{1},\dots,u_{5}),&
      \ssg{2}{0}{4}(\cth{2})&=\sigma_{2}(v_{1},v_{2},v_{3}).
    \end{align}
    \item
    $\ssg{2}{0}{6}:\UE{2}{0}{6}\to\TE{2}{0}{2}$:
    \begin{align}
      \ssg{2}{0}{6}(\cfi{3})&=\sigma_{3}(u_{1},\dots,u_{5}),&
      \ssg{2}{0}{6}(\cth{3})&=\sigma_{3}(v_{1},v_{2},v_{3}).
    \end{align}
    \item
    $\ssg{2}{3}{0}:\UE{2}{3}{0}\to\TE{2}{3}{0}$:
    \begin{align}
    \ssg{2}{3}{0}(x_{1})=x_{1}
    \end{align}
\end{itemize}
Since $\Ud{2}{p}{q}=0$ and $\Td{2}{p}{q}=0$, it turns out that $\ssg{3}{p}{q}=\ssg{2}{p}{q}$.

Now, we identify the differential maps $\Ud{3}{p}{q}$ for relevant $p$ and $q$ in turn, and investigate $\Ker\Ud{3}{p}{q}$ and $\Im\Ud{3}{p}{q}$.
\begin{itemize}
    \item $\Ud{3}{0}{2}:\UE{3}{0}{2}\to\UE{3}{3}{0}$:\\
    From the commutativity of the diagram~\eqref{eq:com_diagram_g}, we have the relations,
    \begin{align}\begin{aligned}
        \ssg{3}{3}{0}\circ\Ud{3}{0}{2}(\cfi{1})&=\Td{3}{0}{2}\circ\ssg{3}{0}{2}(\cfi{1})=\Td{3}{0}{2}(\sigma_{1}(u_{1},\dots,u_{5}))=25x_{1},\\
        \ssg{3}{3}{0}\circ\Ud{3}{0}{2}(\cth{1})&=\Td{3}{0}{2}\circ\ssg{3}{0}{2}(\cth{1})=\Td{3}{0}{2}(\sigma_{1}(v_{1},v_{2},v_{3}))=-12x_{1}.
    \end{aligned}\end{align}
    To obtain these expressions, we have used the formulae~\eqref{eq:Td302} and \eqref{eq:g202}.
    These relations imply that
    \begin{align}\label{eq:Gd302}
        \Ud{3}{0}{2}(\cfi{1})&=25x_{1},&
        \Ud{3}{0}{2}(\cth{1})&=-12x_{1}.
    \end{align}
    Then, we conclude that
    \begin{align}
        \Ker\Ud{3}{0}{2}&\cong\bZ(12\cfi{1}+25\cth{1}), &
        \Im\Ud{3}{0}{2}&\cong\UE{3}{3}{0}.
    \end{align}
    \item $\Ud{3}{0}{4}:\UE{3}{0}{4}\to\UE{3}{3}{2}$:\\
    Let us determine the image of $\cfi{2}$ under $\Ud{3}{0}{4}$.
    From the commutativity of the diagram~\eqref{eq:com_diagram_g} and the formula~\eqref{eq:g204}, we have
    \begin{align}
    \ssg{3}{3}{2}\circ\Ud{3}{0}{4}(\cfi{2})=\Td{3}{0}{4}\circ\ssg{3}{0}{4}(\cfi{2})=\Td{3}{0}{4}(\sigma_{2}(u_{1},\dots,u_{5})).
    \end{align}
    To proceed the calculation, note that the differential map $\Td{3}{0}{4}$ satisfies the Leibniz rule.
    For instance, it holds that
    \begin{align}
    \Td{3}{0}{4}(u_{1}u_{2})=\Td{3}{0}{2}(u_{1})u_{2}+u_{1}\Td{3}{0}{2}(u_{2})=5x_{1}(u_{1}+u_{2}).
    \end{align}
    By applying the Leibniz rule, we can obtain $\ssg{3}{3}{2}\circ\Ud{3}{0}{4}(\cfi{2})=20x_{1}\sigma_{1}(u_{1},\dots,u_{5})$.
    This result implies that $\Ud{3}{0}{4}(\cfi{2})=20x_{1}\cfi{1}$.
    In the same way, we can determine the image of $\cth{2}$ under $\Ud{3}{0}{4}$ as $\Ud{3}{0}{4}(\cth{2})=-8x_{1}\cth{1}$.
    We can also determine the images of $(\cfi{1})^{2}$, $\cfi{1}\cth{1}$, and $(\cth{1})^{2}$ under $\Ud{3}{0}{4}$ by using the Leibniz rule and the formula~\eqref{eq:Gd302}.
    For instance, we have
    \begin{align}
      \Ud{3}{0}{4}(\cfi{1}\cth{1})=\Ud{3}{0}{2}(\cfi{1})\cth{1}+\Ud{3}{0}{2}(\cth{1})\cfi{1}=x_{1}(-12\cfi{1}+25\cth{1}).
    \end{align}
    The results are summarized as follows:
    \begin{align}\begin{cases}\label{eq:Gd304}
        \Ud{3}{0}{4}(\cfi{2})=20x_{1}\cfi{1},\\
        \Ud{3}{0}{4}(\cth{2})=-8x_{1}\cth{1},\\
        \Ud{3}{0}{4}((\cfi{1})^{2})=50x_{1}\cfi{1},\\
        \Ud{3}{0}{4}(\cfi{1}\cth{1})=x_{1}(-12\cfi{1}+25\cth{1}),\\
        \Ud{3}{0}{4}((\cth{1})^{2})=-24x_{1}\cth{1}.
    \end{cases}\end{align}
    After some calculations,\footnote{One convenient way to determine $\Ker{\Ud{3}{0}{4}}$ and $\Im{\Ud{3}{0}{4}}$ is as follows. Since $\UE{3}{3}{2}$ is a free $\bZ$-module generated by $x_{1}\cfi{1}$ and $x_{1}\cth{1}$, the map $\Ud{3}{0}{4}$ can be represented by a $2\times 5$ matrix with integer entries, as encoded in~\eqref{eq:Gd304}. Thus, the problem reduces to computing the Smith normal form of this matrix.} it can be checked that
    \begin{align}\begin{aligned}
        \Ker\Ud{3}{0}{4}&=\bZ(-5\cfi{2}+2(\cfi{1})^{2})\oplus\bZ(24\cfi{2}+125\cth{2}+40\cfi{1}\cth{1})\oplus\bZ(-3\cth{2}+(\cth{1})^{2}),\\
        \Im\Ud{3}{0}{4}&=\bZ x_{1}(2\cfi{1}-\cth{1})\oplus\bZ(-2x_{1}\cfi{1}).
    \end{aligned}\end{align}
    \item $\Ud{3}{0}{6}:\UE{3}{0}{6}\to\UE{3}{3}{4}$:\\
    In the same way as before, we can determine the differential map $\Ud{3}{0}{6}$ by using the formulae so far.
    The results are summarized as follows:
    \begin{align}\begin{cases}
        \Ud{3}{0}{6}(\cfi{3})=15x_{1}\cfi{2},\\
        \Ud{3}{0}{6}(\cth{3})=-4x_{1}\cth{2},\\
        \Ud{3}{0}{6}(\cfi{2}\cfi{1})=x_{1}(25\cfi{2}+20(\cfi{1})^{2}),\\
        \Ud{3}{0}{6}(\cfi{2}\cth{1})=x_{1}(-12\cfi{2}+20\cfi{1}\cth{1}),\\
        \Ud{3}{0}{6}(\cth{2}\cfi{1})=x_{1}(25\cth{2}-8\cfi{1}\cth{1}),\\
        \Ud{3}{0}{6}(\cth{2}\cth{1})=x_{1}(-12\cth{2}-8(\cth{1})^{2}),\\
        \Ud{3}{0}{6}((\cfi{1})^{3})=75x_{1}(\cfi{1})^{2},\\
        \Ud{3}{0}{6}((\cfi{1})^{2}\cth{1})=x_{1}(-12(\cfi{1})^{2}+50\cfi{1}\cth{1}),\\
        \Ud{3}{0}{6}(\cfi{1}(\cth{1})^{2})=x_{1}(-24\cfi{1}\cth{1}+25(\cth{1})^{2}),\\
        \Ud{3}{0}{6}((\cth{1})^{3})=-36x_{1}(\cth{1})^{2}.
    \end{cases}\end{align}
    It can be also verified that
    \begin{align}\begin{aligned}
        \Ker\Ud{3}{0}{6}
        \cong&\bZ(32\cfi{3}+-275\cth{3}+40\cfi{2}\cth{1}+4\cth{2}\cfi{1}+100\cth{2}\cth{1}+32\cfi{1}(\cth{1})^{2})\\
        &\oplus\bZ(25\cfi{3}-15\cfi{2}\cfi{1}+4(\cfi{1})^{3})\\
        &\oplus\bZ(-30\cfi{3}+6\cfi{2}\cfi{1}-25\cfi{2}\cth{1}+10(\cfi{1})^{2}\cth{1})\\
        &\oplus\bZ(96\cfi{3}-924\cth{3}+120\cfi{2}\cth{1}+308\cth{2}\cth{1}+100\cfi{1}(\cth{1})^{2}+(\cth{1})^{3})\\
        &\oplus\bZ(27\cth{3}-9\cth{2}\cth{1}+2(\cth{1})^{3})
    \end{aligned}\end{align}
    \item $\Ud{3}{3}{2}:\UE{3}{3}{2}\to\UE{3}{6}{0}$:\\
    As before, it follows from the commutativity of the diagram~\eqref{eq:com_diagram_g} that
    \begin{align}\begin{aligned}
        \ssg{3}{6}{0}\circ\Ud{3}{3}{2}(x_{1}\cfi{1})&=\Td{3}{3}{2}\circ\ssg{3}{3}{2}(x_{1}\cfi{1})=\Td{3}{3}{2}(x_{1}\sigma_{1}(u_{1},\dots,u_{5}))=-25x_{1}^{2}\equiv y_{2,0},\\
        \ssg{3}{6}{0}\circ\Ud{3}{3}{2}(x_{1}\cth{1})&=\Td{3}{3}{2}\circ\ssg{3}{3}{2}(x_{1}\cth{1})=\Td{3}{3}{2}(x_{1}\sigma_{1}(v_{1},v_{2},v_{3}))=12x_{1}^{2}\equiv 0.
    \end{aligned}\end{align}
    Since $\ssg{3}{0}{6}$ is the identify map, we obtain
    \begin{align}
        \Ud{3}{3}{2}(x_{1}\cfi{1})&=y_{2,0},&
        \Ud{3}{3}{2}(x_{1}\cth{1})&=0.
    \end{align}
    Then, the kernel and the image of $\Ud{3}{3}{2}$ are given by
    \begin{align}
        \Ker\Ud{3}{3}{2}&\cong\bZ 2x_{1}\cfi{1}\oplus\bZ x_{1}\cth{1},&
        \Im\Ud{3}{3}{2}&\cong\UE{3}{6}{0}.
    \end{align}
\end{itemize}

\subsubsection{Cohomology groups of \texorpdfstring{$BG$}{}}
Here, we calculate the cohomology groups of $BG$ up to degree six.
To this end, we apply the Serre spectral sequence to the fibration~\eqref{eq:fibration_first}.
See Table~\ref{ss:cohom_BA4A3T1} for the situation of the $E_{3}$-page.
From the properties of the differential maps $\Ud{3}{p}{q}$ derived in Section~\ref{subsubsec:Gd3pq}, the following results are immediately obtained:
\begin{itemize}
    \item $H^{1}(BG;\bZ)\cong 0$.
    \item $H^{2}(BG;\bZ)\cong \UE{\infty}{0}{2}\cong\UE{4}{0}{2}\cong\Ker\Ud{3}{0}{2}\cong\bZ$.
    \item $H^{3}(BG;\bZ)\cong \UE{\infty}{3}{0}\cong\UE{4}{3}{0}\cong\Ker\Ud{3}{3}{0}/\Im\Ud{3}{0}{2}\cong 0$.
    \item $H^{4}(BG;\bZ)\cong \UE{\infty}{0}{4}\cong\UE{4}{0}{4}\cong\Ker\Ud{3}{0}{4}\cong\bZ^{\oplus 3}$.
    \item $H^{5}(BG;\bZ)\cong \UE{\infty}{3}{2}\cong\UE{4}{3}{2}\cong\Ker\Ud{3}{3}{2}/\Im\Ud{3}{0}{4}\cong 0$.
    \item $H^{6}(BG;\bZ)\cong \UE{\infty}{0}{6}\cong\UE{4}{0}{6}\cong\Ker\Ud{3}{0}{6}\cong\bZ^{\oplus 5}$.
\end{itemize}
\begin{table}[thp]\begin{center}
	\begin{tabular}{ccc}
		\begin{tabular}{c|ccccccc}
			6 & $\bZ^{\oplus 10}$ & $\hphantom{\bZ_2}$ & $\hphantom{\bZ_2}$ & $\ast$ & $\hphantom{\bZ_2}$ & $\hphantom{\bZ_2}$ & $\ast$\\
			5 & & & & & & & \\
			4 & $\bZ^{\oplus 5}$ & & & $\bZ^{\oplus 5}$ & $\hphantom{\bZ_2}$ & & $\ast$ \\
			3 & & & & & & & \\
			2 & $\bZ^{\oplus 2}$ & & & $\bZ^{\oplus 2}$ & & & $\ast$ \\
			1 & & & & & & & \\
			0 & $\bZ$ & & & $\bZ$ & & & $\bZ_{2}$ \\
			\hline
			& 0 & 1 & 2 & 3 & 4 & 5 & 6
		\end{tabular}
		& $\quad \Longrightarrow$ &
		\begin{tabular}{c|c}
			6 & $\bZ^{\oplus 5}$ \\
			5 & \\
			4 & $\bZ^{\oplus 3}$ \\
			3 & \\
			2 & $\bZ$ \\
			1 & \\
			0 & $\bZ$ \\
			\hline
			&
		\end{tabular}
	\end{tabular}
\caption{The $E_{3}$-page of the Serre spectral sequence associated with the fibration~\eqref{eq:fibration_first}.}
\label{ss:cohom_BA4A3T1}
\end{center}\end{table}

\subsection{Bordism group \texorpdfstring{$\Omega^{\Spin}_{5}(BG)$}{}}
\label{subsec:bordism_BG}
Based on the above analysis, let us calculate the spin bordism group $\Omega^{\Spin}_{5}(BG)$, which captures the global anomaly of the gauge theory with the structure group $G$ in four dimensions.
We apply the Atiyah-Hirzebruch spectral sequence to the fibration,~\footnote{See e.g.~\cite{Garcia-Etxebarria:2018ajm}, for the review of the Atiyah-Hirzebruch spectral sequence aimed at physisist.}
\begin{align}\label{eq:fibration_trivial_BG}
    \mathrm{pt} \to BG \to BG.
\end{align}
The homology groups required to write down the $E_{2}$-page of the spectral sequence are summarized in Table~\ref{tab:hom_cohom_BA4A3T1}.
The homology groups with integer coefficients can be read off from the cohomology group with integer coefficients via the universal coefficient theorem,
\begin{align}
    H^{p}(BG;\bZ)\cong\Ext(H_{p-1}(BG;\bZ),\bZ)\oplus\Hom(H_{p}(BG;\bZ),\bZ).
\end{align}
Since $H^{p}(BG;\bZ)$ is torsion free up to degree $6$, it turns out that $H_{p}(BG;\bZ)\cong H^{p}(BG;\bZ)$ for $p\le 6$.
Then, we can also determine the cohomology and homology groups with $\bZ_{2}$ coefficients by the universal coefficient theorem.
\begin{table}[thp]\begin{center}
	\begin{tabular}{c||ccccccc}
		$p$ & $0$ & $1$ & $2$ & $3$ & $4$ & $5$ & $6$ \\
        \hline
        $H_{p}(BG;\bZ)$ & $\bZ$ & $0$ & $\bZ$ & $0$ & $\bZ^{\oplus 3}$ & $0$ & $\bZ^{\oplus 5}$ \\
        $H^{p}(BG;\bZ)$ & $\bZ$ & $0$ & $\bZ$ & $0$ & $\bZ^{\oplus 3}$ & $0$ & $\bZ^{\oplus 5}$ \\
        $H_{p}(BG;\bZ_{2})$ & $\bZ_{2}$ & $0$ & $\bZ_{2}$ & $0$ & $\bZ_{2}^{\oplus 3}$ & $0$ & $\bZ_{2}^{\oplus 5}$ \\
        $H^{p}(BG;\bZ_{2})$ & $\bZ_{2}$ & $0$ & $\bZ_{2}$ & $0$ & $\bZ_{2}^{\oplus 3}$ & $0$ & $\bZ_{2}^{\oplus 5}$ \\
	\end{tabular}
\caption{Homology and cohomology groups of $BG$ up to degree six.}
\label{tab:hom_cohom_BA4A3T1}
\end{center}\end{table}

The $E^{2}$-page of the Atiyah-Hirzebruch spectral sequence that converges to $\widetilde{\Omega}^{\Spin}_{\ast}(BG)$ is written as
\begin{align}
    E^{2}_{p,q}=\widetilde{H}_{p}(BG;\Omega^{\Spin}_{q}({\rm pt})),
\end{align}
where $\widetilde{\Omega}^{\Spin}_{\ast}(BG)$ is the reduced spin bordism group of $BG$.
In Table~\ref{tab:bordism_pt}, the spin bordism groups of a point are summarized for lower degrees.
\begin{table}[thp]\begin{center}
	\begin{tabular}{c||ccccccc}
		$q$ & $0$ & $1$ & $2$ & $3$ & $4$ & $5$ & $6$ \\
        \hline
        $\Omega^{\Spin}_{p}({\rm pt})$ & $\bZ$ & $\bZ_{2}$ & $\bZ_{2}$ & $0$ & $\bZ$ & $0$ & $0$ \\
	\end{tabular}
\caption{The spin bordism groups of a point.}
\label{tab:bordism_pt}
\end{center}\end{table}
In Table~\ref{ss:bordism_BA4A3T1}, the situation in the $E^{2}$-page is exhibited.
\begin{table}[thp]\begin{center}
	\begin{tabular}{c|ccccccc}
		5 & $\hphantom{\bZ_2}\cellcolor{lightyellow}$ & $\hphantom{\bZ_2}$ & $\hphantom{\bZ_2}$ & $\hphantom{\bZ_2}$ & $\hphantom{\bZ_2}$ & $\hphantom{\bZ_2}$ & $\hphantom{\bZ_2}$\\
		4 & & \cellcolor{lightyellow} & $\bZ$ & & $\bZ^{\oplus 3}$ & & $\bZ^{\oplus 5}$ \\
		3 & & & \cellcolor{lightyellow} & & & & \\
		2 & & & $\bZ_{2}$ & $\cellcolor{lightyellow}$ & $\bZ_{2}^{\oplus 3}$ & & $\bZ_{2}^{\oplus 5}$ \\
		1 & & & $\bZ_{2}$ & & $\bZ_{2}^{\oplus 3}\cellcolor{lightyellow}$ & & $\bZ_{2}^{\oplus 5}$ \\
		0 & & & $\bZ$ & & $\bZ^{\oplus 3}$ & $\cellcolor{lightyellow}$ & $\bZ^{\oplus 5}$ \\
		\hline
		& 0 & 1 & 2 & 3 & 4 & 5 & 6
	\end{tabular}
\caption{The $E_{2}$-page of the Atiyah-Hirzebruch spectral sequence for the spin bordism associated with the fibration~\eqref{eq:fibration_trivial_BG}.}
\label{ss:bordism_BA4A3T1}
\end{center}\end{table}
In order to determine the fifth bordism group, we need to investigate the differential maps $d^{2}_{4,1}:E^{2}_{4,1}\to E^{2}_{2,2}$ and $d^{2}_{6,0}:E^{2}_{6,0}\to E^{2}_{4,1}$.
The map $d^{2}_{4,1}$ is given by the dual of the second Steenrod square, $\Sqtwo:H^{2}(BG;\bZ_{2})\to H^{4}(BG;\bZ_{2})$.
Since $H^{2}(BG;\bZ)\cong\bZ(12\cfi{1}+25\cth{1})$, the group $H^{2}(BG;\bZ_{2})$ is generated by $\cth{1}$.
On the other hand, the group $H^{4}(BG;\bZ_{2})$ is generated by $\cfi{2}$, $\cth{2}$, and $(\cth{1})^{2}$.
The action of $\Sqtwo$ on the generator of $H^{2}(BG;\bZ_{2})$ is given by
\begin{align}
    \Sqtwo(\cth{1})=(\cth{1})^{2}-2\cth{2}\equiv(\cth{1})^{2}.
\end{align}
This result implies that
\begin{align}
    \Ker d^{2}_{4,1}&\cong\bZ_{2}\widetilde{\cfi{2}}\oplus\bZ_{2}\widetilde{\cth{2}},&
    \Im d^{2}_{4,1}&\cong\bZ_{2}\widetilde{\cth{1}},
\end{align}
where $\widetilde{\cfi{2}}$, $\widetilde{\cth{2}}$, and $\widetilde{\cth{1}}$ are dual to $\cfi{2}$, $\cth{2}$, and $\cth{1}$, respectively.
The differential map $d^{2}_{6,0}:E^{2}_{6,0}\to E^{2}_{4,1}$ is the composition of the dual of the Steenrod square $\Sqtwo:H^{4}(BG;\bZ_{2})\to H^{6}(BG;\bZ_{2})$ with the homomorphism $H_{6}(BG;\bZ)\to H_{6}(BG;\bZ)$ appearing the homology long exact sequence associated with the short exact sequence $0\to\bZ\to\bZ\to\bZ_{2}\to 0$.
Note that the group $H^{6}(BG;\bZ_{2})\cong\bZ_{2}^{\oplus 5}$ is generated by $\cfi{3}+\cfi{2}\cfi{1}$, $\cth{3}$, $\cth{2}\cth{1}$, $\cfi{2}\cth{1}$, and $(\cth{1})^{3}$.
Then, the actions of $\Sqtwo$ on the generators of $H^{4}(BG;\bZ_{2})$ are directly calculated as
\begin{align}\begin{aligned}
    \Sqtwo(\cfi{2})&=\cfi{2}\cfi{1}-3\cfi{3}\equiv\cfi{2}\cfi{1}+\cfi{3},\\
    \Sqtwo(\cth{2})&=\cth{2}\cth{1}-3\cth{3}\equiv\cth{2}\cth{1}+\cth{3},\\
    \Sqtwo((\cth{1})^{2})&=2(\cth{1})^{3}\equiv 0.
\end{aligned}\end{align}
From these results, it can be checked that
\begin{align}
    \Im d^{2}_{6,0}\cong\bZ_{2}\widetilde{\cfi{2}}\oplus\bZ_{2}\widetilde{\cth{2}}.
\end{align}
Hence, we conclude that
\begin{align}
    E^{3}_{4,1}\cong\Ker d^{2}_{4,1}/\Im d^{2}_{6,0}\cong 0,
\end{align}
and
\begin{align}
    \Omega^{\Spin}_{5}(BG)\cong 0.
\end{align}

\section{Topological data of \texorpdfstring{$E_{7}/G$}{}}
\label{sec:E7G}
The purpose of this appendix is to compute the bordism group relevant to the global sigma-model anomaly in the family unification model with target space $E_{7}/G$.
To this end, we apply the Atiyah-Hirzebruch spectral sequence.
In Section~\ref{subsec:cohom_E7G}, we determine the cohomology groups of $E_{7}/G$, and in Section~\ref{subsec:bordism_E7G} we compute the bordism groups $\Omega^{\Spin}_{\ast}(E_{7}/G)$.

\subsection{Cohomology ring \texorpdfstring{$H^{\ast}(E_{7}/G;\bZ)$}{} up to degree six}
\label{subsec:cohom_E7G}
Following the methods developed in~\cite{MR358847,MR380857}, we compute the cohomology ring of the coset space $E_{7}/G$ with integer coefficients.
For our purposes, it suffices to determine these groups up to degree six.

We consider the fibration
\begin{align}
  G/T \xrightarrow{i_{G}} E_{7}/T \xrightarrow{p_{G}} E_{7}/G,
\end{align}
where $T$ is a maximal torus of $E_{7}$.
This fibration induces ring homomorphisms,
\begin{align}
  &p_{G}^{\ast}: H^{\ast}(E_{7}/G;\bZ) \to H^{\ast}(E_{7}/T;\bZ),&
  &i_{G}^{\ast}: H^{\ast}(E_{7}/T;\bZ) \to H^{\ast}(G/T;\bZ).
\end{align}
It is known that the cohomology groups of $G/T$, $E_{7}/T$, and $E_{7}/G$ are torsion free and vanish in odd degrees.
These properties imply that $p_{G}^{\ast}$ is injective, $i_{G}^{\ast}$ is surjective, and
\begin{align}
  \Ker i_{G}^{\ast}=(p_{G}^{\ast} H^{+}(E_{7}/G;\bZ)),
\end{align}
where $(p_{G}^{\ast}H^{+}(E_{7}/G;\bZ))$ is the ideal generated by $p_{G}^{\ast}H^{+}(E_{7}/G;\bZ)$, and $H^{+}(E_{7}/G;\bZ)$ is the subgroup of $H^{\ast}(E_{7}/G;\bZ)$ consisting of elements with positive degree.~\footnote{For a space Y and an Abel group $\bA$, we denote by $H^{+}(Y;\bA)$ the subgroup of $H^{\ast}(Y;\bA)$ obtained by removing the degree zero part.}

To compute $H^{\ast}(E_{7}/G;\bZ)$, we first regard it as a subring of $H^{\ast}(E_{7}/T;\bZ)$ via the injection $p_{G}^{\ast}$.
We then determine the ideal $(p_{G}^{\ast}H^{+}(E_{7}/G;\bZ))$ by studying the kernel of $i_{G}^{\ast}$.
Finally, to reconstruct $p_{G}^{\ast}H^{\ast}(E_{7}/G;\bZ)$ from this information, we also consider the rational cohomology ring $H^{\ast}(E_{7}/G;\bQ)$.
Since $H^{\ast}(E_{7}/G;\bZ)$ is torsion free, it embeds as a lattice in $H^{\ast}(E_{7}/G;\bQ)$.
This fact allows us to represent generators of $H^{\ast}(E_{7}/G;\bZ)$ as the elements of $H^{\ast}(E_{7}/G;\bQ)$ so that they generate the ideal $(p_{G}^{\ast}H^{+}(E_{7}/G;\bZ))$.

\subsubsection{Kernel of \texorpdfstring{$i_{G}^{\ast}$}{}}
To analyze the kernel of $i_{G}^{\ast}$, we begin by describing the
rings $H^{\ast}(E_{7}/T;\bZ)$ and $H^{\ast}(G/T;\bZ)$ in terms of
generators.
Note that there is an isomorphism
\begin{align}
  H^{2}(E_{7}/T;\bZ)\cong H^{2}(BT;\bZ),
\end{align}
and $H^{2}(BT;\bZ)$ can be identified with the weight lattice of
$\e_{7}$.
Instead of using the fundamental weights $\omega_{i}\,(i=1,\dots,7)$, we can represent $H^{2}(BT;\bZ)$ in terms of the coordinates
\begin{align}\begin{aligned}
  t_{1}&=-\omega_{1}+\omega_{7},&
  t_{2}&=\omega_{1}-\omega_{2}+\omega_{7},&
  t_{3}&=\omega_{2}-\omega_{3}+\omega_{7},&
  t_{4}&=\omega_{3}-\omega_{4},\\
  t_{5}&=\omega_{4}-\omega_{5},&
  t_{6}&=\omega_{5}-\omega_{6},&
  t_{7}&=\omega_{6},&
  \gamma_{1}&=\omega_{7},
\end{aligned}\end{align}
subject to the relation
\begin{align}
  \ce{1}=3\gamma_{1}.
\end{align}
In the following, we use the notation $\ce{i}=\sigma_{i}(t_{1},\dots,t_{7})$.

According to Ref.~\cite{MR380857}, the cohomology ring of $E_{7}/T$ with integer coefficients is given by
\begin{align}
  H^{\ast}(E_{7}/T;\bZ)
  \cong
  \bZ[t_{1},\cdots,t_{7},\gamma_{1},\gamma_{3},\gamma_{4},\gamma_{5},\gamma_{9}]
  /(\rho_{1},\rho_{2},\rho_{3},\rho_{4},\rho_{5},\rho_{9}),
\end{align}
where the generators $\gamma_{j}$ for $j=3,4,5,9$ have degree $2j$.
The relations in low degrees are given by
\begin{align}
  \rho_{1}&=\ce{1}-3\gamma_{1},&
  \rho_{2}&=\ce{2}-4\gamma_{1}^{2},&
  \rho_{3}&=\ce{3}-2\gamma_{3}.
\end{align}
Since we are interested only in degrees up to six, we omit relations of higher degree.

Note that the space $G/T$ is the direct product of two flag manifolds:
\begin{align}
  G/T
  \cong\frac{(\U(5)\times\U(3))/\U(1)}{\U(1)^{8}/\U(1)}
  \cong\frac{\U(5)}{\U(1)^{5}}\times\frac{\U(3)}{\U(1)^{3}}.
\end{align}
Based on this description of $G/T$, the cohomology ring of $G/T$ with integer coefficients is given by
\begin{align}
  H^{\ast}(G/T;\bZ)
  \cong
  \bZ[y_{1},y_{2},y_{3},y_{4},y_{5},z_{1},z_{2},z_{3}]
  /(\cfi{1},\cfi{2},\cfi{3},\cfi{4},\cfi{5},\cth{1},\cth{2},\cth{3}),
\end{align}
where $y_{j},z_{k}\in H^{2}(G/T;\bZ)$,
$\cfi{j}=\sigma_{j}(y_{1},\dots,y_{5})$, and
$\cth{k}=\sigma_{k}(z_{1},z_{2},z_{3})$.~\footnote{For the cohomology
ring of a flag manifold, see e.g.~Theorem~5.5 in Chapter~III of~\cite{Mimura:1991}.}
On the other hand, we have another description as
\begin{align}
    G/T\cong\frac{\SU(5)}{T^{4}}\times\frac{\SU(3)}{T^{2}},
\end{align}
where, $T^{4}$ and $T^{2}$ are the maximal tori of $\SU(5)$ and $\SU(3)$, respectively.
In particular, this description implies that
\begin{align}
    H^{2}(G/T;\bZ)
    \cong H^{2}(\SU(5)/T^{4};\bZ)\oplus H^{2}(\SU(3)/T^{2};\bZ)
    \cong H^{2}(BT^{4};\bZ)\oplus H^{2}(BT^{2};\bZ).
\end{align}
From this expression, it turns out that as the $\bZ$-module, $H^{2}(G/T;\bZ)$ is generated by the fundamental weights of $\su_{5}$ and those of $\su_{3}$.
The set of generators $\fweight{j} (j=1,2,3,4)$ and $\tweight{k} (k=1,2)$ are related to the other set of generators $y_{j}\,(j=1,2,3,4,5)$ and $z_{k}\,(k=1,2,3)$ as
\begin{align}
    \fweight{j}&=\sum_{l=1}^{j}y_{l},&
    \tweight{k}&=\sum_{l=1}^{k}z_{l}.
\end{align}

From the fact in 14.2 of~\cite{MR102800}, it turns out that the fundamental weights $\omega_{i}$ of $\e_{7}$ are mapped as
\begin{align}\begin{aligned}
    i_{G}^{\ast}\omega_{1}&=\fweight{1},&
    i_{G}^{\ast}\omega_{2}&=\fweight{2},&
    i_{G}^{\ast}\omega_{3}&=\fweight{3},&
    i_{G}^{\ast}\omega_{4}&=0,\\
    i_{G}^{\ast}\omega_{5}&=\tweight{1},&
    i_{G}^{\ast}\omega_{6}&=\tweight{2},&
    i_{G}^{\ast}\omega_{7}&=\fweight{4}.&
\end{aligned}\end{align}
Then, the generators $t_{i}\,(i=1,\dots,7)$ are mapped by $i_{G}^{\ast}$ as
\begin{align}\begin{aligned}\label{eq:iG_t}
    i_{G}^{\ast}t_{1}&=-(y_{1}+y_{5}),&
    i_{G}^{\ast}t_{2}&=-(y_{2}+y_{5}),&
    i_{G}^{\ast}t_{3}&=-(y_{3}+y_{5}),&
    i_{G}^{\ast}t_{4}&=-(y_{4}+y_{5}),\\
    i_{G}^{\ast}t_{5}&=-z_{1},&
    i_{G}^{\ast}t_{6}&=-z_{2},&
    i_{G}^{\ast}t_{7}&=-z_{3}.&&
\end{aligned}\end{align}
By using the relations $\rho_{1}$ and $\rho_{3}$, one obtains
\begin{align}
    i_{G}^{\ast}(3\gamma_{1})&=i_{G}^{\ast}(\ce{1})=-3y_{5},&
    i_{G}^{\ast}(2\gamma_{3})&=i_{G}^{\ast}(\ce{3})=-2y_{5}^{3}.
\end{align}
Since $H^{\ast}(G/T;\bZ)$ is torsion free, these results mean that
\begin{align}
    i_{G}^{\ast}\gamma_{1}&=-y_{5},&
    i_{G}^{\ast}\gamma_{3}&=-y_{5}^{3}.
\end{align}
To make the expressions simpler, let us introduce the notations $\bth{k}=\sigma_{k}(t_{5},t_{6},t_{7})$ for $k=1,2,3$.
In terms of them, the kernel of $i_{G}^{\ast}$ is written as
\begin{align}\label{eq:ker_iG}
    \Ker i_{G}^{\ast}=(\bth{1},\bth{2},\bth{3},\gamma_{3}-\gamma_{1}^{3}),
\end{align}
up to degree six.
To demonstrate this result, we check that $b_{k}\in\ker i_{G}^{\ast}$ for $k=1,2,3$.
From the expressions~\eqref{eq:iG_t}, it holds that
\begin{align}
    i_{G}^{\ast}b_{i}=\sigma_{k}(i_{G}^{\ast}t_{5},i_{G}^{\ast}t_{6},i_{G}^{\ast}t_{7})=(-1)^{k}\sigma_{i}(z_{1},z_{2},z_{3})=\cth{k}\equiv 0\in H^{\ast}(G/T;\bZ).
\end{align}

\subsubsection{Cohomology ring \texorpdfstring{$H^{\ast}(E_{7}/G;\bQ)$}{}}
\label{subsubsec:cohom_E7G_Q}
The cohomology ring of $E_{7}/G$ with rational coefficient is obtained by  the formula~\cite{MR102800},
\begin{align}\label{eq:rational_cohom_formula}
    H^{\ast}(E_{7}/G;\bQ)\cong H^{\ast}(BG;\bQ)/(H^{+}(BE_{7};\bQ)).
\end{align}
The Weyl group of $\g=\su_{5}\oplus\su_{3}\oplus\u_{1}$ acts on the roots of $\e_{7}$, and we can extend this action into $H^{\ast}(BT;\bQ)$.
The ring $H^{\ast}(BG;\bQ)$ is identified with the subring of $H^{\ast}(BT;\bQ)$ consisting of invariants under the Weyl group of $\g$.
On the other hand, $H^{\ast}(BE_{7};\bQ)$ is the subring of $H^{\ast}(BT;\bQ)$ consisting of invariants under the Weyl group of $\e_{7}$.

In order to describe $H^{\ast}(BG;\bQ)$ in terms of the Weyl invariants, we introduce the coordinates of $H^{2}(BT;\bQ)$ as
\begin{align}\label{eq:coordinate_byz}
\begin{aligned}
    \bth{1}&=\sigma_{1}(t_{5},t_{6},t_{7}),&
    \zp{1}&=-t_{5}+\frac{1}{3}\bth{1}=-\frac{2}{3}\omega_{4}+\omega_{5},\\
    \zp{2}&=-t_{6}+\frac{1}{3}\bth{1}=\frac{1}{3}\omega_{4}-\omega_{5}+\omega_{6},&
    \zp{3}&=-t_{7}+\frac{1}{3}\bth{1}=\frac{1}{3}\omega_{4}-\omega_{6},\\
    \yp{1}&=-t_{1}+\gamma_{1}-\frac{2}{5}\bth{1}=\omega_{1}-\frac{2}{5}\omega_{4},&
    \yp{2}&=-t_{2}+\gamma_{1}-\frac{2}{5}\bth{1}=-\omega_{1}+\omega_{2}-\frac{2}{5}\omega_{4},\\
    \yp{3}&=-t_{3}+\gamma_{1}-\frac{2}{5}\bth{1}=-\omega_{2}+\omega_{3}-\frac{2}{5}\omega_{4},&
    \yp{4}&=-t_{4}+\gamma_{1}-\frac{2}{5}\bth{1}=-\omega_{3}+\omega_{4}+\frac{3}{5}\omega_{4},\\
    \yp{5}&=-\gamma_{1}+\frac{3}{5}\bth{1}=-\omega_{7}+\frac{3}{5}\omega_{4}.&&\\
\end{aligned}\end{align}
We need to remember that these coordinates satisfy the relations,
\begin{align}
    \sum_{j=1}^{5}\yp{j}&=0,&
    \sum_{k=1}^{3}\zp{k}&=0.
\end{align}
Let $R_{i}$ be the Weyl reflection respect to the simple root $\alpha_{i}$.
The Weyl group of $\g$ is generated by $R_{i}$ for $i=1,2,3,5,6,7$.
The actions of the Weyl reflections $R_{i}$ on the coordinates~\eqref{eq:coordinate_byz} are summarized in Fig.~\ref{tab:Weyl_a4a2t1}.
From these results, we can find that the following polynomials are invariant under the Weyl group of $\g$:
\begin{align}\begin{aligned}
    \cpfi{j}&=\sigma_{j}(\yp{1},\yp{2},\yp{3},\yp{4},\yp{5})\quad (j=1,\dots,5),\\
    \cpth{k}&=\sigma_{k}(\zp{1},\zp{2},\zp{3})\quad (k=1,2,3).
    \label{eq:symmetric_polynomial_byz}
\end{aligned}\end{align}
Thus, the cohomology ring of $BH$ with rational coefficient can be written as
\begin{align}
    H^{\ast}(BG;\bQ)\cong\bQ[\bth{1},\cpfi{1},\cpfi{2},\cpfi{3},\cpfi{4},\cpfi{5},\cpth{1},\cpth{2},\cpth{3}]/(\cpfi{1},\cpth{1}).
\end{align}
\begin{table}[thp]\begin{center}
	\begin{tabular}{c||ccccccccc}
	 & $\bth{1}$ & $\yp{1}$ & $\yp{2}$ & $\yp{3}$ & $\yp{4}$ & $\yp{5}$ & $\zp{1}$ & $\zp{2}$ & $\zp{3}$ \\
    \hline
    $R_{1}$ & & $\yp{2}$ & $\yp{1}$ & & & & & & \\
    $R_{2}$ & & & $\yp{3}$ & $\yp{2}$ & & & & & \\
    $R_{3}$ & & & & $\yp{4}$ & $\yp{3}$ & & & & \\
    $R_{5}$ & & & & & & & $\zp{2}$ & $\zp{1}$ & \\
    $R_{6}$ & & & & & &  & & $\zp{3}$ & $\zp{2}$ \\
    $R_{7}$ & & & & & $\yp{5}$ & $\yp{4}$ & & & \\
	\end{tabular}
\caption{The Weyl reflection on the coordinates~\eqref{eq:coordinate_byz}. The blanks indicate the trivial actions.}
\label{tab:Weyl_a4a2t1}
\end{center}\end{table}

As discussed in Ref.~\cite{MR380857}, the cohomology ring of $BE_{7}$ with rational coefficient is given by
\begin{align}
    H^{\ast}(BE_{7};\bQ)=\bQ[I_{2},I_{6},I_{8},I_{10},I_{12},I_{14},I_{18}],
\end{align}
where $I_{i}\in H^{2i}(BT;\bQ)$.
In terms of the coordinate~\eqref{eq:coordinate_byz}, the explicit forms of $I_{2}$ is given by
\begin{align}
    I_{2}=\frac{32}{5}\bth{1}^{2}+48\left((\cpfi{1})^{2}-2\cpfi{2}+(\cpth{1})^{2}-2\cpth{2}\right).
    \label{eq:I2}
\end{align}

By applying the formula~\eqref{eq:rational_cohom_formula}, we obtain the ring isomorphism,
\begin{align}\label{eq:cohom_E7G_Q_mid}
    H^{\ast}(E_{7}/G;\bQ)\cong\bQ[\bth{1},\cpfi{1},\cpfi{2},\cpfi{3},\cpth{1},\cpth{2},\cpth{3}]/(\cpfi{1},\cpth{1},I_{2}),
\end{align}
up to degree six.
In order to compare this ring with $\Ker i_{G}^{\ast}$, we rewrite down the generators $\cpfi{j}$ and $\cpth{k}$ in terms of the polynomials
\begin{align}\begin{aligned}\label{eq:a_generators}
    \afi{1}&=\sigma_{1}(t_{1},t_{2},t_{3},t_{4})+\bth{1}-3\gamma_{1},\\
    \afi{2}&=\sigma_{2}(t_{1},t_{2},t_{3},t_{4})-4\gamma_{1}^{2}+3\gamma_{1}\bth{1},\\
    \afi{3}&=\sigma_{3}(t_{1},t_{2},t_{3},t_{4})+\gamma_{1}\bth{2}-\gamma_{1}\bth{1}^{2}+2\gamma_{1}^{2}\bth{1}-2\gamma_{1}^{3},
\end{aligned}\end{align}
and $\bth{i}=\sigma_{k}(t_{5},t_{6},t_{7})$ for $k=1,2,3$.
After some calculations, it turns out that the relations in the expression~\eqref{eq:cohom_E7G_Q_mid} are given by
\begin{align}
    \cpfi{1}&=-\afi{1},&
    \cpth{1}&=0,&
    I_{2}&=96(\bth{1}^{2}-\afi{2}-\bth{2})\mod\,\afi{1},
\end{align}
and the other invariant polynomials are expressed as
\begin{align}\begin{aligned}
    \cpfi{2}&=\afi{2}-\frac{3}{5}\bth{1}^{2}\mod\,(\afi{1}),&
    \cpfi{3}&=-\afi{3}+\frac{1}{5}\bth{2}\bth{1}-\frac{3}{25}\bth{3}^{3}\mod\,(\afi{1},I_{2}),\\
    \cpth{2}&=\bth{2}-\frac{1}{3}\bth{1}^{2},&
    \cpth{3}&=-\bth{3}+\frac{1}{3}\bth{2}\bth{1}-\frac{2}{27}\bth{3}.
\end{aligned}\end{align}
Then, the ring $H^{\ast}(E_{7}/G;\bQ)$ up to degree six can be written as
\begin{align}\label{eq:cohom_E7G_Q}
    H^{\ast}(E_{7}/G;\bQ)
    \cong\bQ[\afi{1},\afi{2},\afi{3},\bth{1},\bth{2},\bth{3}]/(\afi{1},\bth{1}^{2}-\afi{2}-\bth{2})
    \cong\bQ[\bth{1},\bth{2},\afi{3},\bth{3}].
\end{align}

\subsubsection{Generators of cohomology ring \texorpdfstring{$H^{\ast}(E_{7}/G;\bZ)$}{}}
Combining the information~\eqref{eq:ker_iG} and~\eqref{eq:cohom_E7G_Q}, we can describe the ring $H^{\ast}(E_{7}/T;\bZ)$ up to degree six in terms of the generators.
As mentioned in the outset of Section~\ref{subsec:cohom_E7G}, we can regard $H^{\ast}(E_{7}/G;\bZ)$ as the lattice in $H^{\ast}(E_{7}/G;\bQ)$.
By comparing the expressions of $\Ker i_{G}^{\ast}$ and $H^{\ast}(E_{7}/G;\bQ)$, we can take $\bth{1}$, $\bth{2}$, and $\bth{3}$ as the generators of $H^{\ast}(E_{7}/G;\bZ)$.

Let us specify the remaining generator in degree six.
This generator can be represented as
\begin{align}
    \acfi{3}=\beta_{1}\afi{3}+\beta_{2}\bth{3}+\beta_{3}\bth{2}\bth{1}+\beta_{4}\bth{1}^{3},
\end{align}
for some rational numbers $\beta_{1},\dots,\beta_{4}$.
On the other hand, we can assume the form,
\begin{align}
    \acfi{3}=\gamma_{3}-\gamma_{1}^{3}+\delta,
\end{align}
where $\delta\in H^{6}(E_{7}/G;\bZ)\cap(\bth{1},\bth{2})$.
By some calculations, one can show that the following choice is consistent with these two expressions:
\begin{align}\begin{aligned}
    \beta_{1}&=\beta_{2}=\beta_{4}=1/2,&
    \beta_{3}&=-1,&
    \delta&=-\gamma_{1}\bth{2}+\gamma_{1}\bth{1}^{2}-\gamma_{1}^{2}\bth{1}.
\end{aligned}\end{align}
The coefficients $\beta_{1},\dots,\beta_{4}$ are determined up to integers.
Then, one can choose the generator as
\begin{align}\begin{aligned}
    \acfi{3}
    &=\frac{1}{2}\afi{3}+\frac{1}{2}\bth{3}-\bth{2}\bth{1}+\frac{1}{2}\bth{1}^{3}\\
    &=\gamma_{3}-\gamma_{1}^{3}-\gamma_{1}\bth{2}+\gamma_{1}\bth{1}^{2}-\gamma_{1}^{2}\bth{1}\mod (\afi{1},\bth{1}^{2}-\afi{2}-\bth{2}).
\end{aligned}\end{align}
In summary, the cohomology ring of $E_{7}/G$ with integer coefficients up to degree six is given by
\begin{align}\label{eq:cohom_E7G}
    H^{\ast}(E_{7}/G;\bZ)\cong\bZ[\bth{1},\bth{2},\bth{3},\acfi{3}].
\end{align}
This description is consistent with $H^{\ast}(E_{7}/G;\bQ)$, and it is obvious that $\Ker i^{\ast}=(\bth{1},\bth{2},\bth{3},\acfi{3})$.

\subsection{Bordism groups \texorpdfstring{$\Omega^{\Spin}_{\ast}(E_{7}/G)$}{} up to degree five}
\label{subsec:bordism_E7G}
We calculate the spin bordism groups $\Omega^{\Spin}_{\ast}(E_{7}/G)$ up to degree five by the Atiyah-Hirzebruch spectral sequence.
In Section~\ref{subsubsec:Steenrod_E7G}, based on the analysis in Section~\ref{subsec:cohom_E7G}, we determine the action of the Steenrod square on $H^{\ast}(E_{7}/G;\bZ_{2})$.
The computation of $\Omega^{\Spin}_{\ast}(E_{7}/G)$ is performed in Section~\ref{subsub:AHSS_E7G}.

\subsubsection{Steenrod square on \texorpdfstring{$H^{\ast}(E_{7}/G;\bZ_{2})$}{}}
\label{subsubsec:Steenrod_E7G}
Since $H^{\ast}(E_{7}/T;\bZ)$ vanishes at the odd degrees, it follows from the universal coefficient theorem that the ring $H^{\ast}(E_{7}/T;\bZ_{2})$ are simply obtained the reductions of $H^{\ast}(E_{7}/T;\bZ)$ and modulo $2$.
The similar statement also holds for the ring $H^{\ast}(E_{7}/G;\bZ_{2})$.
Concretely, the cohomology groups with $\bZ_{2}$ coefficients up to degree six are given by
\begin{align}\begin{aligned}
    H^{2}(E_{7}/G,\bZ_{2})&=\bZ_{2}\bth{1},\\
    H^{4}(E_{7}/G,\bZ_{2})&=\bZ_{2}\bth{2}\oplus\bZ_{2}\bth{1}^{2},\\
    H^{6}(E_{7}/G,\bZ_{2})&=\bZ_{2}\bth{3}\oplus\bZ_{2}\acfi{3}\oplus\bZ_{2}\bth{2}\bth{1}\oplus\bZ_{2}\bth{1}^{3}.
\end{aligned}\end{align}

The Steenrod square on $H^{\ast}(E_{7}/G;\bZ_{2})$ is obtained by restricting that on $H^{\ast}(E_{7}/T;\bZ_{2})$, since $H^{\ast}(E_{7}/G;\bZ_{2})$ can be regarded as a subring of $H^{\ast}(E_{7}/T;\bZ)$ through the injection $p_{G}^{\ast}$.
Then, by some calculations, the actions of the Steenrod square $\Sqtwo$ on the generators of $H^{2}(E_{7}/G;\bZ_{2})$ and $H^{4}(E_{7}/G;\bZ_{2})$ are determined as
\begin{align}\begin{aligned}
    \Sqtwo(\bth{1})&=\bth{1}^{2}-2\bth{2}\equiv\bth{1}^{2},\\
    \Sqtwo(\bth{2})&=\bth{2}\bth{1}-3\bth{3}\equiv\bth{2}\bth{1}+\bth{3},\\
    \Sqtwo(\bth{1}^{2})&=2\bth{1}^{2}\equiv 0.
\end{aligned}\end{align}

\subsubsection{Atiyah-Hirzebruch spectral sequence for \texorpdfstring{$\Omega^{\Spin}_{\ast}(E_{7}/G)$}{}}
\label{subsub:AHSS_E7G}
In order to compute the spin bordism groups of $E_{7}/G$, we apply the Atiyah-Hirzebruch spectral sequence to the fibration,
\begin{align}\label{eq:fibration_trivial_E7_A4A2T1}
    \mathrm{pt}\to E_{7}/G\to E_{7}/G.
\end{align}
The $E^{2}$-page of the Atiyah-Hirzebruch spectral sequence that converges to the reduced spin bordism group $\widetilde{\Omega}^{\Spin}_{\ast}(E_{7}/G)$ is given by
\begin{align}
    E^{2}_{p,q}=\widetilde{H}_{p}(E_{7}/G;\Omega^{\Spin}_{q}(\mathrm{pt})).
\end{align}
The homology groups necessary for the calculation are summarized in Table~\ref{tab:hom_cohom_E7_A4A3T1}.
The spin bordism groups of a point are exhibited in Table~\ref{tab:bordism_pt}.
The situation of the $E^{2}$-page is shown in Table~\ref{ss:bordism_E7_A4A3T1}.
\begin{table}[thp]\begin{center}
	\begin{tabular}{c||ccccccc}
		$p$ & $0$ & $1$ & $2$ & $3$ & $4$ & $5$ & $6$ \\
        \hline
        $H_{p}(BG;\bZ)$ & $\bZ$ & $0$ & $\bZ$ & $0$ & $\bZ^{\oplus 2}$ & $0$ & $\bZ^{\oplus 4}$ \\
        $H^{p}(BG;\bZ)$ & $\bZ$ & $0$ & $\bZ$ & $0$ & $\bZ^{\oplus 2}$ & $0$ & $\bZ^{\oplus 4}$ \\
        $H_{p}(BG;\bZ_{2})$ & $\bZ_{2}$ & $0$ & $\bZ_{2}$ & $0$ & $\bZ_{2}^{\oplus 2}$ & $0$ & $\bZ_{2}^{\oplus 4}$ \\
        $H^{p}(BG;\bZ_{2})$ & $\bZ_{2}$ & $0$ & $\bZ_{2}$ & $0$ & $\bZ_{2}^{\oplus 2}$ & $0$ & $\bZ_{2}^{\oplus 4}$ \\
	\end{tabular}
\caption{Homology and cohomology groups of $E_{7}/G$ up to degree six.}
\label{tab:hom_cohom_E7_A4A3T1}
\end{center}\end{table}
\begin{table}[thp]\begin{center}
	\begin{tabular}{c|ccccccc}
		5 & $\hphantom{\bZ_2}$ & $\hphantom{\bZ_2}$ & $\hphantom{\bZ_2}$ & $\hphantom{\bZ_2}$ & $\hphantom{\bZ_2}$ & $\hphantom{\bZ_2}$ & $\hphantom{\bZ_2}$\\
		4 & & & $\bZ$ & & $\bZ^{\oplus 2}$ & & $\bZ^{\oplus 4}$ \\
		3 & & & & & & & \\
		2 & & & $\bZ_{2}$ & & $\bZ_{2}^{\oplus 2}$ & & $\bZ_{2}^{\oplus 4}$ \\
		1 & & & $\bZ_{2}$ & & $\bZ_{2}^{\oplus 2}$ & & $\bZ_{2}^{\oplus 4}$ \\
		0 & & & $\bZ$ & & $\bZ^{\oplus 2}$ & & $\bZ^{\oplus 4}$ \\
		\hline
		& 0 & 1 & 2 & 3 & 4 & 5 & 6
	\end{tabular}
\caption{The $E^{2}$-page of the Atiyah-Hirzebruch spectral sequence for the spin bordism associated with the fibration~\eqref{eq:fibration_trivial_E7_A4A2T1}.}
\label{ss:bordism_E7_A4A3T1}
\end{center}\end{table}

Our current goal is to compute the spin bordism groups of $E_{7}/G$ up to degree five.
To this end, we need to investigate the differentials $d^{2}_{p,q}:E^{2}_{p,q}\to E^{2}_{p-2,q+1}$ for $(p,q)=(4,0),(4,1),(6,0)$.
Since the differential $d^{2}_{4,1}$ is given by the dual of the Steenrod square $\Sqtwo$, it is straightforward to check that
\begin{align}
    \Ker d^{2}_{4,1}&\cong\bZ_{2}\widetilde{\bth{2}},&
    \Im d^{2}_{4,1}&\cong\bZ_{2}\widetilde{\bth{1}},
\end{align}
where $\widetilde{\bth{1}}$ and $\widetilde{\bth{2}}$ are dual to $\bth{1}$ and $\bth{2}$, respectively.
The differential $d^{2}_{4,0}$ is the composition of the reduction modulo $2$ and $d^{2}_{4,1}$.
Then, we obtain
\begin{align}
    \Ker d^{2}_{4,0}&\cong\bZ\widetilde{\bth{2}}\oplus\bZ 2\widetilde{\bth{1}^{2}},&
    \Im d^{2}_{4,0}&\cong\bZ_{2}\widetilde{\bth{1}}.
\end{align}
For the differential $d^{2}_{6,0}$, we can similarly find that
\begin{align}
    \Im d^{2}_{6,0}\cong\bZ_{2}\widetilde{\bth{2}}.
\end{align}
By combining these results, the reduced bordism groups are calculated as
\begin{align}\begin{aligned}
    \widetilde{\Omega}^{\Spin}_{1}(E_{7}/G)&\cong 0,\\
    \widetilde{\Omega}^{\Spin}_{2}(E_{7}/G)&\cong E^{2}_{2,0}\cong\bZ,\\
    \widetilde{\Omega}^{\Spin}_{3}(E_{7}/G)&\cong E^{2}_{2,1}/\Im d^{2}_{4,0}\cong 0,\\
    \widetilde{\Omega}^{\Spin}_{4}(E_{7}/G)&\cong \Ker d^{2}_{4,0}\cong\bZ^{\oplus 2},\\
    \widetilde{\Omega}^{\Spin}_{5}(E_{7}/G)&\cong \Ker d^{2}_{4,1}/\Im d^{2}_{6,0}\cong 0.
\end{aligned}\end{align}
For later convenience, we exhibit the spin bordism groups of $E_{7}/G$ in Table~\ref{tab:bordism_E7_A4A3T1}.
The fact $\Omega^{\Spin}_{5}(E_{7}/G)\cong 0$ implies that there is no global sigma model anomaly in the theory with target space $E_{7}/G$.
\begin{table}[thp]\begin{center}
	\begin{tabular}{c||cccccc}
		$q$ & $0$ & $1$ & $2$ & $3$ & $4$ & $5$ \\
        \hline
        $\Omega^{\Spin}_{q}(E_{7}/G)$ & $\bZ$ & $\bZ_{2}$ & $\bZ\oplus\bZ_{2}$ & $0$ & $\bZ^{\oplus 3}$ & $0$ \\
        $\widetilde{\Omega}^{\Spin}_{q}(E_{7}/G)$ & $0$ & $0$ & $\bZ$ & $0$ & $\bZ^{\oplus 2}$ & $0$ \\
	\end{tabular}
\caption{Spin bordism groups of $E_{7}/G$ up to degree six.}
\label{tab:bordism_E7_A4A3T1}
\end{center}\end{table}

\section{Bordism group of \texorpdfstring{$(E_{7}/G)_{G}$}{}}
\label{sec:Borel_E7G}
Based on the analysis in Section~\ref{sec:BG} and Section~\ref{sec:E7G}, we calculate the fifth spin bordism group of $(E_{7}/G)_{G}=EG\times_{G}E_{7}/G$.
The topological space $(E_{7}/G)_{G}$ fits the fibration,
\begin{align}\label{eq:fibration_E7_G_BG}
    E_{7}/G\to (E_{7}/G)_{G}\to BG.
\end{align}
We apply the  Atiyah-Hirzebruch spectral sequence to the fibration~\eqref{eq:fibration_E7_G_BG}.
We denote the element in $E^{2}$-page by $\CE{2}{p}{q}$, that is,
\begin{align}
    \CE{2}{p}{q}=H_{p}(BG;\Omega^{\Spin}_{q}(E_{7}/G)).
\end{align}
The data required to write down the $E^{2}$-page are summarized in Table~\ref{tab:hom_cohom_E7_A4A3T1} and Table~\ref{tab:bordism_E7_A4A3T1}.
For the situation of the $E^{2}$-page, see Table~\ref{ss:bordism_E7GG}.
\begin{table}[thp]\begin{center}
	\begin{tabular}{c|ccccccc}
		5 & $\hphantom{\bZ_2}\cellcolor{lightyellow}$ & $\hphantom{\bZ\oplus\bZ_2}$ & $\hphantom{\bZ_2}$ & $\hphantom{\bZ\oplus\bZ_2}$ & $\hphantom{\bZ_2}$ & $\hphantom{\bZ\oplus\bZ_2}$ & $\hphantom{\bZ_2}$\\
		4 & $\bZ^{\oplus 3}$ & \cellcolor{lightyellow} & $\bZ^{\oplus 3}$ & & $\bZ^{\oplus 9}$ & & $\bZ^{\oplus 15}$ \\
		3 & & & \cellcolor{lightyellow} & & & & \\
		2 & $\bZ\oplus\bZ_{2}$ & & $\bZ\oplus\bZ_{2}$ & $\cellcolor{lightyellow}$ & $\bZ^{\oplus 2}\oplus\bZ_{2}^{\oplus 2}$ & & $\bZ^{\oplus 4}\oplus\bZ_{2}^{\oplus 4}$ \\
		1 & $\bZ_{2}$ & & $\bZ_{2}$ & & $\bZ_{2}^{\oplus 3}\cellcolor{lightyellow}$ & & $\bZ_{2}^{\oplus 5}$ \\
		0 & $\bZ$ & & $\bZ$ & & $\bZ^{\oplus 3}$ & $\cellcolor{lightyellow}$ & $\bZ^{\oplus 5}$ \\
		\hline
		& 0 & 1 & 2 & 3 & 4 & 5 & 6
	\end{tabular}
\caption{The $E^{2}$-page of the Atiyah-Hirzebruch spectral sequence for the spin bordism associated with the fibration~\eqref{eq:fibration_E7_G_BG}.}
\label{ss:bordism_E7GG}
\end{center}\end{table}

In order to identify the differential maps $\Cd{2}{p}{q}:\CE{2}{p}{q}\to\CE{2}{p-2}{q+1}$, we utilize the following commutative diagram:
\begin{align}\label{eq:com_diagram_bordism}
\xymatrix{
\mathrm{pt} \ar[d]^{}  \ar[r]^{} & EG\times_{G}\mathrm{pt} \ar[d]^{} \ar[r]^{} & BG \ar[d]^{\rotatebox{-90}{$\cong$}}\\
E_{7}/G \ar[r]^{} & EG\times_{G}E_{7}/G \ar[r] & BG
}
\end{align}
Note that the first row in the diagram~\eqref{eq:com_diagram_bordism} is identical to the fibration~\eqref{eq:fibration_trivial_BG}, since $BGH\cong EG\times_{G}\mathrm{pt}$.
The $E^{2}$-page associated with the fibration~\eqref{eq:fibration_trivial_BG} that converges to $\Omega^{\Spin}_{\ast}(BG)$ is referred to as
\begin{align}
    \PE{2}{p}{q}=H_{p}(BG;\Omega^{\Spin}_{q}(\mathrm{pt})),
\end{align}
and the differential maps are denoted by $\Pd{2}{p}{q}:\PE{2}{p}{q}\to\PE{2}{p-2}{q+1}$.
The vertical maps in the diagram~\eqref{eq:com_diagram_bordism} induce the homomorphisms $\PE{2}{p}{q}\to\CE{2}{p}{q}$ such that
\begin{align}\label{eq:com_diagram_induced_bordism}
\xymatrix{
\PE{2}{p}{q} \ar[r]^-{\Pd{n}{p}{q}}  \ar[d]^-{} & \PE{2}{p-2}{q+1} \ar[d]_-{}\\
\CE{2}{p}{q} \ar[r]^-{\Cd{n}{p}{q}} & \CE{2}{p-2}{q+1}
}
\end{align}
Now, $\CE{2}{p}{q}$ can be decomposed as
\begin{align}
    \CE{2}{p}{q}\cong H_{p}(BG;\Omega^{\Spin}_{q}(\mathrm{pt}))\oplus H_{p}(BG;\widetilde{\Omega}^{\Spin}_{q}(E_{7}/G)).
\end{align}
Then, the induced homomorphism is given by the inclusion of $\PE{2}{p}{q}$ into the first factor of $\CE{2}{p}{q}$ in this decomposition.

In order to determine the bordism group $\Omega^{\Spin}_{5}((E_{7}/G)_{G})$, we need to study the differential maps $\Cd{2}{4}{1}$ and $\Cd{2}{6}{0}$.
Since we have $\Omega^{\Spin}_{0}(E_{7}/G)\cong\Omega^{\Spin}_{0}(\mathrm{pt})$ and $\Omega^{\Spin}_{1}(E_{7}/G)\cong\Omega^{\Spin}_{1}(\mathrm{pt})$, the induced homomorphisms $\PE{2}{4}{1}\to\CE{2}{4}{1}$ and $\PE{2}{6}{0}\to\CE{2}{6}{0}$ are isomorphisms.
Thus, the differential homomorphism $\Cd{2}{4}{1}$ is the composition of $\Pd{2}{4}{1}$ with the induced homomorphism $\PE{2}{2}{2}\to\CE{2}{2}{2}$, whereas $\Cd{2}{6}{0}$ is identified with $\Pd{2}{6}{0}$.
Then, it follows that
\begin{align}
    \Omega^{\Spin}_{5}((E_{7}/G)_{G})\cong \Ker\Cd{2}{4}{1}/\Im\Cd{2}{6}{0}\cong 0.
\end{align}
This result implies that there is no global anomaly in the sigma model with the target space~$E_{7}/G$ after gauging the left $G$-action.

\section{Topological data of \texorpdfstring{$BH$}{}}
\label{sec:BH}
We compute the cohomology groups and the bordism group of the classifying space $BH$.
By applying the strategy similar to Appendix~\ref{subsec:cohom_BG}, we determine the cohomology groups $H^{\ast}(BH;\bZ)$ up to degree six in Section~\ref{subsec:cohom_BH}.
The bordism group $\Omega^{\Spin}_{5}(BH)$ is calculated in Section~\ref{subsec:bordism_BH}.

\subsection{Cohomology groups \texorpdfstring{$H^{\ast}(BH;\bZ)$}{} up to degree six}
\label{subsec:cohom_BH}

From the discussion in Section~\ref{subsec:isom_G}, we also have the group isomorphism
\begin{align}
    H\cong\frac{\U(5)\times\U(1)^{3}}{\U(1)}.
\end{align}
The $\U(1)$ subgroup in this expression is embedded into $\U(5)\times\U(1)^{3}$ as
\begin{align}
\begin{array}{cccc}
    & \U(1) & \longrightarrow & \U(5)\times\U(1)^{3}\\
    &
    \rotatebox{90}{$\in$}&& \rotatebox{90}{$\in$} \\
    & w & \longmapsto & (w^{5}I_{5},w^{-4},w^{-4},w^{-4}),
\end{array}
\end{align}
where $I_{5}$ is the identity matrix of rank $5$.
In other words, we have the short exact sequence,
\begin{align}
    0\to \U(1) \to \U(5)\times\U(1)^{3} \to H \to 0.
\end{align}
From this short exact sequence, we can obtain the fibration,
\begin{align}\label{eq:fibration_first_H}
    B\U(5)\times B\U(3) \to BG \to K(\bZ,3),
\end{align}
which is analogous to the fibration~\eqref{eq:fibration_first} for $BG$.
We denote by $\HE{n}{p}{q}$ the element in the $E_{n}$-page of the Serre spectral sequence that converges to $H^{\ast}(BH;\bZ)$ associated with the fibration~\eqref{eq:fibration_first_H}.

In order to identify the differential maps $\Hd{n}{p}{q}:\HE{n}{p}{q}\to\HE{n}{p+n}{q-n+1}$, we consider the following commutative diagram:
\begin{align}\label{eq:com_diagram_cohom_H}
\xymatrix{
B\U(1)^{5}\times B\U(1)^{3} \ar[d]^{}  \ar[r]^{} & B\frac{\U(1)^{5}\times\U(1)^{3}}{\U(1)} \ar[d]^{} \ar[r]^{} & K(\bZ,3) \ar[d]^{\rotatebox{-90}{$\cong$}}\\
B\U(5)\times B\U(1)^{3} \ar[r]^{} & BH \ar[r] & K(\bZ,3)
}
\end{align}
The first row in this diagram is the same as the second row in the diagram~\eqref{eq:com_diagram_cohom}.
The vertical maps from the first row to the second row in the diagram~\eqref{eq:com_diagram_cohom_H} induce the homomorphisms $\ssh{n}{p}{q}:\HE{n}{p}{q}\to\TE{n}{p}{q}$.
Recall that $\TE{n}{p}{q}$ is referred to as the element of the $E_{2}$-page of the Serre spectral sequence associated with the first row in the diagram~\eqref{eq:com_diagram_cohom_H}.
The corresponding differential maps $\Td{n}{p}{q}$ are determined in Section~\ref{subsubsec:Gd3pq}.
Then, we can study the differential maps $\Hd{n}{p}{q}$ via the commutative diagram:
\begin{align}\label{eq:com_diagram_h}
\xymatrix{
\TE{n}{p}{q} \ar[r]^-{\Td{n}{p}{q}} & \TE{n}{p+n}{q-n+1}\\
\HE{n}{p}{q} \ar[u]^-{\ssh{n}{p}{q}} \ar[r]^-{\Hd{n}{p}{q}} & \HE{n}{p+n}{q-n+1} \ar[u]_-{\ssh{n}{p+n}{q-n+1}}
}
\end{align}
Thus, our task is to determine the homomorphisms $\ssh{n}{p}{q}$ for relevant values of $n$, $p$, and $q$.

The cohomology ring of $B\U(5)\times B\U(1)^{3}$ is written as
\begin{align}
    H^{\ast}(B\U(5)\times B\U(3);\bZ)\cong\bZ[\cfi{1},\dots,\cfi{5},v_{1},v_{2},v_{3}],
\end{align}
where $\cfi{j}\,(j=1,2,3,4,5)$ are the $j$-th Chern class of $\U(5)$, whereas $v_{k}\,(k=1,2,3)$ are the first Chern classes of $\U(1)^{3}$.
The expression of $H^{\ast}(\U(1)^{5}\times\U(1)^{3};\bZ)$ in terms of the generators $u_{j}\,(j=1,2,3,4,5)$ and $v_{k}\,(k=1,2,3)$ is exhibited in Eq.~\eqref{eq:cohom_T8}.
The induced homomorphism $\ssh{2}{p}{q}$ for the required values of $p$ and $q$ are given by
\begin{align}\begin{aligned}
    \ssh{2}{0}{2j}(\cfi{j})&=\sigma_{j}(u_{1},u_{2},u_{3},u_{4},u_{5})\quad(j=1,2,3),\\
    \ssh{2}{0}{2}(v_{k})&=v_{k}\quad(k=1,2,3),\\
    \ssh{2}{3}{0}(x_{1})&=x_{1}.
\end{aligned}\end{align}
As before, the convention $\sigma_{j}(u_{1},\dots,u_{5})$ stands for the $j$-th symmetric polynomial.
Since the differential maps $\Td{2}{p}{q}$ and $\Hd{2}{p}{q}$ are trivial, it holds that $\ssh{3}{p}{q}=\ssh{2}{p}{q}$.

\subsubsection{Differential maps \texorpdfstring{$\Hd{3}{p}{q}$}{}}
\label{subsubsec:Hd3pq}
Applying the commutativity of the diagram~\eqref{eq:com_diagram_h}, we can determine the differential maps $\Hd{3}{p}{q}$ in turn.
The kernel and the image of $\Hd{3}{p}{q}$ are also determined.
In the following, we summarize the results.

\begin{itemize}
\item $\Hd{3}{0}{2}:\HE{3}{0}{2}\to\HE{3}{3}{0}$\\
\begin{align}
    \Hd{3}{0}{2}(\cfi{1})&=25x_{1},&
    \Hd{3}{0}{2}(v_{k})&=-4x_{1}\quad(k=1,2,3).
\end{align}
\begin{align}\begin{aligned}
    &\Ker\Hd{3}{0}{2}\cong\bZ(4\cfi{1}+v_{1}+24v_{3})\oplus\bZ(4\cfi{1}+v_{2}+24v_{3})\oplus\bZ(4\cfi{1}+25v_{3}).\\
    &\Im\Hd{3}{0}{2}\cong \HE{3}{3}{0}.
\end{aligned}\end{align}

\item $\Hd{3}{0}{4}:\HE{3}{0}{4}\to\HE{3}{3}{2}$
\begin{align}\label{eq:Hd304}
    \Hd{3}{0}{4}(\cfi{2})=20x_{1}\cfi{1}.
\end{align}    
\begin{align}\begin{aligned}
    \Ker\Hd{3}{0}{4}\cong&
    \bZ(-5\cfi{2}+2(\cfi{1})^{2})\\
    &\oplus\bZ(8\cfi{2}+4\cfi{1}v_{1}+36\cfi{1}v_{3}+v_{1}^{2}+23v_{1}v_{3}+101v_{3}^{2})\\
    &\oplus\bZ(8\cfi{2}+4\cfi{1}v_{1}+4\cfi{1}v_{2}+32\cfi{1}v_{3}+v_{1}v_{2}+24v_{1}v_{3}+24v_{2}v_{3}+76v_{3}^{2})\\
    &\oplus\bZ(8\cfi{2}+4\cfi{1}v_{1}+36\cfi{1}v_{3}+25v_{1}v_{3}+100v_{3}^{2})\\
    &\oplus\bZ(8\cfi{2}+4\cfi{1}v_{2}+36\cfi{1}v_{3}+v_{2}^{2}+23v_{2}v_{3}+101v_{3}^{2})\\
    &\oplus\bZ(8\cfi{2}+4\cfi{1}v_{2}+36\cfi{1}v_{3}+25v_{2}v_{3}+100v_{3}^{2})\\
    &\oplus\bZ(8\cfi{2}+40\cfi{1}v_{3}+125v_{3}^{2}).
\end{aligned}\end{align}
\begin{align}\begin{aligned}
    \Im\Hd{3}{0}{4}\cong
    \bZ x_{1}v_{3}
    &\oplus\bZ x_{1}(-2\cfi{1}+v_{1}+v_{3})\\
    &\oplus\bZ x_{1}(-2\cfi{1}+v_{2}+v_{3})
    \oplus\bZ 2x_{1}(-\cfi{1}+v_{3}).
\end{aligned}\end{align}

\item $\Hd{3}{0}{6}:\HE{3}{0}{6}\to\HE{3}{3}{4}$\\
\begin{align}
        \Hd{3}{0}{6}(\cfi{3})=15x_{1}\cfi{2}.
\end{align}
\begin{align}\begin{aligned}
    \Ker\Hd{3}{0}{6}\cong&
    \bZ(-10\cfi{3}+2\cfi{2}\cfi{1}-20\cfi{2}v_{1}-5\cfi{2}v_{2}+8(\cfi{1})^{2}v_{1}+2(\cfi{1})^{2}v_{2})\\
    &\oplus\bZ(-10\cfi{3}+2\cfi{2}\cfi{1}-20\cfi{2}v_{1}-5\cfi{2}v_{3}+8(\cfi{1})^{2}v_{1}+2(\cfi{1})^{2}v_{3})\\
    &\oplus\bZ(25\cfi{3}-15\cfi{2}\cfi{1}+4(\cfi{1})^{3})\\
    &\oplus\bZ(-10\cfi{3}+2\cfi{2}\cfi{1}-25\cfi{2}v_{1}+10(\cfi{1})^{2}v_{1})\\
    &\oplus\bZ(32\cfi{3}+8\cfi{2}v_{1}+112\cfi{2}v_{3}+4\cfi{1}v_{1}^{2}+32\cfi{1}v_{1}v_{3}+264\cfi{1}v_{3}^{2}\\
    &\qquad+v_{1}^{3}+22v_{1}^{2}v_{3}+78v_{1}v_{3}^{2}+524v_{3}^{3})\\
    &\oplus\bZ(32\cfi{3}+8\cfi{2}v_{1}+8\cfi{2}v_{2}+104\cfi{2}v_{3}+4\cfi{1}v_{1}^{2}+4\cfi{1}v_{1}v_{2}\\
    &\qquad+28\cfi{1}v_{1}v_{3}+36\cfi{1}v_{2}v_{3}+228\cfi{1}v_{3}^{2}+v_{1}^{2}v_{2}+24v_{1}^{2}v_{3}+24v_{1}v_{2}v_{3}\\
    &\qquad+52v_{1}v_{3}^{2}+101v_{2}v_{3}^{2}+424v_{3}^{3})\\
    &\oplus\bZ(32\cfi{3}+8\cfi{2}v_{1}+112\cfi{2}v_{3}+4\cfi{1}v_{1}^{2}+32\cfi{1}v_{1}v_{3}+264\cfi{1}v_{3}^{2}+25v_{1}^{2}v_{3}\\
    &\qquad+75v_{1}v_{3}^{2}+525v_{3}^{3})\\
    &\oplus\bZ(32\cfi{3}+8\cfi{2}v_{1}+8\cfi{2}v_{2}+104\cfi{2}v_{3}+4\cfi{1}v_{1}v_{2}+36\cfi{1}v_{1}v_{3}\\
    &\qquad+4\cfi{1}v_{2}^{2}+28\cfi{1}v_{2}v_{3}+228\cfi{1}v_{3}^{2}+v_{1}v_{2}^{2}+23v_{1}v_{2}v_{3}+101v_{1}v_{3}^{2}\\
    &\qquad+24v_{2}^{2}v_{3}+52v_{2}v_{3}^{2}+424v_{3}^{3})\\
    &\oplus\bZ(32\cfi{3}+8\cfi{2}v_{1}+8\cfi{2}v_{2}+104\cfi{2}v_{3}+4\cfi{1}v_{1}v_{2}+36\cfi{1}v_{1}v_{3}\\
    &\qquad+36\cfi{1}v_{2}v_{3}+224\cfi{1}v_{3}^{2}+25v_{1}v_{2}v_{3}+100v_{1}v_{3}^{2}+100v_{2}v_{3}^{2}+400v_{3}^{3})\\
    &\oplus\bZ(32\cfi{3}+8\cfi{2}v_{1}+112\cfi{2}v_{3}+40\cfi{1}v_{1}v_{3}+260\cfi{1}v_{3}^{2}+125v_{1}v_{3}^{2}+500v_{3}^{3})\\
    &\oplus\bZ(32\cfi{3}+8\cfi{2}v_{2}+112\cfi{2}v_{3}+4\cfi{1}v_{2}^{2}+32\cfi{1}v_{2}v_{3}+264\cfi{1}v_{3}^{2}+v_{2}^{3}\\
    &\qquad+22v_{2}^{2}v_{3}+78v_{2}v_{3}^{2}+524v_{3}^{3})\\
    &\oplus\bZ(32\cfi{3}+8\cfi{2}v_{2}+112\cfi{2}v_{3}+4\cfi{1}v_{2}^{2}+32\cfi{1}v_{2}v_{3}+264\cfi{1}v_{3}^{2}\\
    &\qquad+25v_{2}^{2}v_{3}+75v_{2}v_{3}^{2}+525v_{3}^{3})\\
    &\oplus\bZ(32\cfi{3}+8\cfi{2}v_{2}+112\cfi{2}v_{3}+40\cfi{1}v_{2}v_{3}+260\cfi{1}v_{3}^{2}+125v_{2}v_{3}^{2}+500v_{3}^{3})\\
    &\oplus\bZ(32\cfi{3}+120\cfi{2}v_{3}+300\cfi{1}v_{3}^{2}+625v_{3}^{3}).
\end{aligned}\end{align}

\item $\Hd{3}{3}{2}:\HE{3}{3}{2}\to\HE{3}{6}{0}$\\
\begin{align}
    \Hd{3}{3}{2}(x_{1}\cfi{1})&=y_{2,0},&
    \Hd{3}{3}{2}(x_{1}v_{k})&=0\quad(k=1,2,3).
\end{align}
\begin{align}\begin{aligned}
    &\Ker\Hd{3}{3}{2}\cong\bZ 2x_{1}\cfi{1}\oplus\bZ x_{1}v_{1}\oplus\bZ x_{1}v_{2}\oplus\bZ x_{1}v_{3},\\
    &\Im\Hd{3}{3}{2}\cong \HE{3}{6}{0}.
\end{aligned}\end{align}
\end{itemize}

\subsubsection{Cohomology groups of \texorpdfstring{$BH$}{}}
We calculate the cohomology groups of $BH$ up to degree six by applying the Serre spectral sequence to the fibration~\eqref{eq:fibration_first_H}.
See Table~\ref{ss:cohom_BA4T3} for the situation of the $E_{3}$-page.
From the analysis in Section~\ref{subsubsec:Hd3pq}, the following results are immediately obtained:
\begin{itemize}
    \item $H^{1}(BH;\bZ)\cong 0$.
    \item $H^{2}(BH;\bZ)\cong \HE{\infty}{0}{2}\cong\HE{4}{0}{2}\cong\Ker\Hd{3}{0}{2}\cong\bZ^{\oplus 3}$.
    \item $H^{3}(BH;\bZ)\cong \HE{\infty}{3}{0}\cong\HE{4}{3}{0}\cong\HE{3}{3}{0}/\Im\Hd{3}{0}{2}\cong 0$.
    \item $H^{4}(BH;\bZ)\cong \HE{\infty}{0}{4}\cong\HE{4}{0}{4}\cong\Ker\Hd{3}{0}{4}\cong\bZ^{\oplus 7}$.
    \item $H^{5}(BH;\bZ)\cong \HE{\infty}{3}{2}\cong\HE{4}{3}{2}\cong\Ker\Hd{3}{3}{2}/\Im\Hd{3}{0}{4}\cong 0$.
    \item $H^{6}(BH;\bZ)\cong \HE{\infty}{0}{6}\cong\HE{4}{0}{6}\cong\Ker\Hd{3}{0}{6}\cong\bZ^{\oplus 14}$.
\end{itemize}
\begin{table}[thp]\begin{center}
	\begin{tabular}{ccc}
		\begin{tabular}{c|ccccccc}
			6 & $\bZ^{\oplus 25}$ & $\hphantom{\bZ_2}$ & $\hphantom{\bZ_2}$ & $\ast$ & $\hphantom{\bZ_2}$ & $\hphantom{\bZ_2}$ & $\ast$\\
			5 & & & & & & & \\
			4 & $\bZ^{\oplus 11}$ & & & $\bZ^{\oplus 11}$ & $\hphantom{\bZ_2}$ & & $\ast$ \\
			3 & & & & & & & \\
			2 & $\bZ^{\oplus 4}$ & & & $\bZ^{\oplus 4}$ & & & $\ast$ \\
			1 & & & & & & & \\
			0 & $\bZ$ & & & $\bZ$ & & & $\bZ_{2}$ \\
			\hline
			& 0 & 1 & 2 & 3 & 4 & 5 & 6
		\end{tabular}
		& $\quad \Longrightarrow$ &
		\begin{tabular}{c|c}
			6 & $\bZ^{\oplus 14}$ \\
			5 & \\
			4 & $\bZ^{\oplus 7}$ \\
			3 & \\
			2 & $\bZ^{\oplus 3}$ \\
			1 & \\
			0 & $\bZ$ \\
			\hline
			&
		\end{tabular}
	\end{tabular}
\caption{The $E_{3}$-page of the Serre spectral sequence associated with the fibration~\eqref{eq:fibration_first_H}.}
\label{ss:cohom_BA4T3}
\end{center}\end{table}

\subsection{Bordism group \texorpdfstring{$\Omega^{\Spin}_{5}(BH)$}{}}
\label{subsec:bordism_BH}
In order to determine the bordism group $\Omega^{\Spin}_{5}(BH)$, we apply the Atiyah-Hirzebruch spectral sequence to the fibration,
\begin{align}\label{eq:fibration_trivial_BH}
    \mathrm{pt} \to BH \to BH.
\end{align}
In particular, we consider the spectral sequence that converges to the reduced spin bordism, as before.
The homology groups required for the $E^{2}$-page are summarized in Table~\ref{tab:hom_cohom_BA4T3}.
\begin{table}[thp]\begin{center}
	\begin{tabular}{c||ccccccc}
		$p$ & $0$ & $1$ & $2$ & $3$ & $4$ & $5$ & $6$ \\
        \hline
        $H_{p}(BH;\bZ)$ & $\bZ$ & $0$ & $\bZ^{\oplus 3}$ & $0$ & $\bZ^{\oplus 7}$ & $0$ & $\bZ^{\oplus 14}$ \\
        $H^{p}(BH;\bZ)$ & $\bZ$ & $0$ & $\bZ^{\oplus 3}$ & $0$ & $\bZ^{\oplus 7}$ & $0$ & $\bZ^{\oplus 14}$ \\
        $H_{p}(BH;\bZ_{2})$ & $\bZ_{2}$ & $0$ & $\bZ_{2}^{\oplus 3}$ & $0$ & $\bZ_{2}^{\oplus 7}$ & $0$ & $\bZ_{2}^{\oplus 14}$ \\
        $H^{p}(BH;\bZ_{2})$ & $\bZ_{2}$ & $0$ & $\bZ_{2}^{\oplus 3}$ & $0$ & $\bZ_{2}^{\oplus 7}$ & $0$ & $\bZ_{2}^{\oplus 14}$ \\
	\end{tabular}
\caption{Homology and cohomology groups of $BH$ up to degree six.}
\label{tab:hom_cohom_BA4T3}
\end{center}\end{table}
In Table~\ref{ss:bordism_BA4T3}, the situation in the $E^{2}$-page is exhibited.
\begin{table}[thp]\begin{center}
	\begin{tabular}{c|ccccccc}
		5 & $\hphantom{\bZ_2}\cellcolor{lightyellow}$ & $\hphantom{\bZ_2}$ & $\hphantom{\bZ_2}$ & $\hphantom{\bZ_2}$ & $\hphantom{\bZ_2}$ & $\hphantom{\bZ_2}$ & $\hphantom{\bZ_2}$\\
		4 & & \cellcolor{lightyellow} & $\bZ^{\oplus 3}$ & & $\bZ^{\oplus 7}$ & & $\bZ^{\oplus 14}$ \\
		3 & & & \cellcolor{lightyellow} & & & & \\
		2 & & & $\bZ_{2}^{\oplus 3}$ & $\cellcolor{lightyellow}$ & $\bZ_{2}^{\oplus 7}$ & & $\bZ_{2}^{\oplus 14}$ \\
		1 & & & $\bZ_{2}^{\oplus 3}$ & & $\bZ_{2}^{\oplus 7}\cellcolor{lightyellow}$ & & $\bZ_{2}^{\oplus 14}$ \\
		0 & & & $\bZ^{\oplus 3}$ & & $\bZ^{\oplus 7}$ & $\cellcolor{lightyellow}$ & $\bZ^{\oplus 14}$ \\
		\hline
		& 0 & 1 & 2 & 3 & 4 & 5 & 6
	\end{tabular}
\caption{The $E_{2}$-page of the Atiyah-Hirzebruch spectral sequence for the spin bordism associated with the fibration~\eqref{eq:fibration_trivial_BH}.}
\label{ss:bordism_BA4T3}
\end{center}\end{table}

Similar to the calculation in Section~\ref{subsec:bordism_BG}, the kernel and the image of the differential maps $d^{2}_{4,1}$ and $d^{2}_{6,0}$ are determined by the action of the Steenrod square $\Sqtwo$ on the cohomology groups of $BH$.
The cohomology groups of $BH$ with $\bZ_{2}$ coefficients are given by
\begin{align}\begin{aligned}
    H^{2}(BH;\bZ_{2})&\cong\bigoplus_{k=1}^{3}\bZ_{2}v_{k},\\
    H^{4}(BH;\bZ_{2})&\cong\bZ_{2}\cfi{2}\oplus\bZ_{2}\bigoplus_{1\le k\le l\le 3}\bZ_{2}v_{k}v_{l},\\
    H^{6}(BH;\bZ_{2})&\cong\bZ_{2}(\cfi{3}+\cfi{2}\cfi{1})\oplus\bigoplus_{k=1}^{3}\cfi{2}v_{k}\oplus\bigoplus_{1\le k\le l\le m\le 3}\bZ_{2}v_{k}v_{l}v_{m}.
\end{aligned}\end{align}
The properties of $d^{2}_{4,1}$ and $d^{2}_{6,0}$ required for the calculation of $\Omega^{\Spin}_{5}(BH)$ are summarized as
\begin{align}\begin{aligned}
    \Ker d^{2}_{4,1}&\cong\bZ_{2}\widetilde{\cfi{2}}\oplus\bigoplus_{1\le k\le l\le 3}\bZ_{2}\widetilde{v_{k}v_{l}},&
    \Im d^{2}_{4,1}&\cong\bigoplus_{k=1}^{3}\bZ_{2}\widetilde{v_{k}},\\
    \Ker d^{2}_{6,0}&\cong\bigoplus_{k=1}^{3}\bZ_{2}\widetilde{\cfi{2}v_{k}}\oplus\bigoplus_{l=1}^{3}\bZ_{2}\widetilde{v_{k}^{3}},&
    \Im d^{2}_{6,0}&\cong\bZ_{2}\widetilde{\cfi{2}}\oplus\bigoplus_{1\le k< l\le 3}\bZ_{2}\widetilde{v_{k}v_{l}}.
\end{aligned}\end{align}
From these results, we have
\begin{align}
    \Omega^{\Spin}_{5}(BH)\cong\Ker d^{2}_{4,1}/\Im d^{2}_{6,0}\cong 0,
\end{align}
which implies that there is no global anomaly in the gauge theory in four dimensions with the structure group $H$.

\section{Topological data of \texorpdfstring{$E_{7}/H$}{}}
\label{sec:E7H}
We compute the bordism groups of $E_{7}/H$ up to degree five, following the same procedure as in Appendix~\ref{sec:E7G}.

\subsection{Cohomology ring \texorpdfstring{$H^{\ast}(E_{7}/H;\bZ)$}{} up to degree six}
\label{subsec:cohom_E7H}

We consider the fibration
\begin{align}
  H/T \xrightarrow{i_{H}} E_{7}/T \xrightarrow{p_{H}} E_{7}/H.
\end{align}
Recall that $T$ is a maximal torus of $E_{7}$.
This fibration induces ring homomorphisms,
\begin{align}
  &p_{H}^{\ast}: H^{\ast}(E_{7}/H;\bZ) \to H^{\ast}(E_{7}/T;\bZ),&
  &i_{H}^{\ast}: H^{\ast}(E_{7}/T;\bZ) \to H^{\ast}(H/T;\bZ).
\end{align}
As before, $p_{H}^{\ast}$ is injective and $i_{H}^{\ast}$ is surjective.
Moreover, we have
\begin{align}
\Ker i_{H}^{\ast}=(p_{H}^{\ast} H^{+}(E_{7}/H;\bZ)).
\end{align}
Therefore, to determine $H^{\ast}(E_{7}/H;\bZ)$, we proceed in three steps.
First, we identify the ideal $\Ker i_{H}^{\ast}$ inside $H^{\ast}(E_{7}/T;\bZ)$.
Second, we compute the rational cohomology ring $H^{\ast}(E_{7}/H;\bQ)$.
Finally, based on this information, we describe $H^{\ast}(E_{7}/H;\bZ)$ in terms of the generators.

\subsubsection{Kernel of \texorpdfstring{$i_{H}^{\ast}$}{}}

The cohomology ring of $H/T$ with integer coefficients is given by
\begin{align}
  H^{\ast}(H/T;\bZ)
  \cong
  \bZ[y_{1},y_{2},y_{3},y_{4},y_{5}]
  /(\cfi{1},\cfi{2},\cfi{3},\cfi{4},\cfi{5}),
\end{align}
where $y_{j}\in H^{2}(H/T;\bZ)$ and $\cfi{j}=\sigma_{j}(y_{1},\dots,y_{5})$.
For the cohomology ring $H^{\ast}(E_{7}/T)$, we use the notation introduced in Section~\ref{subsec:cohom_E7G}.
Under the map $i_{H}^{\ast}$, the generators of $H^{\ast}(E_{7}/T;\bZ)$ in low degrees are mapped as follows:
\begin{align}\begin{aligned}\label{eq:iH_t}
    i_{H}^{\ast}t_{1}&=-(y_{1}+y_{5}),&
    i_{H}^{\ast}t_{2}&=-(y_{2}+y_{5}),&
    i_{H}^{\ast}t_{3}&=-(y_{3}+y_{5}),&
    i_{H}^{\ast}t_{4}&=-(y_{4}+y_{5}),\\
    i_{H}^{\ast}t_{5}&=0,&
    i_{H}^{\ast}t_{6}&=0,&
    i_{H}^{\ast}t_{7}&=0,&&
\end{aligned}\end{align}
and
\begin{align}
    i_{H}^{\ast}\gamma_{1}&=-y_{5},&
    i_{H}^{\ast}\gamma_{3}&=-y_{5}^{3}.
\end{align}
From these results, we conclude that
\begin{align}\label{eq:ker_iH}
    \Ker i_{H}^{\ast}=(t_{5},t_{6},t_{7},\gamma_{3}-\gamma_{1}^{3}),
\end{align}
up to degree six.

\subsubsection{Cohomology ring \texorpdfstring{$H^{\ast}(E_{7}/H;\bQ)$}{}}
We determine the cohomology ring of $E_{7}/G$ with rational coefficient by the formula~\cite{MR102800},
\begin{align}\label{eq:rational_cohom_formula_H}
    H^{\ast}(E_{7}/H;\bQ)\cong H^{\ast}(BH;\bQ)/(H^{+}(BE_{7};\bQ)).
\end{align}
Since the ring $H^{\ast}(BH;\bQ)$ is identified with the subring of $H^{\ast}(BT;\bQ)$ consisting of invariants under the Weyl group of $\h$, it can be written as
\begin{align}
    H^{\ast}(BH;\bQ)\cong\bQ[t_{5},t_{6},t_{7},\cpfi{1},\cpfi{2},\cpfi{3},\cpfi{4},\cpfi{5}]/(\cpfi{1}),
\end{align}
in terms of the notations introduced in Section~\ref{subsec:cohom_E7G}.
Then, by applying the formula~\eqref{eq:rational_cohom_formula_H}, we obtain the expression,
\begin{align}\label{eq:cohom_E7H_Q_mid}
    H^{\ast}(E_{7}/H;\bQ)\cong\bQ[t_{5},t_{6},t_{7},\cpfi{1},\cpfi{2},\cpfi{3},\cpfi{4},\cpfi{5}]/(\cpfi{1},I_{2}),
\end{align}
up to degree six.
In terms of the generators~\eqref{eq:a_generators}, the ring is rewritten as
\begin{align}\begin{aligned}\label{eq:cohom_E7H_Q}
    H^{\ast}(E_{7}/H;\bQ)
    &\cong\bQ[t_{5},t_{6},t_{7},\afi{1},\afi{2},\afi{3}]/(\afi{1},\sigma_{1}(t_{5},t_{6},t_{7})^{2}-\afi{2}-\sigma_{2}(t_{5},t_{6},t_{7}))\\
    &\cong\bQ[t_{5},t_{6},t_{7},\afi{3}].
\end{aligned}\end{align}

\subsubsection{Generators of cohomology ring \texorpdfstring{$H^{\ast}(E_{7}/H;\bZ)$}{}}
Now, we can describe the ring $H^{\ast}(E_{7}/T;\bZ)$ up to degree six in terms of the generators.
As before, we can regard $H^{\ast}(E_{7}/H;\bZ)$ as the lattice in $H^{\ast}(E_{7}/H;\bQ)$.
By comparing the expressions~\eqref{eq:ker_iH} and~\eqref{eq:cohom_E7H_Q_mid}, we can take $t_{5}$, $t_{6}$, and $t_{7}$ as the generators of $H^{\ast}(E_{7}/H;\bZ)$.
The remaining generator in degree six is chosen as
\begin{align}\begin{aligned}
    \acfi{3}
    &=\frac{1}{2}\afi{3}+\frac{1}{2}\sigma_{3}(t_{5},t_{6},t_{7})-\sigma_{2}(t_{5},t_{6},t_{7})\sigma_{1}(t_{5},t_{6},t_{7})+\frac{1}{2}\sigma_{1}(t_{5},t_{6},t_{7})^{3}
\end{aligned}\end{align}
Then, the cohomology ring of $E_{7}/H$ with integer coefficients up to degree six is given by
\begin{align}\label{eq:cohom_E7H}
    H^{\ast}(E_{7}/H;\bZ)\cong\bZ[t_{5},t_{6},t_{7},\acfi{3}].
\end{align}

\subsection{Bordism groups \texorpdfstring{$\Omega^{\Spin}_{\ast}(E_{7}/H)$}{} up to degree five}
\label{subsec:bordism_E7H}

In order to compute $\Omega^{\Spin}_{\ast}(E_{7}/H)$, we apply the Atiyah-Hirzebruch spectral sequence to the fibration,
\begin{align}\label{eq:fibration_trivial_E7_A4T3}
    \mathrm{pt}\to E_{7}/H\to E_{7}/H.
\end{align}
The $E^{2}$-page of the Atiyah-Hirzebruch spectral sequence that converges to $\widetilde{\Omega}^{\Spin}_{\ast}(E_{7}/H)$ is given by
\begin{align}
    E^{2}_{p,q}=\widetilde{H}_{p}(E_{7}/H;\Omega^{\Spin}_{q}(\mathrm{pt})).
\end{align}
The homology groups required for the calculation are summarized in Table~\ref{tab:hom_cohom_E7_A4T3}.
The situation of the $E^{2}$-page is shown in Table~\ref{ss:bordism_E7_A4T3}.
\begin{table}[thp]\begin{center}
	\begin{tabular}{c||ccccccc}
		$p$ & $0$ & $1$ & $2$ & $3$ & $4$ & $5$ & $6$ \\
        \hline
        $H_{p}(BH;\bZ)$ & $\bZ$ & $0$ & $\bZ^{\oplus 3}$ & $0$ & $\bZ^{\oplus 6}$ & $0$ & $\bZ^{\oplus 11}$ \\
        $H^{p}(BH;\bZ)$ & $\bZ$ & $0$ & $\bZ^{\oplus 3}$ & $0$ & $\bZ^{\oplus 6}$ & $0$ & $\bZ^{\oplus 11}$ \\
        $H_{p}(BH;\bZ_{2})$ & $\bZ_{2}$ & $0$ & $\bZ_{2}^{\oplus 3}$ & $0$ & $\bZ_{2}^{\oplus 6}$ & $0$ & $\bZ_{2}^{\oplus 11}$ \\
        $H^{p}(BH;\bZ_{2})$ & $\bZ_{2}$ & $0$ & $\bZ_{2}^{\oplus 3}$ & $0$ & $\bZ_{2}^{\oplus 6}$ & $0$ & $\bZ_{2}^{\oplus 11}$ \\
	\end{tabular}
\caption{Homology and cohomology groups of $E_{7}/H$ up to degree six.}
\label{tab:hom_cohom_E7_A4T3}
\end{center}\end{table}
\begin{table}[thp]\begin{center}
	\begin{tabular}{c|ccccccc}
		5 & $\hphantom{\bZ_2}$ & $\hphantom{\bZ_2}$ & $\hphantom{\bZ_2}$ & $\hphantom{\bZ_2}$ & $\hphantom{\bZ_2}$ & $\hphantom{\bZ_2}$ & $\hphantom{\bZ_2}$\\
		4 & & & $\bZ^{\oplus 3}$ & & $\bZ^{\oplus 6}$ & & $\bZ^{\oplus 11}$ \\
		3 & & & & & & & \\
		2 & & & $\bZ_{2}^{\oplus 3}$ & & $\bZ_{2}^{\oplus 6}$ & & $\bZ_{2}^{\oplus 11}$ \\
		1 & & & $\bZ_{2}^{\oplus 3}$ & & $\bZ_{2}^{\oplus 6}$ & & $\bZ_{2}^{\oplus 11}$ \\
		0 & & & $\bZ^{\oplus 3}$ & & $\bZ^{\oplus 6}$ & & $\bZ^{\oplus 11}$ \\
		\hline
		& 0 & 1 & 2 & 3 & 4 & 5 & 6
	\end{tabular}
\caption{The $E^{2}$-page of the Atiyah-Hirzebruch spectral sequence for the spin bordism associated with the fibration~\eqref{eq:fibration_trivial_E7_A4T3}.}
\label{ss:bordism_E7_A4T3}
\end{center}\end{table}

From the analysis in Section~\ref{subsubsec:Steenrod_E7G}, the properties of the differential maps $d^{2}_{4,0}$ and $d^{2}_{4,1}$ in the $E^{2}$-page are determined as
\begin{align}
    \Ker d^{2}_{4,0}&\cong\bigoplus_{5\le k<l\le 7}\bZ\widetilde{t_{k}t_{l}}\oplus\bigoplus_{m=5}^{7}\bZ 2\widetilde{t_{m}^{2}},&
    \Im d^{2}_{4,0}&\cong\bigoplus_{k=5}^{7}\bZ_{2}\widetilde{t_{k}},\\
    \Ker d^{2}_{4,1}&\cong\bigoplus_{5\le k<l\le 7}\bZ_{2}\widetilde{t_{k}t_{l}},&
    \Im d^{2}_{4,1}&\cong\bigoplus_{k=5}^{7}\bZ_{2}\widetilde{t_{k}}.
\end{align}
Similarly, the image of the $d^{2}_{6,0}$ is given by
\begin{align}
    \Im d^{2}_{6,0}&\cong\bigoplus_{5\le k<l\le 7}\bZ_{2}\widetilde{t_{k}t_{l}}.
\end{align}
Consequently, the reduced bordism groups are calculated as
\begin{align}\begin{aligned}
    \widetilde{\Omega}^{\Spin}_{1}(E_{7}/H)&\cong 0,\\
    \widetilde{\Omega}^{\Spin}_{2}(E_{7}/H)&\cong E^{2}_{2,0}\cong\bZ^{\oplus 3},\\
    \widetilde{\Omega}^{\Spin}_{3}(E_{7}/H)&\cong E^{2}_{2,1}/\Im d^{2}_{4,0}\cong 0,\\
    \widetilde{\Omega}^{\Spin}_{4}(E_{7}/H)&\cong \Ker d^{2}_{4,0}\cong\bZ^{\oplus 6},\\
    \widetilde{\Omega}^{\Spin}_{5}(E_{7}/H)&\cong \Ker d^{2}_{4,1}/\Im d^{2}_{6,0}\cong 0.
\end{aligned}\end{align}
In summary, we exhibit the spin bordism groups of $E_{7}/G$ in Table~\ref{tab:bordism_E7_A4T3}.
\begin{table}[thp]\begin{center}
	\begin{tabular}{c||cccccc}
		$q$ & $0$ & $1$ & $2$ & $3$ & $4$ & $5$ \\
        \hline
        $\Omega^{\Spin}_{q}(E_{7}/H)$ & $\bZ$ & $\bZ_{2}$ & $\bZ^{\oplus 3}\oplus\bZ_{2}$ & $0$ & $\bZ^{\oplus 7}$ & $0$ \\
        $\widetilde{\Omega}^{\Spin}_{q}(E_{7}/H)$ & $0$ & $0$ & $\bZ^{\oplus 3}$ & $0$ & $\bZ^{\oplus 6}$ & $0$ \\
	\end{tabular}
\caption{Spin bordism groups of $E_{7}/H$ up to degree six.}
\label{tab:bordism_E7_A4T3}
\end{center}\end{table}

\section{Bordism group of \texorpdfstring{$(E_{7}/H)_{H}$}{}}
\label{sec:Borel_E7H}
We calculate the fifth spin bordism group of $(E_{7}/H)=EG\times_{H}E_{7}/H$.
From the construction, the topological space $(E_{7}/H)_{H}$ fits the fibration,
\begin{align}\label{eq:fibration_E7_H_BH}
    E_{7}/H\to (E_{7}/H)_{H}\to BH.
\end{align}
We apply the  Atiyah-Hirzebruch spectral sequence to the fibration~\eqref{eq:fibration_E7_H_BH}.
The $E^{2}$-page is given by
\begin{align}
    \DE{2}{p}{q}=H_{p}(BH;\Omega^{\Spin}_{q}(E_{7}/H)),
\end{align}
and the situation is illustrated in Table~\ref{ss:bordism_E7HH}.
\begin{table}[thp]\begin{center}
	\begin{tabular}{c|ccccccc}
		5 & $\hphantom{\bZ_2}\cellcolor{lightyellow}$ & $\hphantom{\bZ\oplus\bZ_2}$ & $\hphantom{\bZ_2}$ & $\hphantom{\bZ\oplus\bZ_2}$ & $\hphantom{\bZ_2}$ & $\hphantom{\bZ\oplus\bZ_2}$ & $\hphantom{\bZ_2}$\\
		4 & $\bZ^{\oplus 7}$ & \cellcolor{lightyellow} & $\bZ^{\oplus 21}$ & & $\bZ^{\oplus 49}$ & & $\bZ^{\oplus 98}$ \\
		3 & & & \cellcolor{lightyellow} & & & & \\
		2 & $\bZ^{\oplus 3}\oplus\bZ_{2}$ & & $\bZ^{\oplus 9}\oplus\bZ_{2}^{\oplus 3}$ & $\cellcolor{lightyellow}$ & $\bZ^{\oplus 21}\oplus\bZ_{2}^{\oplus 7}$ & & $\bZ^{\oplus 42}\oplus\bZ_{2}^{\oplus 14}$ \\
		1 & $\bZ_{2}$ & & $\bZ_{2}^{\oplus 3}$ & & $\bZ_{2}^{\oplus 7}\cellcolor{lightyellow}$ & & $\bZ_{2}^{\oplus 14}$ \\
		0 & $\bZ$ & & $\bZ^{\oplus 3}$ & & $\bZ^{\oplus 7}$ & $\cellcolor{lightyellow}$ & $\bZ^{\oplus 14}$ \\
		\hline
		& 0 & 1 & 2 & 3 & 4 & 5 & 6
	\end{tabular}
\caption{The $E^{2}$-page of the Atiyah-Hirzebruch spectral sequence for the spin bordism associated with the fibration~\eqref{eq:fibration_E7_G_BG}.}
\label{ss:bordism_E7HH}
\end{center}\end{table}

We consider the following commutative diagram that is analogous to the diagram~\eqref{eq:com_diagram_bordism}:
\begin{align}\label{eq:com_diagram_bordism_H}
\xymatrix{
\mathrm{pt} \ar[d]^{}  \ar[r]^{} & EH\times_{H}\mathrm{pt} \ar[d]^{} \ar[r]^{} & BH \ar[d]^{\rotatebox{-90}{$\cong$}}\\
E_{7}/H \ar[r]^{} & EH\times_{H}E_{7}/H \ar[r] & BH
}
\end{align}
The $E^{2}$-page associated with the first row in the diagram that converges to $\Omega^{\Spin}_{\ast}(BH)$ is referred to as
\begin{align}
    \QE{2}{p}{q}=H_{p}(BH;\Omega^{\Spin}_{q}(\mathrm{pt})),
\end{align}
and the differential maps are denoted by $\Qd{2}{p}{q}:\QE{2}{p}{q}\to\QE{2}{p-2}{q+1}$.
The vertical maps in the diagram~\eqref{eq:com_diagram_bordism_H} induce the homomorphisms $\QE{2}{p}{q}\to\DE{2}{p}{q}$ such that the following diagram commutes:
\begin{align}\label{eq:com_diagram_induced_bordism_H}
\xymatrix{
\QE{2}{p}{q} \ar[r]^-{\Qd{n}{p}{q}}  \ar[d]^-{} & \QE{2}{p-2}{q+1} \ar[d]_-{}\\
\DE{2}{p}{q} \ar[r]^-{\Dd{n}{p}{q}} & \DE{2}{p-2}{q+1}
}
\end{align}
Here, we denote by $\Dd{2}{p}{q}$ the differential map in the $E^{2}$-page of the spectral sequence associated with the fibration~\eqref{eq:fibration_E7_H_BH}.
By the same argument in Appendix~\ref{sec:Borel_E7G}, we can determine the differential maps $\Dd{2}{4}{1}$ and $\Dd{2}{6}{1}$.
As a result, $\Dd{2}{6}{0}$ is identified with $\Qd{2}{6}{0}$, whereas $\Dd{2}{4}{1}$ is the composition of the map $\Qd{2}{4}{1}$ with the inclusion,
\begin{align}
H_{2}(BH;\Omega^{\Spin}_{2}({\rm pt}))\to H_{2}(BH;\Omega^{\Spin}_{2}(E_{7}/H)).
\end{align}
Finally, we can conclude that
\begin{align}
    \Omega^{\Spin}_{5}((E_{7}/H)_{H})\cong \Ker\Dd{2}{4}{1}/\Im\Dd{2}{6}{0}\cong 0.
\end{align}
This result implies that there is no global anomaly in the sigma model with the target space~$E_{7}/H$ after gauging the left $H$-action.

\bibliographystyle{ytphys}
\bibliography{Unification}
\end{document}